\documentclass[notitlepage,a4paper,aps,prd,onecolumn,superscriptaddress,nofootinbib,groupedaddress]{revtex4}

\usepackage{amsmath,comment}
\usepackage{amsfonts,color}
\usepackage{amsmath}
\usepackage{amsthm}
\usepackage{pdfpages}
\usepackage{amsmath}
\usepackage{empheq}
\usepackage{amsfonts,color}
\usepackage{amssymb,float}
\usepackage{physics}
\usepackage{booktabs,multirow}
\usepackage{titlesec}

\titleformat{\paragraph}
{\normalfont\normalsize\bfseries}{\theparagraph}{1em}{}
\titlespacing*{\paragraph}
{0pt}{3.25ex plus 1ex minus .2ex}{1.5ex plus .2ex}

\usepackage{amsmath}
\usepackage[utf8]{inputenc}
\allowdisplaybreaks
\usepackage{color}
\usepackage{hyperref}
\hypersetup{
    colorlinks=true,
    linkcolor=blue,
    filecolor=magenta,      
   citecolor=blue
}

\usepackage{enumitem}

\usepackage{tikz}
\usetikzlibrary{shapes.geometric}
\usetikzlibrary{arrows.meta,arrows}

\begin{document}
\title{Gravitational waves in Cubic Metric-Affine Gravity}

\author{Sebastian Bahamonde}
\email{sbahamondebeltran@gmail.com}
\affiliation{Cosmology, Gravity, and Astroparticle Physics Group, Center for Theoretical Physics of the Universe,
Institute for Basic Science (IBS), Daejeon 34126, Korea.}
\affiliation{Kavli Institute for the Physics and Mathematics of the Universe (WPI), The University of Tokyo Institutes
for Advanced Study (UTIAS), The University of Tokyo, Kashiwa, Chiba 277-8583, Japan.}

\author{Jorge Gigante Valcarcel}
\email{jorgevalcarcel@ibs.re.kr}
\affiliation{Center for Geometry and Physics, Institute for Basic Science (IBS), Pohang 37673, Korea.}

\author{José M.M. Senovilla}
\affiliation{Departamento de Física, Facultad de Ciencia y Tecnología, Universidad del País Vasco UPV/EHU, Apartado 644, 48080 Bilbao, Spain.}
\affiliation{EHU Quantum Center, Universidad del País Vasco UPV/EHU, Bilbao, Spain.}
\begin{abstract}

We derive new exact gravitational wave solutions with dynamical torsion and nonmetricity tensors in the framework of cubic Metric-Affine Gravity (MAG). For this purpose, we consider the full algebraic classification of the gravitational field in general metric-affine geometries and impose a set of Type N conditions on the field strength tensors that implement the kinetics of torsion and nonmetricity in a particular cubic MAG model, recently considered to eliminate ghostly instabilities from the vector and axial sectors of the theory. The new solutions represent pp-waves characterised by a metric function that includes the dynamical contributions of the torsion and nonmetricity tensors provided by the field equations of the model. In particular, these quantities induce a scalar polarisation mode in the gravitational-wave spectrum, thus offering a distinctive phenomenological signature beyond the ordinary tensor polarisation modes of General Relativity.

\end{abstract}

\maketitle

\section{Introduction}

Metric-Affine Gravity (MAG) constitutes a natural extension of General Relativity (GR) that includes, on top of the curvature tensor, an enriched space-time geometry with torsion and nonmetricity. In fact, these two quantities generally arise when considering a gauge characterisation of the space-time geometry, which leads to a gravitational action containing the Einstein-Hilbert term and a large number of gauge invariants that introduce their dynamics~\cite{Hehl:1976kj,Obukhov:1987tz,Hehl:1994ue,Gronwald:1995em,Blagojevic:2013xpa,ponomarev2017gauge,Cabral:2020fax,obukhov2023poincare}. In particular, at least the presence of quadratic curvature invariants in the action is needed to allow the propagation of these fields, which at the same time generally requires further constraints to avoid different types of instabilities~\cite{Neville:1978bk,Sezgin:1979zf,Sezgin:1981xs,Miyamoto:1983bf,Fukui:1984gn,Fukuma:1984cz,Battiti:1985mu,Kuhfuss:1986rb,Blagojevic:1986dm,Baikov:1992uh,Yo:1999ex,Yo:2001sy,Lin:2018awc,BeltranJimenez:2019acz,Jimenez:2019qjc,Percacci:2020ddy,BeltranJimenez:2020sqf,Lin:2020phk,Marzo:2021iok,Baldazzi:2021kaf,Jimenez-Cano:2022sds,Barker:2024dhb,Marzo:2024pyn,Bahamonde:2024sqo,Bahamonde:2024efl}.

Thereby, it is clear that the gravitational interaction enhanced by torsion and nonmetricity will be mediated by waves that in general can transmit their dynamical effects. This possibility has motivated a variety of theoretical studies on the gravitational wave solutions of quadratic MAG (e.g. see~\cite{adamowicz1980plane,Babourova:1998ct,Garcia:2000yi,Obukhov:2006gy,Blagojevic:2017wzf,Obukhov:2017pxa,Blagojevic:2017ssv,Jimenez-Cano:2020lea,Jimenez-Cano:2022arz,Khodadi:2026zoi} and references therein), which provide generalisations of the gravitational waves predicted by GR~\cite{Einstein:1937qu,misner1973gravitation,Stephani:2003tm,Griffiths:2009dfa}.

Following these lines, in this work we perform a comprehensive study on the gravitational waves described by a cubic MAG model, which has been recently considered to eliminate fundamental instabilities present in the vector and axial sectors of quadratic MAG~\cite{Bahamonde:2024sqo,Bahamonde:2024efl}. Indeed, it is reasonable to expect that the corresponding interactions offering a safer context for the propagation of the torsion and nonmetricity fields of cubic MAG may also provide an interesting phenomenology to the spectrum of gravitational waves, in comparison with the respective solutions of the quadratic actions. 

This paper is organised as follows. In Sec.~\ref{sec:defs}, we introduce the main definitions and conventions for the curvature, torsion and nonmetricity tensors defined in metric-affine geometry, as well as the cubic MAG model under consideration. With these ingredients, in Sec.~\ref{sec:wave_settings} we set up the gravitational wave profile that will form the basis of our investigation. Specifically, we consider the Brinkmann form for the metric tensor describing a pp-wave in a flat transverse space, alongside Type N algebraic conditions that can be straightforwardly imposed by taking into account the full algebraic classification of the gravitational field in metric-affine geometry~\cite{Bahamonde:2023piz,Bahamonde:2024svi}. Then, in Sec.~\ref{sec:sol} we apply these conditions for our model, which significantly reduces the complexity of the field equations and allows us to find new exact gravitational wave solutions with dynamical torsion and nonmetricity. For the sake of clarity in the presentation, we solve first the field equations in Sec.~\ref{subsecRC} for the simplest case, given by Riemann-Cartan geometry. The first extension of the solutions is then carried out in Sec.~\ref{subsecWC} for the case of Weyl-Cartan geometry, while in Sec.~\ref{subsecGeneralMAG} we complete our analysis by focusing on the general metric-affine geometry, highlighting the most significant differences with respect to the standard case of GR. Finally, we present the main conclusions of our work in Sec.~\ref{sec:conclusions}, while some technical details and cumbersome expressions are relegated to the appendices.

We work in natural units $c=G=1$ and consider the metric signature $(+,-,-,-)$. On the other hand, we use a tilde accent to denote those quantities that are defined from the general affine connection, in contrast to their unaccented counterparts constructed from the Levi-Civita connection. In addition, we denote with a diagonal arrow traceless and pseudotraceless pieces of tensors (e.g. ${\nearrow\!\!\!\!\!\!\!Q}\,^{\lambda}{}_{\mu\nu}$ and ${\nearrow\!\!\!\!\!\!\!\tilde{R}}\,^{\lambda}{}_{[\rho\mu\nu]}$). Latin and Greek indices run from $0$ to $3$, referring to anholonomic and coordinate bases, respectively.

\section{Definitions and conventions}\label{sec:defs}

The description of gravity within an affinely connected metric space-time introduces the antisymmetric part of the affine connection and the covariant derivative of the metric tensor as additional properties of the gravitational field:
\begin{equation}
    T^{\lambda}\,_{\mu \nu}=2\tilde{\Gamma}^{\lambda}\,_{[\mu \nu]}\,, \quad Q_{\lambda \mu \nu}=\tilde{\nabla}_{\lambda}g_{\mu \nu}\,,
\end{equation}
which generalise the covariant derivative acting on an arbitrary vector $v^{\lambda}$ as
\begin{equation}
\tilde{\nabla}_{\mu}v^{\lambda}=\nabla_{\mu}v^{\lambda}+N^{\lambda}\,_{\rho\mu}v^{\rho}\,,
\end{equation}
with
\begin{equation}
    N^{\lambda}\,_{\rho\mu}=\frac{1}{2}\left(T^{\lambda}\,_{\rho \mu}-T_{\rho}\,^{\lambda}\,_{\mu}-T_{\mu}\,^{\lambda}\,_{\rho}\right)+\frac{1}{2}\left(Q^{\lambda}\,_{\rho \mu}-Q_{\rho}\,^{\lambda}\,_{\mu}-Q_{\mu}\,^{\lambda}\,_{\rho}\right).
\end{equation}

The corresponding curvature tensor can then be expressed as the sum of the Riemann tensor and further post-Riemannian corrections
\begin{equation}\label{totalcurvature}
\tilde{R}^{\lambda}\,_{\rho\mu\nu}=R^{\lambda}\,_{\rho\mu\nu}+\nabla_{\mu}N^{\lambda}\,_{\rho \nu}-\nabla_{\nu}N^{\lambda}\,_{\rho \mu}+N^{\lambda}\,_{\sigma \mu}N^{\sigma}\,_{\rho \nu}-N^{\lambda}\,_{\sigma \nu}N^{\sigma}\,_{\rho \mu}\,,
\end{equation}
whose algebraic symmetries allow the definition of three independent traces; namely, the Ricci, co-Ricci tensors
\begin{align}\label{Riccitensor}
\tilde{R}_{\mu\nu}&=\tilde{R}^{\lambda}{}_{\mu \lambda \nu}\,,\\
\label{co-Riccitensor}
\hat{R}_{\mu\nu}&=\tilde{R}_{\mu}{}^{\lambda}{}_{\nu\lambda}\,,
\end{align}
as well as the homothetic curvature tensor
\begin{equation}
    \bar{R}_{\mu\nu}=\tilde{R}^{\lambda}{}_{\lambda\mu\nu}\,.
\end{equation}
Furthermore, the trace of the Ricci and co-Ricci tensors provides a unique independent scalar curvature
\begin{equation}
    \tilde{R}=g^{\mu\nu}\tilde{R}_{\mu\nu}\,,
\end{equation}
whereas the pseudotrace of the curvature tensor gives rise to the so-called Holst pseudoscalar
\begin{equation}
\ast\tilde{R}=\varepsilon^{\lambda\rho\mu\nu}\tilde{R}_{\lambda\rho\mu\nu}\,.
\end{equation}

For stability purposes and other phenomenological aspects, it is important to consider the irreducible decomposition of these tensors under the four-dimensional pseudo-orthogonal group~\cite{McCrea:1992wa}. First, for the torsion tensor, the aforementioned decomposition includes vector, axial and tensor modes as
\begin{equation}
    T^{\lambda}\,_{\mu \nu}=\frac{1}{3}\left(\delta^{\lambda}\,_{\nu}T_{\mu}-\delta^{\lambda}\,_{\mu}T_{\nu}\right)+\frac{1}{6}\,\varepsilon^{\lambda}\,_{\rho\mu\nu}S^{\rho}+t^{\lambda}\,_{\mu \nu}\,,
\end{equation}
where
\begin{align}\label{Tdec1}
    T_{\mu}&=T^{\nu}\,_{\mu\nu}\,,\\
    S_{\mu}&=\varepsilon_{\mu\lambda\rho\nu}T^{\lambda\rho\nu}\,,\\
    t_{\lambda\mu\nu}&=T_{\lambda\mu\nu}-\frac{2}{3}g_{\lambda[\nu}T_{\mu]}-\frac{1}{6}\,\varepsilon_{\lambda\rho\mu\nu}S^{\rho}\,,\label{Tdec3}
\end{align}
while the nonmetricity tensor can be separated into trace and traceless components
\begin{equation}
    Q_{\lambda\mu\nu}=g_{\mu\nu}W_{\lambda}+{\nearrow\!\!\!\!\!\!\!Q}_{\lambda\mu\nu}\,,
\end{equation}
with ${\nearrow\!\!\!\!\!\!\!Q}_{\lambda\mu\nu}=g_{\lambda(\mu}\Lambda_{\nu)}-\frac{1}{4}g_{\mu\nu}\Lambda_{\lambda}+\frac{1}{3}\varepsilon_{\lambda\rho\sigma(\mu}\Omega_{\nu)}\,^{\rho\sigma}+q_{\lambda\mu\nu}$, thus displaying further vector and tensor modes
\begin{align}
    W_{\mu}&=\frac{1}{4}\,Q_{\mu\nu}\,^{\nu}\,,\label{Qdec1}\\
    \Lambda_{\mu}&=\frac{4}{9}\left(Q^{\nu}\,_{\mu\nu}-W_{\mu}\right),\label{Lambdavector}\\
    \Omega_{\lambda}\,^{\mu\nu}&=-\left[\varepsilon^{\mu\nu\rho\sigma}Q_{\rho\sigma\lambda}+\varepsilon^{\mu\nu\rho}\,_{\lambda}\left(\frac{3}{4}\Lambda_{\rho}-W_{\rho}\right)\right],\label{Omegatensor}\\
    q_{\lambda\mu\nu}&=Q_{(\lambda\mu\nu)}-g_{(\mu\nu}W_{\lambda)}-\frac{3}{4}g_{(\mu\nu}\Lambda_{\lambda)}\,.\label{qtensor}
\end{align}

On the other hand, the respective irreducible parts of the curvature tensor can be grouped into its antisymmetric and symmetric components:
\begin{equation}
    \tilde{W}_{\lambda\rho\mu\nu}:=\tilde{R}_{[\lambda\rho]\mu\nu}\,, \quad \tilde{Z}_{\lambda\rho\mu\nu}:=\tilde{R}_{(\lambda\rho)\mu\nu}\,,
\end{equation}
displaying six irreducible parts $\tilde{W}_{\lambda\rho\mu\nu}=\displaystyle\sum_{i=1}^{6}{}^{(i)}\tilde{W}_{\lambda\rho\mu\nu}$ in the antisymmetric component:
\begin{eqnarray}
{}^{(1)}\tilde{W}_{\lambda\rho\mu\nu}&=&\tilde{W}_{\lambda\rho\mu\nu}-\sum_{i=2}^{6}{}^{(i)}\tilde{W}_{\lambda\rho\mu\nu}\,,\label{first_irreducible_antisymmetric_piece}\\
{}^{(2)}\tilde{W}_{\lambda\rho\mu\nu}&=&\frac{3}{4}\Big({\nearrow\!\!\!\!\!\!\!\tilde{R}}^{(T)}_{\lambda[\rho\mu\nu]}+{\nearrow\!\!\!\!\!\!\!\tilde{R}}^{(T)}_{\nu[\lambda\rho\mu]}-{\nearrow\!\!\!\!\!\!\!\tilde{R}}^{(T)}_{\rho[\lambda\mu\nu]}-{\nearrow\!\!\!\!\!\!\!\tilde{R}}^{(T)}_{\mu[\lambda\rho\nu]}\Bigr)+\frac{1}{2}\Bigl({\nearrow\!\!\!\!\!\!\!\tilde{R}}^{(Q)}_{\mu[\lambda\rho\nu]}-{\nearrow\!\!\!\!\!\!\!\tilde{R}}^{(Q)}_{\nu[\lambda\rho\mu]}\Big)\,,\\
    {}^{(3)}\tilde{W}_{\lambda\rho\mu\nu}&=&-\,\frac{1}{24}\left.\ast\tilde{R}\right. \varepsilon_{\lambda\rho\mu\nu}\,,\\
     {}^{(4)}\tilde{W}_{\lambda\rho\mu\nu}&=&\frac{1}{4}\Big[g_{\lambda\mu}\left(2{\nearrow\!\!\!\!\!\!\!\tilde{R}}_{(\rho\nu)}+{\nearrow\!\!\!\!\!\!\!\hat{R}}^{(Q)}_{(\rho \nu) }\right)+g_{\rho\nu}\left(2{\nearrow\!\!\!\!\!\!\!\tilde{R}}_{(\lambda\mu)} +{\nearrow\!\!\!\!\!\!\!\hat{R}}^{(Q)}_{(\lambda\mu)}\right)-g_{\lambda\nu}\left(2 {\nearrow\!\!\!\!\!\!\!\tilde{R}}_{(\rho\mu)}+{\nearrow\!\!\!\!\!\!\!\hat{R}}^{(Q)}_{(\rho\mu)}\right)-g_{\rho\mu}\left(2{\nearrow\!\!\!\!\!\!\!\tilde{R}}_{(\lambda\nu)}+{\nearrow\!\!\!\!\!\!\!\hat{R}}^{(Q)}_{(\lambda\nu)}\right)\Big]\,,\nonumber \\ \\
{}^{(5)}\tilde{W}_{\lambda\rho\mu\nu}&=&\frac{1}{4}\Big[g_{\lambda\mu}\Bigl(2\tilde{R}^{(T)}_{[\rho\nu]}+\hat{R}^{(Q)}_{[\rho\nu]}\Bigr)+g_{\rho\nu}\Bigl(2\tilde{R}^{(T)}_{[\lambda\mu]}+\hat{R}^{(Q)}_{[\lambda\mu]}\Bigr)-g_{\lambda\nu}\Bigl(2\tilde{R}^{(T)}_{[\rho\mu]}+\hat{R}^{(Q)}_{[\rho\mu]}\Bigr)-g_{\rho\mu}\Bigl(2\tilde{R}^{(T)}_{[\lambda\nu]}+\hat{R}^{(Q)}_{[\lambda\nu]}\Bigr)\nonumber\\
 &&+\,\tilde{R}^{\sigma}{}_{\sigma\lambda[\mu}g_{\nu]\rho}-\tilde{R}^{\sigma}{}_{\sigma\rho[\mu}g_{\nu]\lambda}\Big]\,,\\
   {}^{(6)}\tilde{W}_{\lambda\rho\mu\nu}&=&\frac{1}{6}\,\tilde{R}\,g_{\lambda[\mu}g_{\nu]\rho}\,,\label{irreducible_pieces_antisymmetric_component}
\end{eqnarray}
and five irreducible parts $\tilde{Z}_{\lambda\rho\mu\nu}=\displaystyle\sum_{i=1}^{5}{}^{(i)}\tilde{Z}_{\lambda\rho\mu\nu}$ in the symmetric one:
\begin{eqnarray}
 {}^{(1)}\tilde{Z}_{\lambda\rho\mu\nu}&=& \tilde{Z}_{\lambda\rho\mu\nu}-\sum_{i=2}^{5}{}^{(i)}\tilde{Z}_{\lambda\rho\mu\nu}\,,\label{first_irreducible_symmetric_piece}\\
       {}^{(2)}\tilde{Z}_{\lambda\rho\mu\nu}&=&\frac{1}{4}\Big({\nearrow\!\!\!\!\!\!\!\tilde{R}}^{(Q)}_{\lambda[\rho\mu\nu]}+{\nearrow\!\!\!\!\!\!\!\tilde{R}}^{(Q)}_{\rho[\lambda\mu\nu]}\Big)\,,\\
       {}^{(3)}\tilde{Z}_{\lambda\rho\mu\nu}&=&\frac{1}{6}\Big(g_{\lambda\nu}\hat{R}^{(Q)}_{[\rho\mu]}+g_{\rho\nu}\hat{R}^{(Q)}_{[\lambda\mu]}-g_{\lambda\mu}\hat{R}^{(Q)}_{[\rho\nu]}-g_{\rho\mu}\hat{R}^{(Q)}_{[\lambda\nu]}+g_{\lambda\rho}\hat{R}^{(Q)}_{[\mu\nu]}\Big)\,,\\
       {}^{(4)}\tilde{Z}_{\lambda\rho\mu\nu}&=&\frac{1}{4}g_{\lambda\rho}\tilde{R}^{\sigma}{}_{\sigma\mu\nu}\,,\\
       {}^{(5)}\tilde{Z}_{\lambda\rho\mu\nu}&=&\frac{1}{8}\Big(g_{\lambda\nu}{\nearrow\!\!\!\!\!\!\!\hat{R}}^{(Q)}_{(\rho\mu)}+g_{\rho\nu} {\nearrow\!\!\!\!\!\!\!\hat{R}}^{(Q)}_{(\lambda\mu)}-g_{\lambda\mu} {\nearrow\!\!\!\!\!\!\!\hat{R}}^{(Q)}_{(\rho\nu)}-g_{\rho\mu} {\nearrow\!\!\!\!\!\!\!\hat{R}}^{(Q)}_{(\lambda\nu)}\Big)\,,\label{irreducible_pieces_symmetric_component}
\end{eqnarray}
where
\begin{eqnarray}
      {\nearrow\!\!\!\!\!\!\!\tilde{R}}_{(\mu\nu)}&=&{\nearrow\!\!\!\!\!\!\!{R}}_{\mu\nu}+\nabla_{\lambda}T_{(\mu\nu)}{}^{\lambda} -\nabla_{(\mu}T^{\lambda}{}_{\nu)\lambda}+\frac{1}{2}g_{\mu\nu}\nabla_{\lambda}T^{\rho\lambda}\,_{\rho}-\nabla_{\lambda}Q_{(\mu\nu)}{}^{\lambda}+\frac{1}{2}\nabla_{(\mu}Q_{\nu)}{}^{\lambda}{}_{\lambda}+\frac{1}{2}\nabla_{\lambda}Q^{\lambda}{}_{\mu\nu}\nonumber\\
    &&+\,\frac{1}{4}g_{\mu\nu}\bigl(\nabla_{\lambda}Q^{\rho}\,_{\rho}\,^{\lambda}-\nabla_{\lambda}Q^{\lambda}\,_{\rho}\,^{\rho}\bigr)+ \frac{1}{2}T^{\rho}{}_{\lambda(\mu}T_{\nu)}{}^{\lambda}{}_{\rho}+T^{\rho\lambda}{}_{\rho}T_{(\mu\nu)\lambda}+\frac{1}{4}T_{\mu\lambda\rho} T_{\nu}{}^{\lambda\rho}\nonumber\\
    &&+\,\frac{1}{4}g_{\mu\nu}\Bigl(T^{\lambda}\,_{\lambda\sigma}T^{\rho}\,_{\rho}\,^{\sigma}-\frac{1}{2}T_{\lambda\rho\sigma}T^{\rho\lambda\sigma}-\frac{1}{4}T_{\lambda\rho\sigma}T^{\lambda\rho\sigma}\Bigr)+Q_{\lambda\mu\rho}Q^{[\lambda}{}_{\nu}{}^{\rho]}+\frac{1}{2}Q^{\lambda \rho}{}_{\rho}Q_{(\mu\nu)\lambda}\nonumber\\
    &&-\,\frac{1}{4}\bigl(Q_{\lambda\mu\nu}Q^{\lambda\rho}{}_{\rho}+Q_{\mu\lambda\rho}Q_{\nu}{}^{\lambda\rho}\bigr)+\frac{1}{16}g_{\mu\nu}\bigl(2Q_{\lambda\rho\sigma}Q^{\rho\lambda\sigma}+Q^{\rho\lambda}\,_{\lambda}Q_{\rho}\,^{\sigma}\,_{\sigma}-Q_{\lambda\rho\sigma}Q^{\lambda\rho\sigma}-2Q^{\sigma\lambda}\,_{\lambda}Q^{\rho}\,_{\rho\sigma}\bigr)\nonumber\\
    &&+\,\frac{1}{2}\bigl(T^{\lambda}{}_{(\mu}{}^{\rho}Q_{\nu)\lambda\rho}-Q_{\lambda\rho(\mu}T^{\lambda}{}_{\nu)}{}^{\rho}-2Q_{(\mu\nu)\lambda}T^{\rho\lambda}{}_{\rho}+Q_{\lambda\rho(\mu}T^{\rho}{}_{\nu)}{}^{\lambda}-2Q_{\lambda\rho(\mu}T_{\nu)}{}^{\lambda\rho}-Q_{\lambda}{}^{\rho}{}_{\rho}T_{(\mu\nu)}{}^{\lambda}+Q_{\lambda\mu\nu}T^{\rho\lambda}{}_{\rho}\bigr)\nonumber\\
    &&+\,\frac{1}{4}g_{\mu\nu}\bigl(T^{\lambda}\,_{\lambda\rho}Q^{\rho\sigma}\,_{\sigma}-T_{\lambda\rho\sigma}Q^{\sigma\lambda\rho}-T^{\lambda}\,_{\lambda\rho}Q^{\sigma\rho}\,_{\sigma}\bigr)\,,\label{BB1}\\
    \tilde{R}^{(T)}_{[\mu\nu]}&=&\tilde{\nabla}_{[\mu}T^{\lambda}\,_{\nu]\lambda}+\frac{1}{2}\tilde{\nabla}_{\lambda}T^{\lambda}\,_{\mu\nu}-\frac{1}{2}T^{\lambda}\,_{\rho\lambda}T^{\rho}\,_{\mu\nu}\,, \quad \tilde{R}^{\lambda}\,_{\lambda\mu\nu}=\nabla_{[\nu}Q_{\mu]\lambda}{}^{\lambda}\,,\label{BB2}\\
    {\nearrow\!\!\!\!\!\!\!\hat{R}}^{(Q)}_{(\mu\nu)}&=&\tilde\nabla_{\lambda}{\nearrow\!\!\!\!\!\!\!Q}_{(\mu \nu )}{}^{\lambda}-\tilde\nabla_{(\mu}{\nearrow\!\!\!\!\!\!\!Q}^\lambda{}_{\nu )\lambda}+{\nearrow\!\!\!\!\!\!\!Q}^{\lambda\rho}\,_{\lambda}{\nearrow\!\!\!\!\!\!\!Q}_{(\mu\nu)\rho}-{\nearrow\!\!\!\!\!\!\!Q}_{\lambda\rho(\mu}{\nearrow\!\!\!\!\!\!\!Q}_{\nu)}{}^{\lambda\rho}-T_{\lambda\rho(\mu}{\nearrow\!\!\!\!\!\!\!Q}^{\lambda\rho}{}_{\nu)}\,,\label{BB3}\\
     \hat{R}^{(Q)}_{[\mu\nu]}&=&\tilde{\nabla}_{[\mu}{\nearrow\!\!\!\!\!\!\!Q}^{\lambda}{}_{\nu]\lambda}-\tilde{\nabla}_{\lambda}{\nearrow\!\!\!\!\!\!\!Q}_{[\mu\nu]}{}^{\lambda}-\frac{1}{2}\tilde{\nabla}_{[\mu}{\nearrow\!\!\!\!\!\!\!Q}_{\nu]\lambda}{}^{\lambda}+{\nearrow\!\!\!\!\!\!\!Q}_{[\nu\mu]\lambda}{\nearrow\!\!\!\!\!\!\!Q}_{\rho}{}^{\lambda\rho}-{\nearrow\!\!\!\!\!\!\!Q}_{\rho\lambda [\mu}{\nearrow\!\!\!\!\!\!\!Q}_{\nu] }{}^{\rho\lambda} + {\nearrow\!\!\!\!\!\!\!Q}_{\lambda\rho[\mu}T^{\lambda}{}_{\nu]}{}^{\rho}+ \frac{1}{4}{\nearrow\!\!\!\!\!\!\!Q}^{\lambda\rho}{}_{\rho} T_{\lambda\mu\nu} \,,\label{BB4}\\
    {\nearrow\!\!\!\!\!\!\!\tilde{R}}^{(T)}_{\lambda[\rho\mu\nu]}&=& \frac{1}{2} g_{\lambda[\rho|}\tilde{\nabla}_{\sigma}T^{\sigma}{}_{|\mu\nu]}+g_{\lambda[\rho}\tilde{\nabla}_{\mu}T^{\sigma}{}_{\nu]\sigma}-g_{\lambda\sigma}\tilde{\nabla}_{[\rho}T^{\sigma}{}_{\mu\nu]}+\frac{1}{24}\varepsilon_{\lambda\rho\mu\nu}\varepsilon_{\sigma}{}^{\alpha\beta\gamma}\bigl(\tilde{\nabla}_{\gamma}T^{\sigma}{}_{\beta\alpha}+T_{\beta\omega\gamma}T^{\omega\sigma}{}_{\alpha}\bigr)\nonumber\\
    &&+\,T_{\lambda\sigma[\rho}T^{\sigma}{}_{\mu\nu]}-\frac{1}{2}g_{\lambda[\rho}T^{\sigma}{}_{\mu\nu]} T^{\omega}{}_{\sigma\omega}\,,\label{BB5}\\
    {\nearrow\!\!\!\!\!\!\!\tilde{R}}^{(Q)}_{\lambda[\rho\mu\nu]} &=&\frac{3}{2}\Big( g_{\lambda[\rho|}\tilde{\nabla}_{\sigma}{\nearrow\!\!\!\!\!\!\!Q}_{|\mu\nu]}{}^{\sigma}-g_{\lambda[\rho}\tilde{\nabla}_{\mu}{\nearrow\!\!\!\!\!\!\!Q}^{\sigma}{}_{\nu]\sigma}-2\tilde{\nabla}_{[\rho}{\nearrow\!\!\!\!\!\!\!Q}_{\mu\nu]\lambda}+g_{\lambda[\rho}{\nearrow\!\!\!\!\!\!\!Q}_{\mu\nu]\sigma}{\nearrow\!\!\!\!\!\!\!Q}_{\omega}{}^{\sigma\omega}+g_{\lambda[\rho}{\nearrow\!\!\!\!\!\!\!Q}^{\sigma }{}_{\mu}{}^{\omega}{\nearrow\!\!\!\!\!\!\!Q}_{\nu]\sigma\omega}+{\nearrow\!\!\!\!\!\!\!Q}_{\sigma\lambda[\rho}T^{\sigma}{}_{\mu\nu]}\nonumber\\
    &&+\,g_{\lambda[\rho|}{\nearrow\!\!\!\!\!\!\!Q}_{\sigma|\mu|}{}^{\omega}T^{\sigma}{}_{\omega|\nu]}+\frac{1}{2}Q_{[\rho|\sigma}{}^{\sigma}{\nearrow\!\!\!\!\!\!\!Q}_{|\mu\nu]\lambda}\Big)\,,\label{BB6}\\  {}^{(1)}\tilde{W}_{\lambda\rho\mu\nu}&=&\tilde{R}_{[\lambda\rho]\mu\nu}-\frac{3}{4}\Big({\nearrow\!\!\!\!\!\!\!\tilde{R}}^{(T)}_{\lambda[\rho\mu\nu]}+{\nearrow\!\!\!\!\!\!\!\tilde{R}}^{(T)}_{\nu[\lambda\rho\mu]}-{\nearrow\!\!\!\!\!\!\!\tilde{R}}^{(T)}_{\rho[\lambda\mu\nu]}-{\nearrow\!\!\!\!\!\!\!\tilde{R}}^{(T)}_{\mu[\lambda\rho\nu]}\Bigr)-\frac{1}{2}\Bigl({\nearrow\!\!\!\!\!\!\!\tilde{R}}^{(Q)}_{\mu[\lambda\rho\nu]}-{\nearrow\!\!\!\!\!\!\!\tilde{R}}^{(Q)}_{\nu[\lambda\rho\mu]}\Big)+\,\frac{1}{24}\left.\ast\tilde{R}\right. \varepsilon_{\lambda\rho\mu\nu}\nonumber\\
&&-\,\frac{1}{4}\Big[g_{\lambda\mu}\left(2{\nearrow\!\!\!\!\!\!\!\tilde{R}}_{(\rho\nu)}+{\nearrow\!\!\!\!\!\!\!\hat{R}}^{(Q)}_{(\rho \nu) }\right)+g_{\rho\nu}\left(2{\nearrow\!\!\!\!\!\!\!\tilde{R}}_{(\lambda\mu)} +{\nearrow\!\!\!\!\!\!\!\hat{R}}^{(Q)}_{(\lambda\mu)}\right)-g_{\lambda\nu}\left(2 {\nearrow\!\!\!\!\!\!\!\tilde{R}}_{(\rho\mu)}+{\nearrow\!\!\!\!\!\!\!\hat{R}}^{(Q)}_{(\rho\mu)}\right)-g_{\rho\mu}\left(2{\nearrow\!\!\!\!\!\!\!\tilde{R}}_{(\lambda\nu)}+{\nearrow\!\!\!\!\!\!\!\hat{R}}^{(Q)}_{(\lambda\nu)}\right)\Big]\nonumber\\
&&-\,\frac{1}{4}\Big[g_{\lambda\mu}\Bigl(2\tilde{R}^{(T)}_{[\rho\nu]}+\hat{R}^{(Q)}_{[\rho\nu]}\Bigr)+g_{\rho\nu}\Bigl(2\tilde{R}^{(T)}_{[\lambda\mu]}+\hat{R}^{(Q)}_{[\lambda\mu]}\Bigr)-g_{\lambda\nu}\Bigl(2\tilde{R}^{(T)}_{[\rho\mu]}+\hat{R}^{(Q)}_{[\rho\mu]}\Bigr)-g_{\rho\mu}\Bigl(2\tilde{R}^{(T)}_{[\lambda\nu]}+\hat{R}^{(Q)}_{[\lambda\nu]}\Bigr)\nonumber\\
 &&+\,\tilde{R}^{\sigma}{}_{\sigma\lambda[\mu}g_{\nu]\rho}-\tilde{R}^{\sigma}{}_{\sigma\rho[\mu}g_{\nu]\lambda}\Big]-\frac{1}{6}\tilde{R}\,g_{\lambda[\mu}g_{\nu]\rho}\,,\label{BB7}\\     
   {}^{(1)}\tilde{Z}_{\lambda\rho\mu\nu} &=&\tilde{R}_{(\lambda\rho)\mu\nu}-\frac{1}{4}\Big({\nearrow\!\!\!\!\!\!\!\tilde{R}}^{(Q)}_{\lambda[\rho\mu\nu]}+{\nearrow\!\!\!\!\!\!\!\tilde{R}}^{(Q)}_{\rho[\lambda\mu\nu]}\Big)-\frac{1}{6}\Big(g_{\lambda\nu}\hat{R}^{(Q)}_{[\rho\mu]}+g_{\rho\nu}\hat{R}^{(Q)}_{[\lambda\mu]}-g_{\lambda\mu}\hat{R}^{(Q)}_{[\rho\nu]}-g_{\rho\mu}\hat{R}^{(Q)}_{[\lambda\nu]}+g_{\lambda\rho}\hat{R}^{(Q)}_{[\mu\nu]}\Big)\nonumber\\
 &&-\,\frac{1}{4}g_{\lambda\rho}\tilde{R}^{\sigma}{}_{\sigma\mu\nu}-\frac{1}{8}\Big(g_{\lambda\nu}{\nearrow\!\!\!\!\!\!\!\hat{R}}^{(Q)}_{(\rho\mu)}+g_{\rho\nu} {\nearrow\!\!\!\!\!\!\!\hat{R}}^{(Q)}_{(\lambda\mu)}-g_{\lambda\mu} {\nearrow\!\!\!\!\!\!\!\hat{R}}^{(Q)}_{(\rho\nu)}-g_{\rho\mu} {\nearrow\!\!\!\!\!\!\!\hat{R}}^{(Q)}_{(\lambda\nu)}\Big)\,,\label{BB8}\\
    \tilde{R}&=& R-2\nabla_{\mu}T^{\nu \mu}\,_{\nu}+\nabla_{\mu}Q^{\mu}\,_{\nu}\,^{\nu}-\nabla_{\mu}Q^{\nu}\,_{\nu}\,^{\mu}+\frac{1}{4}T_{\lambda \mu \nu}T^{\lambda \mu \nu}+\frac{1}{2}T_{\lambda \mu \nu}T^{\mu \lambda \nu}-T^{\lambda}\,_{\lambda\nu}T^{\mu}\,_{\mu}\,^{\nu}+T_{\lambda\mu\nu}Q^{\nu\lambda\mu}\nonumber\\
    &&+\,T^{\lambda}\,_{\lambda\nu}Q^{\mu\nu}\,_{\mu}-T^{\lambda}\,_{\lambda\nu}Q^{\nu\mu}\,_{\mu}+\frac{1}{4}Q_{\lambda\mu\nu}Q^{\lambda\mu\nu}-\frac{1}{2}Q_{\lambda\mu\nu}Q^{\mu\lambda\nu}+\frac{1}{2}Q^{\nu\lambda}\,_{\lambda}Q^{\mu}\,_{\mu\nu}-\frac{1}{4}Q^{\nu\lambda}\,_{\lambda}Q_{\nu}\,^{\mu}\,_{\mu}\,,\label{BB9}\\
\ast\tilde{R}&=&\varepsilon^{\lambda\rho\mu\nu}\Bigl(\nabla_{\lambda}T_{\rho\mu\nu}+\frac{1}{2}T^{\sigma}{}_{\lambda\rho}T_{\sigma\mu\nu}-Q_{\lambda\sigma\rho}T^{\sigma}{}_{\mu\nu}\Bigr)\,.\label{BB10}
\end{eqnarray}

Thereby, the resulting eleven irreducible parts of the curvature tensor can be included in the general action of MAG to endow the torsion and nonmetricity fields with dynamics~\cite{Hehl:1994ue}. In this sense, different restrictions imposed on the Lagrangian coefficients lead to a large class of gravitational models, for which an extensive number of fundamental properties and phenomenological implications may arise. Thus, one of the most relevant aspects to be considered in this construction concerns stability, which has been recently analysed in the framework of cubic MAG to provide a safer context for the propagation of the torsion and nonmetricity fields around general backgrounds~\cite{Bahamonde:2024sqo,Bahamonde:2024efl}.

Following these lines, in the present work we focus on the following cubic MAG model, which has shown to eliminate the main gravitational instabilities affecting the vector and axial sectors of quadratic MAG\footnote{For simplicity in the calculations, we assume vanishing mass terms for torsion and nonmetricity.}:
\begin{align}
    S=&\,\frac{1}{16\pi}\int
    \Bigl[
    -\,R-\frac{1}{2}\left(2c_{1}+c_{2}\right)\tilde{R}_{\lambda\rho\mu\nu}\tilde{R}^{\mu\nu\lambda\rho}+\left(a_{2}-c_{1}\right)\tilde{R}_{\lambda\rho\mu\nu}\tilde{R}^{\rho\lambda\mu\nu}+a_{2}\tilde{R}_{\lambda\rho\mu\nu}\tilde{R}^{\lambda\rho\mu\nu}+a_{5}\tilde{R}_{\lambda\rho\mu\nu}\tilde{R}^{\lambda\mu\rho\nu}
    \Bigr.
    \nonumber\\
    \Bigl.
    &+a_{6}\tilde{R}_{\lambda\rho\mu\nu}\tilde{R}^{\rho\mu\lambda\nu}+\left(c_{2}-a_{5}+a_{6}\right)\tilde{R}_{\lambda\rho\mu\nu}\tilde{R}^{\mu\rho\lambda\nu}+\left(d_{1}-a_{10}-a_{12}\right)\tilde{R}_{\mu\nu}\tilde{R}^{\mu\nu}+a_{9}\tilde{R}_{\mu\nu}\tilde{R}^{\nu\mu}+a_{10}\hat{R}_{\mu\nu}\hat{R}^{\mu\nu}
    \Bigr.
    \nonumber\\
    \Bigl.
    &+a_{11}\hat{R}_{\mu\nu}\hat{R}^{\nu\mu}-\left(d_{1}+a_{9}+a_{11}\right)\tilde{R}_{\mu\nu}\hat{R}^{\nu\mu}+a_{12}\tilde{R}_{\mu\nu}\hat{R}^{\mu\nu}+a_{14}\tilde{R}^{\lambda}{}_{\lambda\mu\nu}\tilde{R}^{\rho}{}_{\rho}{}^{\mu\nu}+a_{15}\tilde{R}_{\mu\nu}\tilde{R}^{\lambda}{}_{\lambda}{}^{\mu\nu}+a_{16}\hat{R}_{\mu\nu}\tilde{R}^{\lambda}{}_{\lambda}{}^{\mu\nu}
    \Bigr.
    \nonumber\\
    \Bigl.
    &+\mathcal{\bar{L}}_{\rm curv-tor}^{(3)}+\mathcal{\bar{L}}_{\rm curv-nonm}^{(3)}+\mathcal{\bar{L}}_{\rm curv-tor-nonm}^{(3)}\Bigr]\sqrt{-g}\,d^4x\,,\label{cubicmodel}
\end{align}
where
$\mathcal{\bar{L}}_{\rm curv-tor}^{(3)}$, $\mathcal{\bar{L}}_{\rm curv-nonm}^{(3)}$ and $\mathcal{\bar{L}}_{\rm curv-tor-nonm}^{(3)}$ refer to Lagrangian densities of cubic order defined from mixing terms of curvature, torsion and/or nonmetricity, whose explicit form can be found in Appendix~\ref{appe1}.

In general, on top of the quadratic order invariants, by construction the model includes up to $209$ cubic order invariants that can be parametrised by a set of coefficients $\{h_{i}\}_{i=1}^{209}$, in such a way that the stability of both vector and axial sectors requires the following conditions:
\begin{align}
    c_{2}&=2c_{1}\,,  \quad a_{11}= -\,\frac{1}{4} \left(2 a_{2}+  a_{6}+4 a_{10}\right), \quad a_{12}= \frac{1}{4} \left(4 d_{1}+2 a_{2} + a_{6} + 4 a_{9}-4 a_{10}\right), \quad h_{3}= -\, \frac{1}{6}\left(c_{1}+6h_{13}\right),\,\label{sti}\\
    h_{4}&=\frac{h_{13}}{2}\,, \quad h_{14}= -\,2h_{13}\,, \quad h_{15}=4h_{13}\,, \quad h_{51}=-\,h_{50}\,, \quad h_{133}=3h_{13}+h_{132}\,, \quad h_{136}=6 h_{13} -  h_{135}\,,\\
    h_{139}&= \frac{1}{12} \left(2 a_{5} -  a_{6} + c_{1} - 9 h_{13} + 12 h_{138} + 36 h_{28}\right), \quad h_{140}= \frac{1}{12} \left(2 a_{5} -  a_{6} - 8 c_{1}{} - 36 h_{13} - 72 h_{28}\right), \\
    h_{142}&= \frac{1}{24} \left(12 a_{10} + 6 a_{2} + 4 a_{5} + a_{6} + 12 a_{9} + 20 c_{1}{} + 36 h_{13} - 24 h_{141} + 288 h_{28}\right),\\
    h_1&=h_2=h_{27}=h_{47}=h_{48}=h_{49}=h_{52}=h_{68}=h_{69}=h_{70}=h_{71}=h_{72}=h_{73}=h_{108}=0\,,\\
    h_{109}&=h_{110}=h_{111}=h_{112}=h_{113}=h_{114}=h_{115}=h_{116}=h_{117}=h_{118}=h_{119}=h_{134}=0\,.\label{stf}
\end{align}
A further restriction on the Lagrangian coefficients provides Reissner-Nordström-like solutions with spin, dilation and shear charges, which simplifies the parameter space of the model~\cite{Bahamonde:2024efl}.

By performing variations in Expression~\eqref{cubicmodel} with respect to the tetrad
field $e^{a}{}_{\mu}$ and the anholonomic connection $\omega^{a}{}_{b\mu}$, both related to the metric tensor and the affine connection as
\begin{align}
    g_{\mu \nu}&=e^{a}{}_{\mu}\,e^{b}{}_{\nu}\,g_{a b}\,,\label{vierbein_def}\\
    \omega^{a}{}_{b\mu}&=e^{a}{}_{\lambda}\,e_{b}{}^{\rho}\,\tilde{\Gamma}^{\lambda}{}_{\rho \mu}+e^{a}{}_{\lambda}\,\partial_{\mu}\,e_{b}{}^{\lambda}\,,\label{anholonomic_connection}
\end{align}
it is cumbersome but straightforward to derive the general field equations of the model
\begin{align}
    E^{\mu\nu}&=0\label{tetrad_eq}\,,\\
    E^{\lambda\mu\nu}&=0\label{connection_eq}\,,
\end{align}
where $E^{\mu\nu}$ and $E^{\lambda\mu\nu}$ are tensor quantities depending on curvature, torsion and nonmetricity, defined in Appendix~\ref{appe2}.

Thus, in the following sections we shall consider the cubic MAG model satisfying the aforementioned stability conditions and the compatibility with Reissner-Nordström-like black holes\footnote{Note the Lagrangian coefficients providing Reissner-Nordström-like solutions in cubic MAG generally include three proportionality constants $N_{1}$, $N_{2}$ and $N_{3}$, but for simplicity in the presentation we assume the case where $N_2=0$.}, in order to obtain new gravitational wave solutions with dynamical torsion and nonmetricity.

\section{Gravitational wave profile}\label{sec:wave_settings}

For the description of gravitational waves with torsion and nonmetricity, we first consider that the corresponding metric structure acquires a pp-wave form in a flat transverse space, hence characterised by a wave null vector that is covariantly constant:
\begin{equation}
    \nabla_{\mu}k^{\nu}=0\,.
\end{equation}
The line element can then be written in terms of Brinkmann coordinates $(u,v,x,y)$ as~\cite{Brinkmann:1925fr,blanco2013structure,Lasenby:2019gmi}:
\begin{equation}
  ds^2 = 2 du dv - dx^2 - dy^2-H(u,x,y) du^2\,,
\end{equation}
where $k^\mu=(0,1,0,0)$. This form of the metric is kept by the following coordinate changes~\cite{blanco2013structure}:
\begin{align}
u'(u,x,y)&= u+\mbox{const.}\,, \hspace{5mm} x'(u,x,y)=x \cos\phi +y\sin\phi +p_{1}(u)\,, \hspace{5mm} y'(u,x,y) = y\cos\phi-x \sin\phi +p_{2}(u)\,,\\
v'(u,x,y) &= v+x \bigl(\dot{p}_{1}(u) \cos\phi -\dot{p}_{2}(u) \sin\phi\bigr)+y\bigl(\dot{p}_{1}(u)\sin\phi +\dot{p}_{2}(u) \cos\phi\bigr)+p_{3}(u)\,,
\end{align}
where $\phi$ is an arbitrary parameter and $p_1, p_2$ and $p_3$ are arbitrary functions of $u$. The new function $H'$ then reads\footnote{Observe that terms linear in $x$ and $y$ in the function $H$ can always be eliminated by a suitable choice of coordinates.}:
\begin{equation}
    H'(u,x,y) = H(u,x,y) + \dot{p}_{1}^{2}(u) +\dot{p}_{2}^{2}(u)-2x \bigl(\ddot{p}_{1}(u) \cos\phi -\ddot{p}_{2}(u) \sin\phi\bigr)-2y\bigl(\ddot{p}_{1}(u)\sin\phi +\ddot{p}_{2}(u) \cos\phi\bigr)-2\dot{p}_{3}(u) \, .
\end{equation}

In order to address the corresponding extension to metric-affine geometry, the invariance defined by the wave vector can be directly imposed on the torsion and nonmetricity tensors, which in turn preserves the associated curvature tensor of the space-time. In that case, the torsion and nonmetricity tensors present the general form:
\begin{align}
    T^{\lambda}{}_{\mu\nu} &= T^{\lambda}{}_{\mu\nu} (u,x,y)\,,\\
    Q_{\lambda\mu\nu} &= Q_{\lambda\mu\nu} (u,x,y)\,.
\end{align}

On the other hand, as is well known, an important property of pp-waves is that they constitute algebraically special configurations, characterised by Type N Riemannian Weyl and Ricci tensors:
\begin{equation}
    W_{\lambda \rho \mu \nu}k^{\mu} = R_{\nu [\lambda}k_{\rho]}=0\,.
\end{equation}
Thereby, it is reasonable to impose the same types of algebraic conditions on the corresponding field strengths of the torsion and nonmetricity tensors of a MAG model. In particular, under the stability conditions~\eqref{sti}-\eqref{stf}, the quantities $\{{\nearrow\!\!\!\!\!\!\!\tilde{R}}^{(T)}_{\lambda[\rho\mu\nu]},\tilde{R}^{(T)}_{[\mu\nu]},\ast\tilde{R},\tilde{R}^{\lambda}{}_{\lambda\mu\nu},{}^{(1)}\tilde{Z}_{\lambda\rho\mu\nu},{\nearrow\!\!\!\!\!\!\!\tilde{R}}^{(Q)}_{\lambda [\rho\mu\nu]},\hat{R}^{(Q)}_{[\mu\nu]}\}$ turn out to provide the kinetic terms to the corresponding vector, axial and tensor modes of torsion and nonmetricity in the gravitational action~\eqref{cubicmodel}, in virtue of their post-Riemannian expansions:
\begin{align}
    {\nearrow\!\!\!\!\!\!\!\tilde{R}}^{(T)}_{\lambda[\rho\mu\nu]}=&- \frac{1}{24}\bigl(\varepsilon_{\rho  \mu  \nu  \sigma  } \nabla_{\lambda  }S^{\sigma} + \varepsilon_{\lambda  \sigma  [\rho  \mu  } \nabla_{\nu]  }S^{\sigma}+2g_{\lambda  [\rho  }\varepsilon_{\mu  \nu]  \omega  }{}^{\sigma  }  \nabla_{\sigma  }S^{\omega}\bigr) +\frac{1}{2}  \nabla_{\sigma  }t^{\sigma}{}_{[\rho  \mu  }g_{\nu ] \lambda  }- g_{\lambda\sigma}\nabla_{[\rho  }t^{\sigma}{}_{\mu\nu]}+\mathcal{O}(T^2,QT)\,,\\
   \tilde{R}^{(T)}_{[\mu\nu]}=&\,\frac{2}{3} \nabla_{[\mu  }T_{\nu]}-\frac{1}{12} \varepsilon_{\mu  \nu  \lambda}{}^{\rho} \nabla_{\rho  }S^{\lambda}+\frac{1}{2} \nabla_{\lambda}t^{\lambda}{}_{[\mu  \nu ] }+\mathcal{O}(T^2,QT)\,,\\
   \ast\tilde{R}=&\,\nabla_{\mu}S^{\mu}+\mathcal{O}(T^2,QT)\,, \quad \tilde{R}^{\lambda}{}_{\lambda\mu\nu} = -\,4\nabla_{[\mu}W_{\nu]}\,,\\
    \hat{R}^{(Q)}_{[\mu\nu]}=&\,\frac{3}{2} \nabla_{[\mu  }\Lambda _{\nu]  }+ \frac{1}{6}\bigl(\varepsilon_{\lambda  \sigma  \omega  [\mu  } \nabla^{\lambda  }\Omega _{\nu]  }{}^{\sigma  \omega  }- \varepsilon_{\mu  \nu\lambda\rho} \nabla_{\sigma  }\Omega ^{\sigma\lambda\rho}\bigr)  +\mathcal{O}({\nearrow\!\!\!\!\!\!\!Q}Q,{\nearrow\!\!\!\!\!\!\!Q}T)\,,\\
   {\nearrow\!\!\!\!\!\!\!\tilde{R}}^{(Q)}_{\lambda [\rho\mu\nu]}  =&\,\frac{3}{4}\bigl(2\nabla_{[\rho  }\Omega _{\mu  }{}^{\sigma  \omega  }\varepsilon_{\nu]  \lambda  \sigma  \omega  }- g_{\lambda  [\rho  } \nabla^{\tau  }\Omega _{\mu  }{}^{\sigma  \omega  }\varepsilon_{\nu]  \sigma  \omega\tau  }\bigr)+\mathcal{O}({\nearrow\!\!\!\!\!\!\!Q}Q,{\nearrow\!\!\!\!\!\!\!Q}T)\,,\\
   {}^{(1)}\tilde{Z}_{\lambda\rho\mu\nu}=&\,\frac{1}{48}\bigl[2\bigl(g_{\lambda  [\mu  } \varepsilon_{\nu]  \sigma  \omega  }{}^{\tau  } \nabla_{\tau  }\Omega _{\rho  }{}^{\sigma  \omega  }+g_{  \rho [\mu }\varepsilon_{\nu]  \sigma  \omega  }{}^{\tau  }  \nabla_{\tau  }\Omega _{\lambda  }{}^{\sigma  \omega}-\varepsilon_{\lambda  \sigma  \omega  [\mu  } \nabla_{\nu  ]}\Omega _{\rho  }{}^{\sigma  \omega  }-\varepsilon_{\rho  \sigma  \omega  [\mu  } \nabla_{\nu  ]}\Omega _{\lambda  }{}^{\sigma  \omega}\bigr)-g_{\lambda\rho}\varepsilon_{\mu  \nu  \omega  \sigma}\nabla_{\tau  }\Omega ^{\tau  \omega  \sigma  }\bigr]\nonumber\\
   &-\frac{1}{4}g_{\tau(\lambda}g_{\rho)\gamma}\varepsilon^{\gamma}{}_{\sigma  \omega [ \mu  }\nabla^{\tau}\Omega _{\nu ] }{}^{\sigma  \omega  } +\frac{1}{4}\bigl(g_{\lambda  [\mu  } \nabla^{\sigma  }q_{\nu]  \rho  \sigma  }+g_{\rho  [\mu  } \nabla^{\sigma  }q_{\nu]  \lambda  \sigma  }-4 \nabla_{[\mu  }q_{\nu]\lambda  \rho}\bigr)+\mathcal{O}({\nearrow\!\!\!\!\!\!\!Q}Q,{\nearrow\!\!\!\!\!\!\!Q}T)\,.
\end{align}
Thus, any of the previous tensor quantities is susceptible to satisfy Type N algebraic conditions if the respective kinetic terms are not trivialised in the solutions:
\begin{align}
    k^{\mu}\tilde{R}^{(T)}_{[\mu\nu]}&=0\,,\label{TypeN_cond1}\\
    k_{[\rho}\tilde{R}^{(T)}_{\mu\nu]}&=0\,,\label{TypeN_condd1}\\
    k^{\mu}\hat{R}^{(Q)}_{[\mu\nu]}&=0\,,\label{TypeN_cond2}\\
    k_{[\rho}\hat{R}^{(Q)}_{\mu\nu]}&=0\,,\label{TypeN_condd2}\\
    k^{\mu}\tilde{R}^{\lambda}{}_{\lambda\mu\nu}&=0\,,\label{TypeN_cond3}\\
    k_{[\rho|}\tilde{R}^{\lambda}{}_{\lambda|\mu\nu]}&=0\,,\label{TypeN_condd3}\\
    k_{[\sigma}{\nearrow\!\!\!\!\!\!\!\tilde{R}}^{(T)}_{\lambda][\rho\mu\nu]}&=0\,,\label{TypeN_cond4}\\
    k_{[\sigma}{\nearrow\!\!\!\!\!\!\!\tilde{R}}^{(Q)}_{\lambda][\rho\mu\nu]}&=0\,,\label{TypeN_cond5}\\
    k_{[\sigma}{}^{(1)}\tilde{Z}_{\lambda]\rho\mu\nu}&=0\,.\label{TypeN_cond6}
\end{align}

A stronger restriction would consist of extending such conditions to the torsion and nonmetricity tensors themselves. However, it is important to note that the metric tensor of the pp-wave space-time is not Type N, but as mentioned previously the relevant Riemannian quantities fulfilling these algebraic conditions are the Weyl and Ricci tensors derived from the latter. Hence, it sounds reasonable to follow the same guiding principle in the presence of torsion and nonmetricity, imposing the corresponding algebraic conditions only on their associated field strengths, but not on these two tensors themselves. Furthermore, it turns out that for Type N torsion and nonmetricity tensors the component $E_{uu}=0$ of the tetrad field equations~\eqref{tetrad_eq} is reduced to the Laplace equation in the transverse space:
\begin{equation}
    \left(\partial_{xx}+\partial_{yy}\right)H(u,x,y)=0\,,
\end{equation}
which strongly restricts the metric function of the solutions to the case of GR in empty space-time. Accordingly, we shall not consider such a restriction in our gravitational wave profile.

Additionally, we impose orthogonality of the wave null vector with torsion and nonmetricity, which can be expressed in terms of their respective vector, axial and tensor modes as:\begin{align}
    k^{\mu}T_{\mu}&=k^{\mu}S_{\mu}=k^{\mu}W_{\mu}=k^{\mu}\Lambda_{\mu}=0\,,\label{ortho1}\\
    k^{\mu}t_{\mu\lambda\rho}&=k^{\mu}\Omega_{\mu\lambda\rho}=k^{\mu}q_{\mu\lambda\rho}=0\,.\label{ortho2}
\end{align}

Finally, we consider that the mentioned wave null vector is recurrent, which can be  expressed in terms of its total covariant derivative as:
\begin{equation}
    \tilde{\nabla}_{\mu}k^{\nu} = \left(\alpha T_{\mu}+\beta W_{\mu}+\gamma\Lambda_{\mu}\right)k^{\nu}\,,\label{covdevk}
\end{equation}
where $\alpha$, $\beta$ and $\gamma$ are constants. This implies that its transport --with the full connection $\tilde\nabla$-- along {\em any} possible curve is always parallel to itself, whereas its null character is also preserved if the orthogonality conditions~\eqref{ortho1} and~\eqref{ortho2} hold. Likewise, in the cases when $\alpha T_{\mu}+\beta W_{\mu}+\gamma\Lambda_{\mu}$ is a gradient, the vector field $k^\mu$ can be rescaled $k^\mu \rightarrow k'^\mu=A k^\mu$ so that $k'^\mu$ is covariantly constant with respect to $\tilde\nabla$ as well. In general, this recurrence property implies that $k^\mu$ is a well-defined preferred null vector field of the space-time.

\section{Gravitational wave solutions with dynamical torsion and nonmetricity}\label{sec:sol}

Imposing the Killing equations on the torsion and nonmetricity tensors:
\begin{equation}
    \mathcal{L}_{k}T^{\lambda}{}_{\mu\nu} = \mathcal{L}_{k}Q_{\lambda\mu\nu} = 0\,,
\end{equation}
the $24$ independent components of torsion can be written as:
\begin{subequations} \label{eq:torsioncomponents}
\begin{align}
T^u{}_{u v}&=t_{1}(u,x,y)\,,\quad& T^v{}_{u v}&=t_{2}(u,x,y)\,,\quad& T^x{}_{u v}&=t_{3}(u,x,y)\,,\quad&  T^y{}_{u v}&=t_{4}(u,x,y)\,,\\
T^u{}_{ux} &= t_{5}(u,x,y)\,, \quad & T^v{}_{ux} &= t_{6}(u,x,y) \,,\quad & T^x{}_{ux} &= t_{7}(u,x,y)\,, \quad & T^y{}_{ux} &= t_{8}(u,x,y)\,,  \\
T^u{}_{uy} &= t_{9}(u,x,y)\,, \quad & T^v{}_{uy} &= t_{10}(u,x,y) \,,\quad & T^x{}_{uy} &= t_{11}(u,x,y)\,, \quad & T^y{}_{uy} &= t_{12}(u,x,y) \,, \\
T^u{}_{vx} &= t_{13}(u,x,y)\,, \quad & T^v{}_{vx} &= t_{14}(u,x,y)\,, \quad & T^x{}_{vx} &= t_{15}(u,x,y)\,, \quad & T^y{}_{vx} &= t_{16}(u,x,y) \,, \\
T^u{}_{vy} &= t_{17}(u,x,y)\,, \quad & T^v{}_{vy} &= t_{18}(u,x,y)\,, \quad & T^x{}_{vy} &= t_{19}(u,x,y)\,, \quad & T^y{}_{vy} &= t_{20}(u,x,y) \,, \\
T^u{}_{xy} &= t_{21}(u,x,y) \,,\quad & T^v{}_{xy} &= t_{22}(u,x,y)\,, \quad & T^x{}_{xy} &= t_{23}(u,x,y)\,, \quad & T^y{}_{xy} &= t_{24}(u,x,y)\,,
\end{align}
\end{subequations}
while the $40$ independent components of nonmetricity take the form:
\begin{subequations} \label{eq:torsioncomponents}
\begin{align}
Q_{u u u}&=q_{1}(u,x,y)\,,\quad& Q_{v u u}&=q_{2}(u,x,y)\,,\quad& Q_{x u u}&=q_{3}(u,x,y)\,,\quad&  Q_{y u u}&=q_{4}(u,x,y)\,,\\
Q_{u u v}&=q_{5}(u,x,y)\,,\quad& Q_{v u v}&=q_{6}(u,x,y)\,,\quad& Q_{x u v}&=q_{7}(u,x,y)\,,\quad&  Q_{y u v}&=q_{8}(u,x,y)\,,\\
Q_{u u x}&=q_{9}(u,x,y)\,,\quad& Q_{v u x}&=q_{10}(u,x,y)\,,\quad& Q_{x u x}&=q_{11}(u,x,y)\,,\quad&  Q_{y u x}&=q_{12}(u,x,y)\,,\\
Q_{u u y}&=q_{13}(u,x,y)\,,\quad& Q_{v u y}&=q_{14}(u,x,y)\,,\quad& Q_{x u y}&=q_{15}(u,x,y)\,,\quad&  Q_{y u y}&=q_{16}(u,x,y)\,,\\
Q_{u v v}&=q_{17}(u,x,y)\,,\quad& Q_{v v v}&=q_{18}(u,x,y)\,,\quad& Q_{x v v}&=q_{19}(u,x,y)\,,\quad&  Q_{y v v}&=q_{20}(u,x,y)\,,\\
Q_{u v x}&=q_{21}(u,x,y)\,,\quad& Q_{v v x}&=q_{22}(u,x,y)\,,\quad& Q_{x v x}&=q_{23}(u,x,y)\,,\quad&  Q_{y v x}&=q_{24}(u,x,y)\,,\\
Q_{u v y}&=q_{25}(u,x,y)\,,\quad& Q_{v v y}&=q_{26}(u,x,y)\,,\quad& Q_{x v y}&=q_{27}(u,x,y)\,,\quad&  Q_{y v y}&=q_{28}(u,x,y)\,,\\
Q_{u x x}&=q_{29}(u,x,y)\,,\quad& Q_{v x x}&=q_{30}(u,x,y)\,,\quad& Q_{x x x}&=q_{31}(u,x,y)\,,\quad&  Q_{y x x}&=q_{32}(u,x,y)\,,\\
Q_{u x y}&=q_{33}(u,x,y)\,,\quad& Q_{v x y}&=q_{34}(u,x,y)\,,\quad& Q_{x x y}&=q_{35}(u,x,y)\,,\quad&  Q_{y x y}&=q_{36}(u,x,y)\,,\\
Q_{u y y}&=q_{37}(u,x,y)\,,\quad& Q_{v y y}&=q_{38}(u,x,y)\,,\quad& Q_{x y y}&=q_{39}(u,x,y)\,,\quad&  Q_{y y y}&=q_{40}(u,x,y)\,.
\end{align}
\end{subequations}

In addition, imposing the orthogonality conditions~\eqref{ortho1}-\eqref{ortho2} between the wave null vector and the irreducible modes of torsion and nonmetricity, the previous functions are constrained as:
\begin{align}
    t_{1}(u,x,y)&=t_{13}(u,x,y)=t_{17}(u,x,y)=t_{21}(u,x,y)=0\,,\\
    t_{19}(u,x,y)&=t_{16}(u,x,y)\,, \quad t_{20}(u,x,y)=-\,t_{15}(u,x,y)\,,\\
    t_{23}(u,x,y)&=t_{9}(u,x,y)-t_{4}(u,x,y)\,, \quad t_{24}(u,x,y)=t_{3}(u,x,y)-t_{5}(u,x,y)\,,\\
    q_{6}(u,x,y)&=q_{17}(u,x,y)=q_{18}(u,x,y)=q_{19}(u,x,y)=q_{20}(u,x,y)=q_{22}(u,x,y)=q_{26}(u,x,y)=0\,,\\
    q_{27}(u,x,y)&=q_{24}(u,x,y)\,, \quad q_{28}(u,x,y)=-\,q_{23}(u,x,y)\,, \quad q_{30}(u,x,y)=-\,2q_{23}(u,x,y)\,,\\
    q_{34}(u,x,y)&=-\,2q_{24}(u,x,y)\,, \quad q_{35}(u,x,y)=q_{8}(u,x,y)+q_{14}(u,x,y)+q_{32}(u,x,y)-2q_{25}(u,x,y)\,,\\
    q_{36}(u,x,y)&=-\left(q_{7}(u,x,y)+q_{10}(u,x,y)+2q_{21}(u,x,y)+q_{31}(u,x,y)\right),\\
    q_{37}(u,x,y)&=-\left(q_{2}(u,x,y)+2q_{5}(u,x,y)+2q_{11}(u,x,y)+2q_{16}(u,x,y)+q_{29}(u,x,y)\right),\\
    q_{38}(u,x,y)&=2q_{23}(u,x,y)\,, \quad q_{39}(u,x,y)=-\left(2q_{7}(u,x,y)+2q_{10}(u,x,y)+q_{31}(u,x,y)\right),\\
    q_{40}(u,x,y)&=-\left(2q_{8}(u,x,y)+2q_{14}(u,x,y)+q_{32}(u,x,y)\right),
\end{align}
which initially reduces the number of independent components of the torsion and nonmetricity tensors to $16$ and $23$, respectively. 

Likewise, the decomposition of the anholonomic connection~\eqref{anholonomic_connection} into antisymmetric and symmetric components ---the latter including both trace and traceless parts--- must properly account for the contributions of the dynamical torsion and nonmetricity tensors of the solutions. Specifically, the antisymmetric component includes the contribution of the dynamical torsion tensor and the symmetric one the contribution of the dynamical trace and traceless parts of the nonmetricity tensor. This means that the torsion tensor of the solution must be split into dynamical and nondynamical pieces, respectively denoted by $\bar{T}^{\lambda}{}_{\mu\nu}$ and $\mathring{T}^{\lambda}{}_{\mu\nu}$, the latter suppressing any contribution from the trace and traceless parts of the nonmetricity tensor in the antisymmetric component of the anholonomic connection. Taking into account the correspondence~\eqref{anholonomic_connection} between the anholonomic connection and the affine connection, these pieces must then satisfy:
\begin{equation}
    N_{[\lambda\rho]\mu} = \bar{K}_{\lambda\rho\mu}\,,
\end{equation}
or, equivalently:
\begin{equation}
    \mathring{K}_{\lambda\rho\mu}+L_{[\lambda\rho]\mu}=0\,.\label{cond_Nantim}
\end{equation}

In terms of the irreducible modes of the torsion and nonmetricity tensors, the condition~\eqref{cond_Nantim} is then fulfilled if:
\begin{equation}
   \mathring{T}_{\mu}=\frac{3}{8}\left(4W_{\mu}-3\Lambda_{\mu}\right), \quad \mathring{S}_{\mu}=0\,, \quad \mathring{t}_{\lambda\mu\nu}=\frac{1}{4}\varepsilon_{\sigma\rho\mu\nu}\Omega_\lambda{}^{\sigma\rho}\,,
\end{equation}
whereas, splitting the torsion functions as $t_{i}=\mathring{t}_{i}+\bar{t}_{i}$, it implies:
\begin{align}
    \mathring{t}_{2}(u,x,y)&=\frac{1}{2}\left(q_{5}(u,x,y)-q_{2}(u,x,y)\right), \quad \mathring{t}_{3}(u,x,y)=\frac{1}{2}\left(q_{10}(u,x,y)-q_{21}(u,x,y)\right),\\
    \mathring{t}_{4}(u,x,y)&=\frac{1}{2}\left(q_{14}(u,x,y)-q_{25}(u,x,y)\right), \quad \mathring{t}_{5}(u,x,y)=\frac{1}{2}\left(q_{21}(u,x,y)-q_{7}(u,x,y)\right),\\
    \mathring{t}_{6}(u,x,y)&=\frac{1}{2}\left[q_{9}(u,x,y)-q_{3}(u,x,y)+H(u,x,y)\left(q_{21}(u,x,y)-q_{7}(u,x,y)\right)\right],\\
    \mathring{t}_{7}(u,x,y)&=\frac{1}{2}\left(q_{11}(u,x,y)-q_{29}(u,x,y)\right), \quad \mathring{t}_{8}(u,x,y)=\frac{1}{2}\left(q_{15}(u,x,y)-q_{33}(u,x,y)\right),\\
    \mathring{t}_{9}(u,x,y)&=\frac{1}{2}\left(q_{25}(u,x,y)-q_{8}(u,x,y)\right), \quad \mathring{t}_{11}(u,x,y)=\frac{1}{2}\left(q_{12}(u,x,y)-q_{33}(u,x,y)\right),\\
    \mathring{t}_{10}(u,x,y)&=\frac{1}{2}\left[q_{13}(u,x,y)-q_{4}(u,x,y)+H(u,x,y)\left(q_{25}(u,x,y)-q_{8}(u,x,y)\right)\right],\\
    \mathring{t}_{12}(u,x,y)&=\frac{1}{2}\left[q_{2}(u,x,y)+2q_{5}(u,x,y)+2q_{11}(u,x,y)+3q_{16}(u,x,y)+q_{29}(u,x,y)\right],\\
    \mathring{t}_{14}(u,x,y)&=\frac{1}{2}\left(q_{10}(u,x,y)-q_{7}(u,x,y)\right), \quad \mathring{t}_{15}(u,x,y)=\frac{3}{2}q_{23}(u,x,y)\,, \quad \mathring{t}_{16}(u,x,y)=\frac{3}{2}q_{24}(u,x,y)\,,\\
    \mathring{t}_{18}(u,x,y)&=\frac{1}{2}\left(q_{14}(u,x,y)-q_{8}(u,x,y)\right), \quad \mathring{t}_{22}(u,x,y)=\frac{1}{2}\left(q_{15}(u,x,y)-q_{12}(u,x,y)\right).
\end{align}

Furthermore, the condition~\eqref{covdevk} imposes additional constraints on the torsion and nonmetricity functions. Focusing on each of these two sectors separately, first of all, the functions related to the dynamical piece of the torsion tensor are constrained as:
\begin{align}
    \bar{t}_{2}(u,x,y)&=\bar{t}_{5}(u,x,y)=\bar{t}_{9}(u,x,y)=\bar{t}_{15}(u,x,y)=\bar{t}_{16}(u,x,y)=0\,,\label{eq1branch1}\\
    \bar{t}_{14}(u,x,y)&=-\,\bar{t}_{3}(u,x,y)\,, \quad \bar{t}_{18}(u,x,y)=-\,\bar{t}_{4}(u,x,y)\,,\label{eq2branch1}
\end{align}
if $\alpha = 0$, whereas if $\alpha \neq 0$ the result is:
\begin{align}
    \bar{t}_{15}(u,x,y)&=\bar{t}_{16}(u,x,y)=0\,,\\
    \bar{t}_{3}(u,x,y)&=\frac{\alpha-1}{2\alpha}\bar{t}_{5}(u,x,y)\,,\\
    \bar{t}_{4}(u,x,y)&=\frac{\alpha-1}{2\alpha}\bar{t}_{9}(u,x,y)\,,\\
    \bar{t}_{7}(u,x,y)&=-\,\bar{t}_{12}(u,x,y)-\frac{\alpha+1}{\alpha}\,\bar{t}_{2}(u,x,y)\,,\\
    \bar{t}_{14}(u,x,y)&=-\,\frac{3\alpha-1}{2\alpha}\,\bar{t}_{5}(u,x,y)\,,\\
    \bar{t}_{18}(u,x,y)&=-\,\frac{3\alpha-1}{2\alpha}\,\bar{t}_{9}(u,x,y)\,.
\end{align}
On the other hand, for the nonmetricity sector, we have:
\begin{align}
    q_{21}(u,x,y)=&\;q_{23}(u,x,y)=q_{24}(u,x,y)=q_{25}(u,x,y)=0\,,\\
    q_{10}(u,x,y)=&-\frac{3\left(2\beta+1\right)}{3\beta+4\gamma}\,q_{7}(u,x,y)\,,\\
    q_{14}(u,x,y)=&-\frac{3\left(2\beta+1\right)}{3\beta+4\gamma}\,q_{8}(u,x,y)\,,\\
     q_{16}(u,x,y)=&-\frac{1}{2\left(3\beta-4\gamma\right)}\left[\left(3\beta+4\gamma\right)q_{2}(u,x,y)+6\left(2\beta+1\right)q_{5}(u,x,y)\right]-q_{11}(u,x,y)\,,
\end{align}
provided that $3\beta \neq \pm \,4\gamma$. In the particular cases where $3\beta = \pm \,4\gamma$, the relations then simply read:
\begin{align}
    q_{21}(u,x,y)=&\;q_{23}(u,x,y)=q_{24}(u,x,y)=q_{25}(u,x,y)=0\,,\\
    q_{2}(u,x,y)=&-\frac{8\gamma+3}{4\gamma}\,q_{5}(u,x,y)\,,\\
    q_{10}(u,x,y)=&-\frac{\left(8\gamma+3\right)}{8\gamma}\,q_{7}(u,x,y)\,,\\
    q_{14}(u,x,y)=&-\frac{\left(8\gamma+3\right)}{8\gamma}\,q_{8}(u,x,y)\,,
\end{align}
for $3\beta = 4\gamma$, whereas for $3\beta = -\,4\gamma$:
\begin{align}
    q_{7}(u,x,y)=&\;q_{8}(u,x,y)=q_{21}(u,x,y)=q_{23}(u,x,y)=q_{24}(u,x,y)=q_{25}(u,x,y)=0\,,\\
     q_{16}(u,x,y)=&-\frac{1}{8\gamma}\left(8\gamma-3\right)q_{5}(u,x,y)-q_{11}(u,x,y)\,.
\end{align}
Finally, the most restricted case with $\beta=\gamma=0$ involves:
\begin{equation}
    q_{5}(u,x,y)=q_{7}(u,x,y)=q_{8}(u,x,y)=q_{21}(u,x,y)=q_{23}(u,x,y)=q_{24}(u,x,y)=q_{25}(u,x,y)=0\,.
\end{equation}

Overall, the previous conditions meaningfully constrain the form and the number of independent torsion and nonmetricity functions to $9$ and $16$, respectively. Thus, in the next subsections we shall consider this form to obtain gravitational wave solutions with dynamical torsion and nonmetricity to the field equations~\eqref{tetrad_eq} and~\eqref{connection_eq}, starting from the simplest case of Riemann-Cartan geometry and proceeding to the extended cases of Weyl-Cartan geometry and general metric-affine geometry.

\subsection{Riemann-Cartan geometry}\label{subsecRC}

In Riemann-Cartan geometry, the nonmetricity tensor identically vanishes, thus restricting the remaining \,geometrical degrees of freedom (dof) to $9$ independent functions. In general, for different combinations of these functions, it is possible to find analytical solutions to the field equations, but given their cumbersome form here we simply present the most relevant solution, which displays an arbitrary dof of torsion in the metric tensor. Specifically, assuming the branch given by Expressions~\eqref{eq1branch1}-\eqref{eq2branch1} with $\alpha=0$ and $t_{7}=t_{8}=t_{11}=t_{12}=0$, the respective Type N algebraic conditions~\eqref{TypeN_cond1},~\eqref{TypeN_condd1} and~\eqref{TypeN_cond4} of the field strengths of torsion lead to the differential equations
\begin{align}
    &\partial_{x}t_3(u, x, y) + \partial_{y}t_4(u, x, y) + t^{2}_{3}(u, x, y)+t^{2}_{4}(u, x, y)=0\,,\\
    &\partial_{x}t_3(u, x, y) - \partial_{y}t_4(u, x, y)- t^{2}_{3}(u, x, y)+ t^{2}_{4}(u, x, y)=0\,,\\    &\partial_{x}t_4(u, x, y) - t_{3}(u, x, y)t_{4}(u, x, y)=0\,,\\
 &   \partial_y t_3(u,x,y)-t_3(u,x,y) t_4(u,x,y)=0\,.
\end{align}
By computing the integrability conditions of these equations, one readily arrives at
\begin{equation}
    t_{3}^{2}(u,x,y) +t_{4}^{2}(u,x,y)=0\,,
\end{equation}
namely
\begin{equation}
    t_{3}(u,x,y)=t_{4}(u,x,y)=0\,.
\end{equation}

The connection field equations in Riemann-Cartan geometry take then the form
\begin{align}
     9d_{1}\left(\partial_{xy}t_{6}(u,x,y)+\partial_{ux}t_{22}(u,x,y)-\partial_{xx}t_{10}(u,x,y)\right)+2\left(d_{1}+h_{25}-3h_{6}\right)t_{22}(u,x,y)\partial_{y}t_{22}(u,x,y)&=0\,,\\
    9d_{1}\left(\partial_{yy}t_{6}(u,x,y)+\partial_{uy}t_{22}(u,x,y)-\partial_{xy}t_{10}(u,x,y)\right)-2\left(d_{1}+h_{25}-3h_{6}\right)t_{22}(u,x,y)\partial_{x}t_{22}(u,x,y)&=0\,,
\end{align}
which for arbitrary values of the function $t_{22}$ directly gives
\begin{align}
    &h_{6}=\frac{1}{3}\left(d_{1}+h_{25}\right),\label{RC}\\
    &t_{6}(u,x,y)=\int\left(\partial_{x}t_{10}(u,x,y)-\partial_{u}t_{22}(u,x,y)\right)dy\,.
\end{align}
On the other hand, the tetrad field equations are reduced to the single differential equation
\begin{align}
    &\frac{1}{3}\left(d_{1}+4h_{25}\right)\left[t_{22}(u,x,y)\left(\partial_{xx}+\partial_{yy}\right)t_{22}(u,x,y)+\left(\partial_{x}t_{22}(u,x,y)\right)^{2}+\left(\partial_{y}t_{22}(u,x,y)\right)^{2}\right]\nonumber\\
    &-\left(\partial_{xx}+\partial_{yy}\right)H(u,x,y)=0\,,
\end{align}
whose general solution is expressed as
\begin{equation}
    H(u,x,y) = \mathring{H}(u,x,y)+\frac{1}{6}\left(d_{1}+4h_{25}\right)t^{2}_{22}(u,x,y)\,,\label{sol_RC}
\end{equation}
where $\mathring{H}(u,x,y)$ is subject to the Laplace equation in the transverse space
\begin{equation}
   \left(\partial_{xx}+\partial_{yy}\right)\mathring{H}(u,x,y)=0\,. 
\end{equation}

Therefore, the presence of dynamical torsion in the space-time geometry corrects the form of the vacuum pp-waves of GR, by including a quadratic contribution of the free function $t_{22}$ in the metric tensor.

\subsection{Weyl-Cartan geometry}\label{subsecWC}

The simplest extension of the previous solutions takes place in Weyl-Cartan geometry, where the Weyl vector of the nonmetricity tensor is switched on. In this case, the remaining nonmetricity functions are restricted as
\begin{align}
	q_{2}(u,x,y)&=q_{9}(u,x,y)=q_{11}(u,x,y)=q_{12}(u,x,y)=q_{13}(u,x,y)=q_{15}(u,x,y)=q_{33}(u,x,y)=0\,,\\
        q_{5}(u,x,y)&=-\,q_{29}(u,x,y)=-\,\frac{q_{1}(u,x,y)}{H(u,x,y)}\,,\\
	q_{7}(u,x,y)&=-\,q_{31}(u,x,y)=-\,\frac{q_{3}(u,x,y)}{H(u,x,y)}\,,\\
	q_{8}(u,x,y)&=-\,q_{32}(u,x,y)=-\,\frac{q_{4}(u,x,y)}{H(u,x,y)}\,,
\end{align}
which sets the form of the Weyl vector as
\begin{equation}
    W_{\mu}=-\,\frac{1}{H(u,x,y)}\left(q_{1}(u,x,y),0,q_{3}(u,x,y),q_{4}(u,x,y)\right),
\end{equation}
and fixes also the constant $\beta=-\,1/2$ of the recurrence condition of the wave null vector.

Thus, the corresponding field strength of the Weyl vector satisfies the Type N algebraic conditions~\eqref{TypeN_cond3} and~\eqref{TypeN_condd3} if
\begin{equation}
    H(u,x,y)\left(\partial_{y}q_{3}(u,x,y)-\partial_{x}q_{4}(u,x,y)\right)+q_{4}(u,x,y)\,\partial_{x}H(u,x,y)-q_{3}(u,x,y)\,\partial_{y}H(u,x,y)=0\,,
\end{equation}
which determines one of the nonmetricity functions as
\begin{equation}
    q_{4}(u,x,y)=H(u,x,y)\int\left(\frac{H(u,x,y)\,\partial_{y}q_{3}(u,x,y)-q_{3}(u,x,y)\,\partial_{y}H(u,x,y)}{H^{2}(u,x,y)}\right)dx\,.
\end{equation}

On the other hand, for general values of the Lagrangian coefficients that remain yet unfixed in our analysis, the field equations include interaction terms between the torsion tensor and the Weyl vector that constrain the form of the torsion functions $t_{10}$ and $t_{22}$. Nevertheless, such terms identically vanish by setting
\begin{equation}
    h_{137}=\frac{1}{8}\left(6h_{34}-3h_{33}-12h_{35}-4h_{132}-4h_{135}\right), \quad h_{162}=h_{160}\,.\label{WC}
\end{equation}
In that case, the connection field equations are significantly simplified to a Maxwell-like equation for the field strength of the Weyl vector
\begin{equation}
\left(d_{1}+4a_{2}+8a_{14}-2a_{6}\right)\nabla_{\mu}\tilde{R}^{\lambda}{}_{\lambda}{}^{\mu\nu}=0\,,
\end{equation}
which, provided that $d_{1}+4a_{2}+8a_{14}-2a_{6} \neq 0$, has a solution that can be expressed as~\cite{Stephani:2003tm}:
\begin{align}
    q_{1}(u,x,y)&=-\,uH(u,x,y)\partial_{u}w(u,x,y)\,,\\
    q_{3}(u,x,y)&=-\,uH(u,x,y)\partial_{x}w(u,x,y)\,,\\
    q_{4}(u,x,y)&=-\,uH(u,x,y)\partial_{y}w(u,x,y)\,,
\end{align}
where $w(u,x,y)$ satisfies the transverse Laplace equation
\begin{equation}
   \left(\partial_{xx}+\partial_{yy}\right)w(u,x,y)=0\,.\label{Laplace_Weyl}
\end{equation}

Finally, the tetrad field equations acquire the form
\begin{align}
    &\frac{1}{3}\left(d_{1}+4h_{25}\right)\left[t_{22}(u,x,y)\left(\partial_{xx}+\partial_{yy}\right)t_{22}(u,x,y)+\left(\partial_{x}t_{22}(u,x,y)\right)^{2}+\left(\partial_{y}t_{22}(u,x,y)\right)^{2}\right]\nonumber\\
    &+2\left(d_{1}+4a_{2}+8a_{14}-2a_{6}\right)\left[\left(\partial_{x}w(u,x,y)\right)^{2}+\left(\partial_{y}w(u,x,y)\right)^{2}\right]-\left(\partial_{xx}+\partial_{yy}\right)H(u,x,y)=0\,,
\end{align}
which extends the solution~\eqref{sol_RC} of Riemann-Cartan geometry as follows:
\begin{equation}
    H(u,x,y) = \mathring{H}(u,x,y)+\frac{1}{6}\left(d_{1}+4h_{25}\right)t^{2}_{22}(u,x,y)+\left(d_{1}+ 4a_{2} + 8a_{14}-2a_{6}\right)w^{2}(u,x,y)\,.\label{sol_WC}
\end{equation}

Hence, in the framework of Weyl-Cartan geometry, the metric structure of the vacuum pp-waves of GR is affected by the dynamics of the torsion tensor and of the Weyl vector, presenting then quadratic contributions of the arbitrary function $t_{22}$ and the function $w$, the latter satisfying the transverse Laplace equation~\eqref{Laplace_Weyl}.

\subsection{General metric-affine geometry}\label{subsecGeneralMAG}

In the case of general metric-affine geometry, on top of the torsion tensor and the Weyl vector, the traceless nonmetricity tensor is also a dynamical quantity. In general, it includes a vector mode~\eqref{Lambdavector}, as well as tensor modes~\eqref{Omegatensor} and~\eqref{qtensor}. Thus, in analogy to the torsion tensor and to the Weyl vector, these modes can potentially induce dynamical effects in the gravitational waves. In fact, we shall show this already occurs in the simplest scenario, where the tensor modes vanish and the vector mode is then determined by the nonmetricity function $q_{2}$; hereafter, denoted by $\lambda$:
\begin{equation}
    \Lambda_{\mu}=(\lambda(u,x,y),0,0,0)\,.
\end{equation}

This trivially fixes the remaining constant of the recurrence condition of the wave null vector as $\gamma=-\,1/8$, whereas the tensor quantity $\hat{R}^{(Q)}_{[\mu\nu]}$ constitutes the only field strength of the model providing kinetic terms for the vector mode, which under the present assumptions turns out to satisfy the Type N algebraic conditions~\eqref{TypeN_cond2} and~\eqref{TypeN_condd2}.

On the other hand, as in the case of Weyl-Cartan geometry, the field equations include interaction terms between this mode, the Weyl vector and the torsion tensor, which in general give rise to additional constraints on the functions $w$, $t_{10}$ and $t_{22}$. Nevertheless, it is possible to avoid these restrictions and significantly simplify the equations by setting a specific relation between the Lagrangian coefficients (see Expressions~\eqref{genMAG1}-\eqref{genMAG16} in Appendix~\ref{appe3}).

Then, by defining parameters $\{b_{i}\}_{i=1}^{3}$ as in~\eqref{par1}-\eqref{par3}, the antisymmetric and symmetric traceless components of the connection field equations~\eqref{connection_eq} provide the following differential equations for the function $\lambda$:
\begin{align}
    &\left(2a_2+a_6\right)\partial_{ux}\lambda(u,x,y)+b_{1}\lambda(u,x,y)\partial_{x}\lambda(u,x,y)=\left(2a_2+a_6\right)\partial_{uy}\lambda(u,x,y)+b_{1}\lambda(u,x,y)\partial_{y}\lambda(u,x,y)=0\,,\label{deq1}\\
    &\left(2a_2+a_6\right)\partial_{xx}\lambda(u,x,y)=\left(2a_2+a_6\right)\partial_{yy}\lambda(u,x,y)=\left(2a_2+a_6\right)\partial_{xy}\lambda(u,x,y)=0\,,\label{deq2}\\
    &b_{2}\partial_{x}\lambda(u,x,y)=b_{2}\partial_{y}\lambda(u,x,y)=b_{3}\partial_{x}\lambda(u,x,y)=b_{3}\partial_{y}\lambda(u,x,y)=0\label{deq3}\,,
\end{align}
whereas the trace component simply gives the transverse Laplace equation~\eqref{Laplace_Weyl} for the function $w$. Thereby, Eqs.~\eqref{deq1}-\eqref{deq3} restrict the form of any nontrivial solution $\lambda$ depending on the transverse coordinates as
\begin{equation}
    \lambda(u,x,y)=\lambda_{0}(u)+\lambda_{1}x+\lambda_{2}y\,,\label{q2sol}
\end{equation}
as well as the last set of Lagrangian coefficients to vanish $b_{1}$, $b_{2}$ and $b_{3}$ (see Expressions~\eqref{genMAG17}-\eqref{genMAG19}), being $\lambda_{0}$ an arbitrary function of $u$, $\lambda_{1}$ and $\lambda_{2}$ two constant parameters.

Finally, the tetrad field equations~\eqref{tetrad_eq} are simply reduced to the single equation
\begin{align}
    &\frac{1}{3}\left(d_{1}+4h_{25}\right)\left[t_{22}(u,x,y)\left(\partial_{xx}+\partial_{yy}\right)t_{22}(u,x,y)+\left(\partial_{x}t_{22}(u,x,y)\right)^{2}+\left(\partial_{y}t_{22}(u,x,y)\right)^{2}\right]\nonumber\\
    &+2\left(d_{1}-2a_{6}+4a_{2}+8a_{14}\right)\left[\left(\partial_{x}w(u,x,y)\right)^{2}+\left(\partial_{y}w(u,x,y)\right)^{2}\right]-\left(\partial_{xx}+\partial_{yy}\right)H(u,x,y)\nonumber\\
    &+\frac{3}{4}\left(2a_{2}+a_{6}\right)\left[\left(\partial_{x}\lambda(u,x,y)\right)^{2}+\left(\partial_{y}\lambda(u,x,y)\right)^{2}\right]=0\,,
\end{align}
whose general solution can be expressed as
\begin{equation}
    H(u,x,y) = \mathring{H}(u,x,y) +l_{1}t^{2}_{22}(u,x,y)+l_{2}w^{2}(u,x,y)+l_{3}\lambda^{2}(u,x,y)\,,\label{gen_sol}
\end{equation}
where $\mathring{H}(u,x,y)$ satisfies the transverse Laplace equation
\begin{equation}
   \left(\partial_{xx}+\partial_{yy}\right)\mathring{H}(u,x,y)=0\,, 
\end{equation}
and the constant parameters $\{l_{i}\}_{i=1}^{3}$ are linear combinations of the Lagrangian coefficients
\begin{align}
    l_{1}&=\frac{1}{6}\left(d_{1}+4h_{25}\right),\label{parameter1}\\
    l_{2}&=d_{1}+ 4a_{2} + 8a_{14}-2a_{6}\,,\label{parameter2}\\
    l_{3}&=\frac{3}{8}\left(2a_{2} + a_{6}\right).\label{parameter3}
\end{align}

Overall, the nonvanishing torsion and nonmetricity functions of the solutions are written below:
\begin{align}
    t_{2}(u,x,y)&=t_7(u,x,y)=t_{12}(u,x,y)=\frac{1}{8}\left(4u\partial_{u}w(u,x,y)-3\lambda(u,x,y)\right),\label{functionsol_i}\\
    t_{5}(u,x,y)&=t_{14}(u,x,y)=-\,t_{24}(u,x,y)= -\,\frac{1}{2}u\partial_{x}w(u,x,y)\,,\\
    t_{6}(u,x,y)&=\int\left(\partial_{x}t_{10}(u,x,y)-\partial_{u}t_{22}(u,x,y)\right)dy\,,\\
     t_{9}(u,x,y)&=t_{18}(u,x,y)=t_{23}(u,x,y)= -\,\frac{1}{2}u\partial_{y}w(u,x,y)\,,\\
    q_{1}(u,x,y)&=-\,\frac{1}{4}H(u,x,y)\left(4u\partial_{u}w(u,x,y)+3\lambda(u,x,y)\right),\\
    q_{2}(u,x,y)&=-\,2q_{11}(u,x,y)=-\,2q_{16}(u,x,y)=\lambda(u,x,y)\,,\\
    q_{3}(u,x,y)&=-\,uH(u,x,y)\partial_{x}w(u,x,y)\,,\\
    q_{4}(u,x,y)&=-\,uH(u,x,y)\partial_{y}w(u,x,y)\,,\\
    q_{5}(u,x,y)&=u\partial_{u}w(u,x,y)+\frac{1}{4}\lambda(u,x,y)\,,\\
    q_{7}(u,x,y)&=-\,q_{31}(u,x,y)=-\,q_{39}(u,x,y)=u\partial_{x}w(u,x,y)\,,\\
    q_{8}(u,x,y)&=-\,q_{32}(u,x,y)=-\,q_{40}(u,x,y)=u\partial_{y}w(u,x,y)\,,\\
    q_{29}(u,x,y)&=q_{37}(u,x,y)=\frac{1}{4}\lambda(u,x,y)-u\partial_{u}w(u,x,y)\,,\label{functionsol_f}
\end{align}
where $t_{10}$ and $t_{22}$ remain arbitrary, $w$ must satisfy the transverse Laplace equation~\eqref{Laplace_Weyl} and $\lambda$ is given by\, Expression~\eqref{q2sol}. Additionally, apart from the stability and compatibility conditions summarised in Sec.~\ref{sec:defs}, a large number of Lagrangian coefficients must also be fixed for the solutions to satisfy the field equations of the model (see Appendix~\ref{appe3}). In this regard, it is important to emphasise the classical nature of our analysis, which accordingly does not investigate potential radiative instabilities that could spoil this tuning in the quantum regime. For this task, it would be especially interesting to search for and characterise possible hidden symmetries that could prevent their emergence.

In terms of the irreducible decomposition of torsion and nonmetricity under the four-dimensional pseudo-orthogonal group, the nonvanishing vector, axial and tensor modes of the solutions read:
\begin{align}
    T_{\mu} &=\frac{3}{8}\left(4W_{\mu}-3\Lambda_{\mu}\right), \quad S_{\mu} =\left(2t_{22}(u,x,y),0,0,0\right),\\
    W_{\mu} &=u\partial_{\mu}w(u,x,y)\,,\quad
    \Lambda_{\mu} =\left(\lambda(u,x,y),0,0,0\right),\\
    t_{\lambda\mu\nu} &= \frac{1}{3} \left(
\begin{array}{cccccc}
0 & 3t_{6}(u,x,y) & 3t_{10}(u,x,y) & 0 & 0 & 2t_{22}(u,x,y) \\
0 & 0 & 0 & 0 & 0 & 0 \\
0 & 0 & t_{22}(u,x,y) & 0 & 0 & 0 \\
0 & -\,t_{22}(u,x,y) & 0 & 0 & 0 & 0
\end{array} \right),
\end{align}
where the columns in the matrix representation refer to antisymmetric indices in the order ($uv$, $ux$, $uy$, $vx$, $vy$, $xy$).

As for the algebraic structure, the field strengths providing kinetic terms for the previous vector, axial and tensor modes of torsion and nonmetricity turn out to be Type N. Specifically, such field strengths are given by the nontrivial components
\begin{align}
    {\nearrow\!\!\!\!\!\!\!\tilde{R}}^{(T)}_{u[uxy]}&=\frac{1}{6}\,t_{22}(u,x,y)\lambda(u,x,y)\,, \quad {\nearrow\!\!\!\!\!\!\!\tilde{R}}^{(Q)}_{u [uxy]}=\frac{1}{2}\,t_{22}(u,x,y)\lambda(u,x,y)\,,\label{fs1}\\
    \tilde{R}^{(T)}_{[ux]}&=\frac{1}{8}\left(4\partial_{x}w(u,x,y)+3\lambda_{1}\right),\quad \hat{R}^{(Q)}_{[ux]}=-\,\frac{3}{4}\lambda_{1}\,, \quad \tilde{R}^{\lambda}{}_{\lambda u x}=-\,2\partial_{x}w(u,x,y)\,,\label{fs2}\\
    \tilde{R}^{(T)}_{[uy]}&=\frac{1}{8}\left(4\partial_{y}w(u,x,y)+3\lambda_{2}\right), \quad \hat{R}^{(Q)}_{[uy]}=-\,\frac{3}{4}\lambda_{2}\,, \quad \tilde{R}^{\lambda}{}_{\lambda u y}=-\,2\partial_{y}w(u,x,y)\,,\label{fs3}
\end{align}
which clearly satisfy their respective Type N algebraic conditions~\eqref{TypeN_cond1}-\eqref{TypeN_cond5}. In this sense, it is important to note that the tensor quantity ${}^{(1)}\tilde{Z}_{\lambda\rho\mu\nu}$ only provides kinetic terms for the tensor modes of nonmetricity, which are zero for the solutions, meaning that this quantity can be excluded from the aforementioned algebraic conditions.

In any case, for completeness, one can show that the tensor ${}^{(1)}\tilde{Z}_{\lambda\rho\mu\nu}$ presents the more general Type L within its algebraic classification~\cite{Bahamonde:2024svi}. Indeed, considering the set of null vectors\begin{align}
    k^{\mu} &= (0, 1, 0, 0)\,, \quad
l^{\mu} = \frac{1}{2}(2, H(u,x,y), 0, 0)\,,\\
m^{\mu} &= \frac{1}{\sqrt{2}}(0, 0, 1, i)\,, \quad
\bar{m}^{\mu} = \frac{1}{\sqrt{2}}(0, 0, 1, -i)\,,
\end{align}it is straightforward to check that its nonvanishing complex scalars are\begin{align}
     \Delta_{0} &= {}^{(1)}\tilde{Z}_{\lambda \rho \mu \nu } l^{\lambda }l^{\rho }l^{\mu }m^{\nu } = \frac{\sqrt{2}}{4}\left(t_{6}(u,x,y)+it_{10}(u,x,y)\right)\lambda(u,x,y)\,,\label{cs0}\\
     \Delta_{1} &= \frac{1}{2} {}^{(1)}\tilde{Z}_{\lambda \rho \mu \nu } \left(l^{\lambda }l^{\rho }l^{\mu }k^{\nu }-l^{\lambda }l^{\rho }m^{\mu }\bar{m}^{\nu }\right) = \frac{i}{8}t_{22}(u,x,y)\lambda(u,x,y)\label{cs1}\,,
\end{align}
in such a way that the following invariant characterisations hold:\begin{align}
    k^{[\tau}k_{[\omega} {}^{(1)}\tilde{Z}^{\lambda]}{}_{\rho]\mu\nu} &= 0\,,\\
    {}^{(1)}\tilde{Z}_{\lambda\rho\mu[\nu}k_{\sigma]}k^\lambda k^\rho k^\mu &= 0\,,\\
    {}^{(1)}\tilde{Z}_{\lambda\rho\mu[\nu}k'_{\sigma]}k'^\lambda k'^\rho k'^\mu &= 0\,,\label{invchar3}
\end{align}
where the null vector $k'_{\mu}$ results from the complex rotation\begin{equation}
    k'_{\mu} = k_\mu+\bar{\epsilon}\,m_\mu+\epsilon\,\bar{m}_\mu+\epsilon\bar{\epsilon} \,l_\mu\,,
\end{equation}by the parameter $\epsilon = -\,4\bar{\Delta}_{1}/\bar{\Delta}_{0}$. Thereby, the tensor has two principal null directions: $k^{\mu}$ with alignment class V and $k'^{\mu}$ with alignment class I, which means that the tensor is Type L, with symbol (V,\,I). Furthermore, it is clear the quantity~\eqref{cs0} is the only nontrivial complex scalar of the tensor if $t_{22}(u,x,y) = 0$, which at the same time vanishes the field strengths ${\nearrow\!\!\!\!\!\!\!\tilde{R}}^{(T)}_{\lambda[\rho\mu\nu]}$ and ${\nearrow\!\!\!\!\!\!\!\tilde{R}}^{(Q)}_{\lambda[\rho\mu\nu]}$ of the solution. In that case, the algebraic type of the tensor ${}^{(1)}\tilde{Z}_{\lambda\rho\mu\nu}$ is reduced to Type N; being then $k^{\mu}$ a unique principal null direction with alignment class VI and further satisfying the invariant characterisation\footnote{Note that the flow diagram shown in Figure I of Ref.~\cite{Bahamonde:2024svi} refers to a different choice of the null tetrad, wherein a direct degeneracy from Type L to Type N does not occur.}:
\begin{equation}
    k_{[\sigma}{}^{(1)}\tilde{Z}_{\lambda]}{}_{\rho\mu\nu} = 0\,.
\end{equation}

Likewise, it is important to note that all of the curvature, torsion and nonmetricity invariants of the solutions have vanishing values. Indeed, considering the pseudo-orthogonality and normalisation conditions of the null vectors
\begin{align}
    k^\mu l_\mu&=-\,m^\mu \bar{m}_\mu=1\,,\label{ort1}\\
    k^\mu m_\mu&= k^\mu \bar{m}_\mu=l^\mu m_\mu= l^\mu \bar{m}_\mu=0\,,\\
    k^\mu k_\mu&= l^\mu l_\mu= m^\mu m_\mu= \bar{m}^\mu \bar{m}_\mu=0\,,\label{ort3}
\end{align}
it is straightforward to check that this fundamental property, already present in GR~\cite{Pravda:2002us}, is a direct consequence of the algebraic structure of the curvature, torsion and nonmetricity modes of the solutions:
\begin{align}
    ^{(1)}\tilde{Z}_{\lambda\rho\mu\nu}=&\;2\bigr[2\Delta_{1}\bigl(k_{(\lambda}k_{\rho)}m_{[\mu}\bar{m}_{\nu]}+k_{(\lambda}m_{\rho)}k_{[\mu}\bar{m}_{\nu]}-k_{(\lambda}\bar{m}_{\rho)}k_{[\mu}m_{\nu]}\bigr)-\bigl(\Delta_{0}k_{(\lambda}k_{\rho)}k_{[\mu}\bar{m}_{\nu]}+\bar{\Delta}_{0}k_{(\lambda}k_{\rho)}k_{[\mu}m_{\nu]}\bigr)\bigr]\,,\\
    W_{\lambda\rho\mu\nu}=&-4\bigl(\bar{\Sigma}^{(0)}_{0}k_{[\lambda}m_{\rho]}k_{[\mu}m_{\nu]}+\Sigma^{(0)}_{0}k_{[\lambda}\bar{m}_{\rho]}k_{[\mu}\bar{m}_{\nu]}\bigr)\,,\\
    ^{(1)}\tilde{W}_{\lambda\rho\mu\nu}=&\;4\bar{\Sigma}^{(1)}_{1}\bigl(k_{[\lambda}l_{\rho]}k_{[\mu}m_{\nu]}+k_{[\lambda}m_{\rho]}k_{[\mu}l_{\nu]}-k_{[\lambda}m_{\rho]}m_{[\mu}\bar{m}_{\nu]}-m_{[\lambda}\bar{m}_{\rho]}k_{[\mu}m_{\nu]}\bigr)\nonumber\\
    &+4\Sigma^{(1)}_{1}\bigl(k_{[\lambda}l_{\rho]}k_{[\mu}\bar{m}_{\nu]}+k_{[\lambda}\bar{m}_{\rho]}k_{[\mu}l_{\nu]}-k_{[\lambda}\bar{m}_{\rho]}\bar{m}_{[\mu}m_{\nu]}-\bar{m}_{[\lambda}m_{\rho]}k_{[\mu}\bar{m}_{\nu]}\bigr)\nonumber\\
    &-4\bigl(\bar{\Sigma}^{(1)}_{0}k_{[\lambda}m_{\rho]}k_{[\mu}m_{\nu]}+\Sigma^{(1)}_{0}k_{[\lambda}\bar{m}_{\rho]}k_{[\mu}\bar{m}_{\nu]}\bigr)\,,\\
    {\nearrow\!\!\!\!\!\!\!\tilde{R}}^{(Q)}_{\lambda[\rho\mu\nu]}=&\;3{\nearrow\!\!\!\!\!\!\!\tilde{R}}^{(T)}_{\lambda[\rho\mu\nu]}=3i\lambda(u,x,y)t_{22}(u,x,y)k_{\lambda}k_{[\rho}m_{\mu}\bar{m}_{\nu]}\,,\\
    {\nearrow\!\!\!\!\!\!\!R}_{\mu\nu}=&-\Phi^{(0)}_{0}k_{\mu}k_{\nu}\,, \quad {\nearrow\!\!\!\!\!\!\!\tilde{R}}_{(\mu\nu)}=-\,\Phi^{(1)}_{0}k_{\mu}k_{\nu}-\bigl(\bar{\Phi}^{(1)}_{1}k_{(\mu}m_{\nu)}+\Phi^{(1)}_{1}k_{(\mu}\bar{m}_{\nu)}\bigr)\,,\\
    {\nearrow\!\!\!\!\!\!\!\hat{R}}^{(Q)}_{(\mu\nu)}=&-\Phi^{(2)}_{0}k_{\mu}k_{\nu}-\bigl(\bar{\Phi}^{(2)}_{1}k_{(\mu}m_{\nu)}+\Phi^{(2)}_{1}k_{(\mu}\bar{m}_{\nu)}\bigr)\,,\\
    \tilde{R}^{(T)}_{[\mu\nu]}=&\;2\bigl(\Omega^{(1)}_{2}k_{[\mu}m_{\nu]}+\bar{\Omega}^{(1)}_{2}k_{[\mu}\bar{m}_{\nu]}\bigr)\,,\\
    \hat{R}^{(Q)}_{[\mu\nu]}=&\;2\bigl(\Omega^{(2)}_{2}k_{[\mu}m_{\nu]}+\bar{\Omega}^{(2)}_{2}k_{[\mu}\bar{m}_{\nu]}\bigr)\,,\\
    \tilde{R}^{\lambda}{}_{\lambda\mu\nu}=&\;2\bigl(\Omega^{(3)}_{2}k_{[\mu}m_{\nu]}+\bar{\Omega}^{(3)}_{2}k_{[\mu}\bar{m}_{\nu]}\bigr)\,,\\
    t^{\lambda}{}_{\mu\nu}=&\;\Theta_{1}\bigr(2k^{\lambda}m_{[\mu}\bar{m}_{\nu]}+m^{\lambda}k_{[\mu}\bar{m}_{\nu]}-\bar{m}^{\lambda}k_{[\mu}m_{\nu]}\bigr)-\bigl(\Theta_{0}k^{\lambda}k_{[\mu}m_{\nu]}+\bar{\Theta}_{0}k^{\lambda}k_{[\mu}\bar{m}_{\nu]}\bigr)\,,\\
    T_{\mu}=&\;\frac{3}{8}\bigl[4\bigl(\Upsilon_{0} k_{\mu}+\Upsilon_{1}m_{\mu}+\bar{\Upsilon}_{1}\bar{m}_{\mu}\bigr)-3\lambda(u,x,y)k_{\mu}\bigr]\,, \\
    S_{\mu}=&\;2it_{22}(u,x,y)\,\varepsilon_{\mu\lambda\rho\nu}k^{\lambda}m^{\rho}\bar{m}^{\nu}\,,\\
    W_{\mu}=&\;\Upsilon_{0} k_{\mu}+\Upsilon_{1}m_{\mu}+\bar{\Upsilon}_{1}\bar{m}_{\mu}\,,\\
    \Lambda_{\mu}=&\;\lambda(u,x,y)k_{\mu}\,,
\end{align}
where
\begin{align}
    \Sigma^{(0)}_{0}&= \frac{1}{4}\left(\partial_{yy}-\partial_{xx}\right)H(u,x,y)-\frac{i}{2}\partial_{xy}H(u,x,y)\,, \\
    \Sigma^{(1)}_{0}&=\Sigma^{(0)}_{0}+\frac{1}{2}\bigr[\partial_{y}t_{10}(u,x,y)-\partial_{x}t_{6}(u,x,y)-i\bigl(\partial_{y}t_{6}(u,x,y)+\partial_{x}t_{10}(u,x,y)\bigr)\bigl]\,,\\
    \Sigma^{(1)}_{1}&=\frac{1}{4\sqrt{2}}\bigl(\partial_{y}t_{22}(u,x,y)-i\partial_{x}t_{22}(u,x,y)\bigr)\,, \\
    \Phi^{(0)}_{0}&=\frac{1}{2}\left(\partial_{xx}+\partial_{yy}\right)H(u,x,y)\,,\\
    \Phi^{(1)}_{0}&=\Phi^{(0)}_{0}+\partial_{x}t_{6}(u,x,y)+\partial_{y}t_{10}(u,x,y)-\frac{1}{8}\lambda^{2}(u,x,y) -\frac{1}{2}t_{22}^{2}(u,x,y)-\partial_{u}\lambda(u,x,y)\,,\\
    \Phi^{(1)}_{1}&=\frac{1}{\sqrt{2}}\left[\partial_{x}\lambda(u,x,y)-\partial_{y}t_{22}(u,x,y)+i\left(\partial_{y}\lambda(u,x,y)+\partial_{x}t_{22}(u,x,y)\right)\right],\\
    \Phi^{(2)}_{0}&=2\partial_{u}\lambda(u,x,y)\,,\\
    \Phi^{(2)}_{1}&=-\,\sqrt{2}\left(\partial_{x}\lambda+i\partial_{y}\lambda\right),\\
    \Omega^{(1)}_{2}&=-\,\frac{1}{8\sqrt{2}}\left[4\partial_{x}w(u,x,y)+3\partial_{x}\lambda(u,x,y)-i\left(4\partial_{y}w(u,x,y)+3\partial_{y}\lambda(u,x,y)\right)\right],\\
    \Omega^{(2)}_{2}&=\frac{3}{4\sqrt{2}}\left(\partial_{x}\lambda(u,x,y)-i\partial_{y}\lambda(u,x,y)\right),\\
    \Omega^{(3)}_{2}&=\sqrt{2}\left(\partial_{x}w(u,x,y)-i\partial_{y}w(u,x,y)\right),\\
    \Theta_{0}&=\sqrt{2}\left(t_{6}(u,x,y)-it_{10}(u,x,y)\right),\\
    \Theta_{1}&=\frac{2i}{3}t_{22}(u,x,y)\,,\\
    \Upsilon_{0}&=u\partial_{u}w(u,x,y)\,,\\
    \Upsilon_{1}&=-\,\frac{u}{\sqrt{2}}\left(\partial_{x}w(u,x,y)-i\partial_{y}w(u,x,y)\right).
\end{align}

Taking into account the nontrivial components of the field strengths~\eqref{fs1}-\eqref{fs3} of the vector, axial and tensor modes of torsion and nonmetricity, as well as the corresponding components of the Riemannian Weyl and Ricci tensors
\begin{align}
    W_{uxux} &=-\,W_{uyuy}=\frac{1}{4}\left(\partial_{xx}-\partial_{yy}\right)H(u,x,y)\,, \quad W_{uxuy} =\frac{1}{2}\partial_{xy}H(u,x,y)\,,\\
    R_{uu} &=-\,\frac{1}{2}\left(\partial_{xx}+\partial_{yy}\right)H(u,x,y)\,,
\end{align}
it is clear that the solutions reduce to plane waves when the aforementioned components do not depend on the transverse coordinates $x$ and $y$. Thus, neglecting redundant terms in the metric, torsion and nonmetricity functions, this condition holds in general if
\begin{subequations}
\begin{align}
    \mathring{H}(u,x,y)&=\mathring{H}_{1}(u)\left(x^2-y^2\right)+\mathring{H}_{2}(u)\,xy\,,\\
    \lambda(u,x,y)&=0\,, \quad t_{22}(u,x,y)=t_{22, 1}(u)\,x+t_{22, 2}(u)\,y\,,\\
    w(u,x,y)&=w_{1}(u)\,x+w_{2}(u)\,y\,,
\end{align}
\end{subequations}
or if
\begin{subequations}
\begin{align}
    \mathring{H}(u,x,y)&=\mathring{H}_{1}(u)\left(x^2-y^2\right)+\mathring{H}_{2}(u)\,xy\,,\\
    \lambda(u,x,y)&=\lambda_{1}x+\lambda_{2}y\,, \quad t_{22}(u,x,y)=0\,,\\
    w(u,x,y)&=w_{1}(u)\,x+w_{2}(u)\,y\,,
\end{align}
\end{subequations}
where $\{\mathring{H}\}_{i=1}^{2}$, $\{w_{i}\}_{i=1}^{2}$ and $\{t_{22,i}\}_{i=1}^{2}$ are arbitrary functions of the coordinate $u$, while a possible inverse relation between the torsion function $t_{22}$ and the nonmetricity function $\lambda$ is discarded to avoid the occurrence of singularities.

Therefore, an immediate conclusion is that the nonlinear interaction between the dynamical modes of torsion and the dynamical vector mode of the traceless nonmetricity tensor  prevents these quantities to propagate as plane waves. In this sense, vanishing any of them trivialises the interaction, which at the same time simplifies the algebraic structure of the tensor $^{(1)}\tilde{Z}_{\lambda \rho \mu \nu}$ to the Type N if $t_{22}=0$ or to the Type O if $\lambda = 0$. 

On the other hand, according to Expression~\eqref{gen_sol}, the metric tensor of the solutions includes then corrections from the square of the torsion and nonmetricity functions $t_{22}$, $w$ and $q_{2}$, namely from the three relevant parts of the independent affine connection. Introducing a local basis with the vector fields
\begin{align}
    \theta^{\hat{0}}=\frac{1}{\sqrt{2}}\left[\bigl(1-H(u,x,y)/2\bigr)du+dv\right], \quad
    \theta^{\hat{1}}=\frac{1}{\sqrt{2}}\left[\bigl(1+H(u,x,y)/2\bigr)du-dv\right],\quad
    \theta^{\hat{2}}=-\,dx\,, \quad \theta^{\hat{3}}=-\,dy\,,
\end{align}
the geodesic deviation equation describing the displacement of a vector $X^{a}$ along a reference timelike geodesic aligned with $\theta^{\hat{0}}$ takes the form~\cite{Pirani:1956tn,Pirani:1956wr,Szekeres:1965ux,Bicak:1999hb,Podolsky:2012he,deReyNeto:2003mp}:
\begin{equation}
    \frac{d^{2}X^{a}}{d\tau^{2}}=
    R^{a}{}_{\hat{0}b\hat{0}}X^{b}\,,
\end{equation}
namely
\begin{align}
    &\frac{d^{2}X^{\hat{0}}}{d\tau^{2}}=\frac{d^{2}X^{\hat{1}}}{d\tau^{2}}=0\,,\\
    &\frac{d^{2}X^{\hat{2}}}{d\tau^{2}}=\left(\mathcal{A}_{\circ}-\mathcal{A}_{+}\right)X^{\hat{2}}+\mathcal{A}_{\cross}X^{\hat{3}}\,,\\
    &\frac{d^{2}X^{\hat{3}}}{d\tau^{2}}=\left(\mathcal{A}_{\circ}+\mathcal{A}_{+}\right)X^{\hat{3}}+\mathcal{A}_{\cross}X^{\hat{2}}\,,
\end{align}
where
\begin{equation}
    \mathcal{A}_{\circ} = -\,\frac{1}{4}\Phi^{(0)}_{0}\,, \quad \mathcal{A}_{+} = -\,\frac{1}{4}\bigl(\Sigma^{(0)}_{0}+\bar{\Sigma}^{(0)}_{0}\bigr), \quad \mathcal{A}_{\cross} = -\,\frac{i}{4}\bigl(\Sigma^{(0)}_{0}-\bar{\Sigma}^{(0)}_{0}\bigr).
\end{equation}

Thereby, it is clear the contributions $\mathcal{A}_{+}$ and $\mathcal{A}_{\cross}$ give rise to the purely transverse effects in the relative motion of nearby test particles with vanishing hypermomentum and reduce to the propagating helicity-2 polarisation modes that arise in the Minkowski background of GR under the weak-field limit approximation. Likewise, the quantity $\mathcal{A}_{\circ}$ is invariant under rotations of the transverse plane, leading to scalar effects and representing a helicity-0 polarisation mode under the mentioned limit, but in this case induced by the dynamical torsion and nonmetricity fields.

\section{Conclusions}\label{sec:conclusions}

In this work, we have performed a thorough study on the gravitational waves based on the metric-affine geometry of cubic MAG. For this task, we have focused on a cubic MAG model that has been recently considered to eliminate ghostly instabilities from the vector and axial sectors of the theory, deriving its field equations and finding exact gravitational wave solutions that extend the form of the pp-wave space-times of GR in the presence of dynamical torsion and nonmetricity.

In general, due to the highly nonlinear character of the field equations and the large number of dof involved, the search of exact solutions in cubic MAG is an arduous challenge. To overcome this issue, we have imposed a set of consistency constraints, which shape the gravitational wave profile with torsion and nonmetricity, and strongly simplify the problem to solve the field equations in a systematic way.

Specifically, we first consider that the corresponding metric structure acquires a pp-wave form in a flat transverse space, whereas the torsion and nonmetricity fields are preserved under the action of the wave null vector defining the propagation direction. In addition, we impose transversality by a set of orthogonality conditions between the wave null vector and the respective vector, axial and tensor modes of torsion and nonmetricity. In any case, these conditions do not guarantee that the wave vector is a well-defined preferred null vector field of the space-time endowed with torsion and nonmetricity, but we impose an additional recurrence condition to ensure this property. In fact, in the framework of GR, the wave null vector represents a principal null direction of the Weyl and Ricci tensors, which accordingly satisfy the Type N algebraic conditions of the Petrov and Segre classifications, respectively. Hence, we consider the full algebraic classification of metric-affine geometry, recently derived in Refs.~\cite{Bahamonde:2023piz,Bahamonde:2024svi}, in order to extend these conditions to the respective field strength tensors providing kinetic terms to the vector, axial and tensor modes of torsion and nonmetricity in our cubic MAG model. Overall, these conditions consistently constrain the geometrical dof of the model, which allows us to find new exact gravitational wave solutions with the three main contributions of the independent affine connection; namely, the torsion tensor, the Weyl vector and the traceless nonmetricity tensor.

Thereby, the solutions represent pp-waves in metric-affine geometry, thus exhibiting certain distinctive features, in comparison to their counterparts of GR. In particular, while the metric tensor can retain its plane-wave structure in the cases where the Riemannian Weyl and Ricci tensors do not depend on the transverse coordinates, the dynamical modes of the torsion and traceless nonmetricity tensors exhibit a nonlinear interaction that prevents these quantities to propagate as plane waves. Furthermore, all of these modes and the Weyl vector, describing the trace part of the nonmetricity tensor, give rise to scalar effects in the relative motion of nearby test particles with vanishing hypermomentum, featuring a helicity-0 polarisation mode in the metric tensor under the weak-field limit approximation.

Current observational data place upper bounds on the amplitude of scalar polarisation modes~\cite{LIGOScientific:2017ycc,TheLIGOScientific:2017qsa,LIGOScientific:2018dkp}, but do not exclude their presence at subdominant levels~\cite{Takeda:2021hgo,Wu:2024yno,Liang:2024sfn}. In this sense, it is worth emphasising that although the parameter space of the cubic MAG model of the solutions includes, on top of the gravitational constant of GR, a total of $64$ independent coefficients, the aforementioned effects of the torsion and nonmetricity tensors in the gravitational-wave spectrum are solely introduced by $3$ parameters (see Expressions~\eqref{parameter1}-\eqref{parameter3}). Hence, from a statistical perspective, the gravitational-wave phenomenology induced by the torsion and nonmetricity tensors under current observational constraints is effectively parametrised by these $3$ quantities, which mitigates the suppression of the marginal likelihood that typically occurs in Bayesian model selection when integrating over large prior volumes.

Further developments such as the improvement in the sensitivity of the interferometers and the expansion of the global network may thus enable the detection of these subtle effects, providing a complementary probe of the extended gravitational dynamics beyond GR.

\bigskip

\noindent
\section*{Acknowledgements}
The work of S.B. is supported by the Institute for Basic Science (IBS-R018-D3), and was also supported in its initial stages by the Agencia Nacional de Investigación y Desarrollo (ANID) through the “Becas Chile Postdoctorado al Extranjero” program, Grant No. 74220006. The work of J.G.V. is \,supported by the Institute for Basic Science (IBS-R003-D1). The work of J.M.M.S. is supported by the Basque Government Grant No. IT1628-22, and by the Grant No. PID2021-123226NB-I00 funded by the Spanish MCIN/AEI/10.13039/501100011033 together with ``ERDF A way of making Europe''.

\newpage

\appendix

\section{Explicit form of the Lagrangian densities with mixing terms of curvature, torsion and/or nonmetricity}\label{appe1}

In general, the Lagrangian densities of the cubic MAG model described by Expression~\eqref{cubicmodel}, which depend on mixing terms of the curvature, torsion and/or nonmetricity tensors, acquire a cumbersome form. Specifically, they can be expressed in terms of the following contributions:
\begin{align}
    \mathcal{\bar{L}}_{\rm curv-tor}^{(3)}&=\mathcal{\bar{L}}^{(3)}_{1,\rm tor}+\mathcal{\bar{L}}^{(3)}_{2,\rm tor}+\mathcal{\bar{L}}^{(3)}_{3,\rm tor}+\mathcal{\bar{L}}^{(3)}_{4,\rm tor}+\mathcal{\bar{L}}^{(3)}_{5,\rm tor}\,,\\
    \mathcal{\bar{L}}_{\rm curv-nonm}^{(3)}&=\mathcal{\bar{L}}^{(3)}_{1,\rm non}+\mathcal{\bar{L}}^{(3)}_{2,\rm non}+\mathcal{\bar{L}}^{(3)}_{3,\rm non}+\mathcal{\bar{L}}^{(3)}_{4,\rm non}+\mathcal{\bar{L}}^{(3)}_{5,\rm non}\,,\\
    \mathcal{\bar{L}}_{\rm curv-tor-nonm}^{(3)}&=\mathcal{\bar{L}}^{(3)}_{1,\rm tor-non}+\mathcal{\bar{L}}^{(3)}_{2,\rm tor-non}+\mathcal{\bar{L}}^{(3)}_{3,\rm  tor-non}+\mathcal{\bar{L}}^{(3)}_{4,\rm  tor-non}+\mathcal{\bar{L}}^{(3)}_{5,\rm tor-non}\,,
\end{align}
where we have defined:


In this parametrisation, the corresponding cubic invariants appear in the Lagrangian density with coefficients $\{\bar{h}_{i}\}_{i=1}^{209}$. An equivalent parametrisation in terms of the irreducible modes~\eqref{Tdec1}-\eqref{Tdec3} and~\eqref{Qdec1}-\eqref{qtensor} of the torsion and nonmetricity tensors can also be introduced by a set of coefficients $\{h_{i}\}_{i=1}^{209}$, leading to the following relations between the two parametrisations:
%

\section{Field equations of the cubic MAG model}\label{appe2}

Variation of the gravitational action~\eqref{cubicmodel} leads to the field equations of our cubic MAG model:
\begin{equation}
    \delta S=\frac{1}{16 \pi}\int{\left(e_{a\mu}E^{\mu\nu}\delta e^{a}\,_{\nu}+e_{a\lambda}e^{b}{}_{\mu}E^{\lambda\mu\nu}\delta \omega^{a}{}_{b\nu}\right)\sqrt{-g}\,d^4x}=0\,.\end{equation}

First of all, the tensor $E^{\lambda\mu\nu}$ provides the connection field equations in the presence of torsion and nonmetricity, presenting the general form
\begin{equation}
    E^{\lambda\mu\nu}=2\big(\nabla_\alpha Y^{\lambda \mu[ \nu \alpha] }+N^{\mu }{}_{\alpha \beta } Y^{\lambda \alpha [\nu \beta] } - N^{\alpha \lambda }{}_{\beta } Y_{\alpha }{}^{\mu [\nu \beta] }\big)-2X^{\lambda[\mu\nu]}-2Z^{(\lambda\mu)\nu}\,,
\end{equation}
where we have defined
\begin{align}
    Y^{\lambda\rho\mu\nu}=&\;2 a_{2}{} \tilde{R}^{\lambda \rho \mu \nu } + 2\left(a_{2}-c_{1}\right) \tilde{R}^{\rho \lambda \mu \nu } - \left(2c_{1}+c_{2}\right) \tilde{R}^{\mu \nu \lambda \rho } -2 a_{5}{} \tilde{R}^{\lambda \mu \nu \rho } + a_{6}{} \bigl( \tilde{R}^{\rho \mu \lambda \nu }- \tilde{R}^{\mu \lambda \nu \rho }\bigr) + 2\left(c_{2}-a_{5}+a_{6}\right) \tilde{R}^{\mu \rho \lambda \nu } \nonumber\\
    &+ a_{11}{} \bigl(g^{\nu \rho } \hat{R}^{\mu \lambda } -  g^{\mu \rho } \hat{R}^{\nu \lambda }\bigr) + a_{9}{} \bigl(g^{\lambda \mu } \tilde{R}^{\nu \rho }- g^{\lambda \nu } \tilde{R}^{\mu \rho }\bigr) + a_{10}{} \bigl(g^{\nu \rho } \hat{R}^{\lambda \mu } -  g^{\mu \rho } \hat{R}^{\lambda \nu }\bigr)+ a_{16}{}\bigl(g^{\lambda \rho } \hat{R}^{\mu \nu } + g^{\nu \rho } \bar{R}^{\lambda \mu }\bigr)\nonumber\\
    & + \left(d_{1}-a_{10}-a_{12}\right) \bigl(g^{\lambda \mu } \tilde{R}^{\rho \nu }- g^{\lambda \nu } \tilde{R}^{\rho \mu }\bigr)- a_{12}{}\bigl(g^{\mu \rho } \tilde{R}^{\lambda \nu } +  g^{\lambda \nu } \hat{R}^{\rho \mu }\bigr) + \left(d_{1}+a_{9}+a_{11}\right)\bigl(g^{\mu \rho } \tilde{R}^{\nu \lambda } + g^{\lambda \nu } \hat{R}^{\mu \rho }\bigr)\nonumber\\
    &+a_{15}{}\bigl(g^{\lambda \rho } \tilde{R}^{\mu \nu } + g^{\lambda \nu } \bar{R}^{\mu \rho }\bigr)+2a_{14}{} g^{\lambda \rho } \bar{R}^{\mu \nu }+\bar{h}_{1}^{} T_{\alpha }{}^{\mu \nu } T^{\alpha \lambda \rho } + \bar{h}_{2}^{} T_{\alpha }{}^{\mu \rho } T^{\alpha \lambda \nu } - \bar{h}_{3}^{} T^{\alpha \mu \nu } T^{\lambda \rho }{}_{\alpha } + \bar{h}_{4}^{} T^{\alpha \nu \rho } T^{\lambda \mu }{}_{\alpha }  \nonumber\\
   &+ \bar{h}_{5}^{} T^{\alpha \nu \rho } T^{\mu \lambda }{}_{\alpha } - \bar{h}_{6}^{} T^{\alpha \lambda \rho } T^{\mu \nu }{}_{\alpha } + \bar{h}_{7}^{} T^{\rho \mu \nu } T^{\lambda }- \bar{h}_{8}^{} T^{\mu \nu \rho } T^{\lambda } - \bar{h}_{9}^{} T^{\lambda \nu \rho } T^{\mu } + \bar{h}_{11}^{} T^{\lambda \rho }{}_{\alpha } T^{\mu \nu \alpha } + \bar{h}_{12}^{} T^{\lambda \mu }{}_{\alpha } T^{\rho \nu \alpha } \nonumber\\
   &+ \bar{h}_{13}^{} T^{\lambda \mu }{}_{\alpha } T^{\nu \rho \alpha } + \bar{h}_{14}^{} T^{\mu \lambda }{}_{\alpha } T^{\nu \rho \alpha }+ \bar{h}_{15}^{} g^{\lambda \mu } T^{\nu }{}_{\alpha \kappa } T^{\rho \alpha \kappa } + \bar{h}_{16}^{} g^{\lambda \mu } T^{\alpha \rho \kappa } T^{\nu }{}_{\alpha \kappa } + \bar{h}_{17}^{} g^{\lambda \mu } T^{\alpha \nu \kappa } T^{\rho }{}_{\alpha \kappa }+ \bar{h}_{18}^{} g^{\lambda \mu } T_{\alpha }{}^{\rho \kappa } T^{\alpha \nu }{}_{\kappa } \nonumber\\
   & + \bar{h}_{19}^{} g^{\lambda \mu } T^{\alpha \nu }{}_{\kappa } T^{\kappa \rho }{}_{\alpha }+ \bar{h}_{20}^{} g^{\lambda \mu } T^{\nu } T^{\rho } + \bar{h}_{21}^{} g^{\lambda \mu } T^{\rho \nu }{}_{\alpha } T^{\alpha } + \bar{h}_{22}^{} g^{\lambda \mu } T^{\nu \rho }{}_{\alpha } T^{\alpha } - \bar{h}_{23}^{} g^{\lambda \mu } T^{\alpha \nu \rho } T_{\alpha } + \bar{h}_{24}^{} g^{\lambda \mu } g^{\nu \rho } T_{\alpha }{}^{\kappa \theta } T^{\alpha }{}_{\kappa \theta }\nonumber\\
   &+ \bar{h}_{10}^{} T^{\nu \lambda \rho } T^{\mu }- \bar{h}_{25}^{} g^{\lambda \mu } g^{\nu \rho } T^{\alpha \kappa }{}_{\theta } T^{\theta }{}_{\alpha \kappa } + \bar{h}_{26}^{} g^{\lambda \mu } g^{\nu \rho } T_{\alpha } T^{\alpha } + \bar{h}_{27}^{} T^{\mu \nu }{}_{\alpha } T^{\rho \lambda \alpha } + \bar{h}_{28}^{} T^{\nu \lambda }{}_{\alpha } T^{\rho \mu \alpha }+ \bar{h}_{29}^{} T^{\alpha \lambda \nu } T^{\rho \mu }{}_{\alpha } \nonumber\\
   & + \bar{h}_{30}^{} T^{\alpha \lambda \nu } T^{\mu \rho }{}_{\alpha }+ \bar{h}_{31}^{} T^{\alpha \mu \nu } T^{\rho \lambda }{}_{\alpha } + \bar{h}_{32}^{} T^{\lambda \mu \nu } T^{\rho } + \bar{h}_{34}^{} T^{\rho \lambda \nu } T^{\mu }-\bar{h}_{35}^{} g^{\mu \rho } T^{\lambda }{}_{\alpha \kappa } T^{\nu \alpha \kappa }- \bar{h}_{36}^{} g^{\mu \rho } T_{\alpha }{}^{\nu \kappa } T^{\alpha \lambda }{}_{\kappa } \nonumber\\
   &+ \bar{h}_{33}^{} T^{\mu \lambda \nu } T^{\rho }- \bar{h}_{37}^{} g^{\mu \rho } T^{\alpha \lambda \kappa } T^{\nu }{}_{\alpha \kappa } - \bar{h}_{38}^{} g^{\mu \rho } T^{\alpha \nu \kappa } T^{\lambda }{}_{\alpha \kappa } - \bar{h}_{39}^{} g^{\mu \rho } T^{\alpha \lambda }{}_{\kappa } T^{\kappa \nu }{}_{\alpha }- \bar{h}_{40}^{} g^{\mu \rho } T^{\lambda } T^{\nu }- \bar{h}_{41}^{} g^{\mu \rho } T^{\lambda \nu }{}_{\alpha } T^{\alpha }\nonumber\\
   &- \bar{h}_{42}^{} g^{\mu \rho } T^{\nu \lambda }{}_{\alpha } T^{\alpha }- \bar{h}_{43}^{} g^{\mu \rho } T^{\alpha \lambda \nu } T_{\alpha } + \bar{h}_{44}^{} g^{\lambda \rho } T^{\alpha \nu \kappa } T^{\mu }{}_{\alpha \kappa } + \bar{h}_{45}^{} g^{\lambda \rho } T^{\alpha \mu \nu } T_{\alpha } - \bar{h}_{46}^{} g^{\lambda \rho } T^{\mu \nu }{}_{\alpha } T^{\alpha }+\bar{h}_{47}^{} Q^{\lambda \rho }{}_{\alpha } Q^{\mu \nu \alpha }\nonumber\\
   &+ \bar{h}_{48}^{} Q^{\alpha \nu \rho } Q^{\lambda \mu }{}_{\alpha } + \bar{h}_{49}^{} Q^{\lambda \mu }{}_{\alpha } Q^{\rho \nu \alpha }+ \bar{h}_{50}^{} Q^{\lambda \mu }{}_{\alpha } Q^{\nu \rho \alpha } + \bar{h}_{51}^{} Q_{\alpha }{}^{\lambda \mu } Q^{\alpha \nu \rho } + \bar{h}_{52}^{} Q_{\alpha }{}^{\lambda \alpha } Q^{\mu \nu \rho }+ \bar{h}_{53}^{} Q^{\mu \nu }{}_{\alpha } Q^{\rho \lambda \alpha }\nonumber\\
   &+\bar{h}_{54}^{} Q^{\alpha \nu \rho } Q^{\mu \lambda }{}_{\alpha } + \bar{h}_{55}^{} Q^{\mu \lambda }{}_{\alpha } Q^{\rho \nu \alpha } + \bar{h}_{56}^{} Q_{\alpha }{}^{\mu \alpha } Q^{\lambda \nu \rho } + \bar{h}_{57}^{} Q^{\mu \lambda }{}_{\alpha } Q^{\nu \rho \alpha } + \bar{h}_{58}^{} Q^{\alpha \lambda \nu } Q^{\rho \mu }{}_{\alpha }+ \bar{h}_{59}^{} Q_{\alpha }{}^{\rho \alpha } Q^{\mu \lambda \nu } \nonumber\\
   &+\bar{h}_{60}^{} Q^{\alpha \lambda \nu } Q^{\mu \rho }{}_{\alpha } + \bar{h}_{61}^{} Q_{\alpha }{}^{\mu \alpha } Q^{\rho \lambda \nu } + \bar{h}_{62}^{} Q^{\alpha \lambda \rho } Q^{\mu \nu }{}_{\alpha } + \bar{h}_{63}^{} Q_{\alpha }{}^{\mu \alpha } Q^{\nu \lambda \rho } + \bar{h}_{64}^{} Q^{\lambda \alpha }{}_{\alpha } Q^{\mu \nu \rho }- \bar{h}_{69}^{} g^{\lambda \nu } Q^{\mu }{}_{\alpha \kappa } Q^{\rho \alpha \kappa }\nonumber\\
   &+ \bar{h}_{65}^{} Q^{\lambda \nu \rho } Q^{\mu \alpha }{}_{\alpha } + \bar{h}_{66}^{} Q^{\mu \lambda \nu } Q^{\rho \alpha }{}_{\alpha } + \bar{h}_{67}^{} Q^{\mu \alpha }{}_{\alpha } Q^{\rho \lambda \nu }+ \bar{h}_{68}^{} Q^{\mu \alpha }{}_{\alpha } Q^{\nu \lambda \rho } - \bar{h}_{70}^{} g^{\lambda \nu } Q^{\alpha \rho \kappa } Q^{\mu }{}_{\alpha \kappa }- \bar{h}_{71}^{} g^{\lambda \nu } Q^{\alpha \mu \kappa } Q^{\rho }{}_{\alpha \kappa }\nonumber\\
   & - \bar{h}_{72}^{} g^{\lambda \nu } Q_{\alpha }{}^{\mu }{}_{\kappa } Q^{\alpha \rho \kappa } - \bar{h}_{73}^{} g^{\lambda \nu } Q_{\alpha }{}^{\mu }{}_{\kappa } Q^{\kappa \rho \alpha } - \bar{h}_{74}^{} g^{\lambda \nu } Q_{\alpha }{}^{\mu \alpha } Q_{\kappa }{}^{\rho \kappa } - \bar{h}_{75}^{} g^{\lambda \nu } Q_{\kappa }{}^{\alpha \kappa } Q^{\rho \mu }{}_{\alpha }- \bar{h}_{76}^{} g^{\lambda \nu } Q_{\kappa }{}^{\alpha \kappa } Q^{\mu \rho }{}_{\alpha } \nonumber\\
   &-\bar{h}_{77}^{} g^{\lambda \nu } Q_{\alpha }{}^{\mu \rho } Q_{\kappa }{}^{\alpha \kappa } - \bar{h}_{78}^{} g^{\lambda \nu } Q_{\kappa }{}^{\mu \kappa } Q^{\rho \alpha }{}_{\alpha } - \bar{h}_{79}^{} g^{\lambda \nu } Q_{\kappa }{}^{\rho \kappa } Q^{\mu \alpha }{}_{\alpha } - \bar{h}_{80}^{} g^{\lambda \nu } Q^{\mu \alpha }{}_{\alpha } Q^{\rho \kappa }{}_{\kappa }- \bar{h}_{81}^{} g^{\lambda \nu } Q^{\alpha \kappa }{}_{\kappa } Q^{\rho \mu }{}_{\alpha }\nonumber\\
   &-\bar{h}_{82}^{} g^{\lambda \nu } Q^{\alpha \kappa }{}_{\kappa } Q^{\mu \rho }{}_{\alpha } - \bar{h}_{83}^{} g^{\lambda \nu } Q_{\alpha }{}^{\mu \rho } Q^{\alpha \kappa }{}_{\kappa } + \bar{h}_{84}^{} g^{\nu \rho } Q^{\lambda }{}_{\alpha \kappa } Q^{\mu \alpha \kappa } + \bar{h}_{85}^{} g^{\nu \rho } Q^{\alpha \lambda \kappa } Q^{\mu }{}_{\alpha \kappa }+ \bar{h}_{86}^{} g^{\nu \rho } Q^{\alpha \mu \kappa } Q^{\lambda }{}_{\alpha \kappa }\nonumber\\
   &+\bar{h}_{87}^{} g^{\nu \rho } Q_{\alpha }{}^{\lambda }{}_{\kappa } Q^{\alpha \mu \kappa } + \bar{h}_{88}^{} g^{\nu \rho } Q_{\alpha }{}^{\lambda }{}_{\kappa } Q^{\kappa \mu \alpha } + \bar{h}_{89}^{} g^{\nu \rho } Q_{\alpha }{}^{\lambda \alpha } Q_{\kappa }{}^{\mu \kappa } + \bar{h}_{90}^{} g^{\nu \rho } Q_{\kappa }{}^{\alpha \kappa } Q^{\lambda \mu }{}_{\alpha } + \bar{h}_{91}^{} g^{\nu \rho } Q_{\kappa }{}^{\alpha \kappa } Q^{\mu \lambda }{}_{\alpha }\nonumber\\
   &+\bar{h}_{92}^{} g^{\nu \rho } Q_{\alpha }{}^{\lambda \mu } Q_{\kappa }{}^{\alpha \kappa } + \bar{h}_{93}^{} g^{\nu \rho } Q_{\kappa }{}^{\mu \kappa } Q^{\lambda \alpha }{}_{\alpha } + \bar{h}_{94}^{} g^{\nu \rho } Q_{\kappa }{}^{\lambda \kappa } Q^{\mu \alpha }{}_{\alpha } + \bar{h}_{95}^{} g^{\nu \rho } Q^{\lambda \alpha }{}_{\alpha } Q^{\mu \kappa }{}_{\kappa }+ \bar{h}_{96}^{} g^{\nu \rho } Q^{\alpha \kappa }{}_{\kappa } Q^{\lambda \mu }{}_{\alpha }\nonumber\\
   &+\bar{h}_{97}^{} g^{\nu \rho } Q^{\alpha \kappa }{}_{\kappa } Q^{\mu \lambda }{}_{\alpha } + \bar{h}_{98}^{} g^{\nu \rho } Q_{\alpha }{}^{\lambda \mu } Q^{\alpha \kappa }{}_{\kappa } + \bar{h}_{99}^{} g^{\lambda \rho } Q^{\alpha \nu \kappa } Q^{\mu }{}_{\alpha \kappa } + \bar{h}_{100}^{} g^{\lambda \rho } Q_{\kappa }{}^{\alpha \kappa } Q^{\mu \nu }{}_{\alpha } + \bar{h}_{101}^{} g^{\lambda \rho } Q^{\alpha \kappa }{}_{\kappa } Q^{\mu \nu }{}_{\alpha }\nonumber\\
   & + \bar{h}_{102}^{} g^{\lambda \rho } Q_{\kappa }{}^{\mu \kappa } Q^{\nu \alpha }{}_{\alpha } + \bar{h}_{103}^{} g^{\lambda \mu } g^{\nu \rho } Q_{\alpha \kappa \theta } Q^{\alpha \kappa \theta } + \bar{h}_{104}^{} g^{\lambda \mu } g^{\nu \rho } Q^{\alpha \kappa \theta } Q_{\kappa \alpha \theta }+\bar{h}_{105}^{} g^{\lambda \mu } g^{\nu \rho } Q_{\alpha }{}^{\alpha \kappa } Q_{\theta \kappa }{}^{\theta }+\bar{h}_{108}^{} Q_{\alpha }{}^{\lambda \rho } T^{\alpha \mu \nu }\nonumber\\
   &+\bar{h}_{106}^{} g^{\lambda \mu } g^{\nu \rho } Q_{\alpha }{}^{\theta }{}_{\theta } Q^{\alpha \kappa }{}_{\kappa } + \bar{h}_{107}^{} g^{\lambda \mu } g^{\nu \rho } Q^{\alpha \kappa }{}_{\kappa } Q_{\theta \alpha }{}^{\theta } - \bar{h}_{109}^{} Q_{\alpha }{}^{\lambda \mu } T^{\alpha \nu \rho }+ \bar{h}_{110}^{} Q_{\alpha }{}^{\mu \rho } T^{\alpha \lambda \nu }+ \bar{h}_{111}^{} Q^{\lambda \rho }{}_{\alpha } T^{\alpha \mu \nu }\nonumber\\
   &-\bar{h}_{112}^{} Q^{\lambda \mu }{}_{\alpha } T^{\alpha \nu \rho } + \bar{h}_{113}^{} Q^{\rho \lambda }{}_{\alpha } T^{\alpha \mu \nu } + \bar{h}_{114}^{} Q_{\alpha }{}^{\lambda \alpha } T^{\rho \mu \nu } - \bar{h}_{115}^{} Q^{\mu \lambda }{}_{\alpha } T^{\alpha \nu \rho } + \bar{h}_{116}^{} Q_{\alpha }{}^{\rho \alpha } T^{\lambda \mu \nu }-\bar{h}_{117}^{} Q_{\alpha }{}^{\mu \alpha } T^{\lambda \nu \rho }\nonumber\\
   &+\bar{h}_{118}^{} Q^{\rho \mu }{}_{\alpha } T^{\alpha \lambda \nu } + \bar{h}_{119}^{} Q^{\mu \rho }{}_{\alpha } T^{\alpha \lambda \nu }+ \bar{h}_{120}^{} Q^{\mu \nu }{}_{\alpha } T^{\alpha \lambda \rho } - \bar{h}_{121}^{} Q_{\alpha }{}^{\lambda \rho } T^{\mu \nu \alpha } - \bar{h}_{122}^{} Q_{\alpha }{}^{\lambda \mu } T^{\rho \nu \alpha }- \bar{h}_{123}^{} Q_{\alpha }{}^{\mu \rho } T^{\lambda \nu \alpha }\nonumber\\
   &-\bar{h}_{124}^{} Q^{\lambda \rho }{}_{\alpha } T^{\mu \nu \alpha } - \bar{h}_{125}^{} Q^{\lambda \mu }{}_{\alpha } T^{\rho \nu \alpha }- \bar{h}_{126}^{} Q^{\rho \lambda }{}_{\alpha } T^{\mu \nu \alpha } - \bar{h}_{127}^{} Q_{\alpha }{}^{\lambda \alpha } T^{\mu \nu \rho } - \bar{h}_{128}^{} Q^{\mu \lambda }{}_{\alpha } T^{\rho \nu \alpha }+\bar{h}_{129}^{} Q_{\alpha }{}^{\rho \alpha } T^{\mu \lambda \nu }\nonumber\\
   &+\bar{h}_{130}^{} Q_{\alpha }{}^{\mu \alpha } T^{\rho \lambda \nu } - \bar{h}_{131}^{} Q^{\rho \mu }{}_{\alpha } T^{\lambda \nu \alpha }- \bar{h}_{132}^{} Q^{\mu \rho }{}_{\alpha } T^{\lambda \nu \alpha } - \bar{h}_{133}^{} Q^{\mu \nu }{}_{\alpha } T^{\lambda \rho \alpha } - \bar{h}_{134}^{} Q_{\alpha }{}^{\lambda \mu } T^{\nu \rho \alpha }-\bar{h}_{135}^{} Q_{\alpha }{}^{\mu \rho } T^{\nu \lambda \alpha }\nonumber\\
   &-\bar{h}_{136}^{} Q^{\lambda \mu }{}_{\alpha } T^{\nu \rho \alpha } - \bar{h}_{137}^{} Q^{\mu \lambda }{}_{\alpha } T^{\nu \rho \alpha } + \bar{h}_{138}^{} Q_{\alpha }{}^{\mu \alpha } T^{\nu \lambda \rho } - \bar{h}_{139}^{} Q^{\rho \mu }{}_{\alpha } T^{\nu \lambda \alpha } - \bar{h}_{140}^{} Q^{\mu \rho }{}_{\alpha } T^{\nu \lambda \alpha }- \bar{h}_{141}^{} Q^{\mu \nu }{}_{\alpha } T^{\rho \lambda \alpha }\nonumber\\
   &+\bar{h}_{142}^{} Q^{\lambda \mu \rho } T^{\nu } + \bar{h}_{143}^{} Q^{\rho \lambda \mu } T^{\nu } + \bar{h}_{144}^{} Q^{\mu \lambda \rho } T^{\nu } + \bar{h}_{145}^{} Q^{\mu \lambda \nu } T^{\rho } + \bar{h}_{146}^{} Q^{\mu \nu \rho } T^{\lambda }+ \bar{h}_{147}^{} Q^{\lambda \alpha }{}_{\alpha } T^{\rho \mu \nu }+\bar{h}_{148}^{} Q^{\rho \alpha }{}_{\alpha } T^{\lambda \mu \nu }\nonumber\\
   &-\bar{h}_{149}^{} Q^{\mu \alpha }{}_{\alpha } T^{\lambda \nu \rho }- \bar{h}_{150}^{} Q^{\lambda \alpha }{}_{\alpha } T^{\mu \nu \rho } + \bar{h}_{151}^{} Q^{\rho \alpha }{}_{\alpha } T^{\mu \lambda \nu } + \bar{h}_{152}^{} Q^{\mu \alpha }{}_{\alpha } T^{\rho \lambda \nu } + \bar{h}_{153}^{} Q^{\mu \alpha }{}_{\alpha } T^{\nu \lambda \rho }+\bar{h}_{154}^{} g^{\lambda \mu } Q_{\alpha }{}^{\alpha }{}_{\kappa } T^{\rho \nu \kappa }\nonumber\\
   &-\bar{h}_{155}^{} g^{\lambda \nu } Q_{\alpha }{}^{\mu }{}_{\kappa } T^{\rho \alpha \kappa } + \bar{h}_{156}^{} g^{\lambda \mu } Q_{\alpha }{}^{\alpha }{}_{\kappa } T^{\nu \rho \kappa } + \bar{h}_{157}^{} g^{\lambda \mu } Q_{\alpha }{}^{\rho }{}_{\kappa } T^{\nu \alpha \kappa } - \bar{h}_{158}^{} g^{\lambda \nu } Q^{\mu }{}_{\alpha \kappa } T^{\alpha \rho \kappa } - \bar{h}_{159}^{} g^{\lambda \nu } Q_{\alpha }{}^{\mu }{}_{\kappa } T^{\alpha \rho \kappa }\nonumber\\
   &+\bar{h}_{160}^{} g^{\lambda \mu } Q^{\rho }{}_{\alpha \kappa } T^{\alpha \nu \kappa } + \bar{h}_{161}^{} g^{\lambda \mu } Q_{\alpha }{}^{\rho }{}_{\kappa } T^{\alpha \nu \kappa } - \bar{h}_{162}^{} g^{\lambda \mu } Q_{\alpha }{}^{\alpha }{}_{\kappa } T^{\kappa \nu \rho } - \bar{h}_{163}^{} g^{\lambda \nu } Q_{\alpha }{}^{\mu }{}_{\kappa } T^{\kappa \rho \alpha }+ \bar{h}_{164}^{} g^{\lambda \mu } Q_{\alpha }{}^{\rho }{}_{\kappa } T^{\kappa \nu \alpha }\nonumber\\
   &-\bar{h}_{165}^{} g^{\lambda \nu } Q_{\alpha }{}^{\mu \alpha } T^{\rho } + \bar{h}_{166}^{} g^{\lambda \mu } Q_{\alpha }{}^{\rho \alpha } T^{\nu } - \bar{h}_{167}^{} g^{\lambda \nu } Q^{\rho \mu }{}_{\alpha } T^{\alpha }- \bar{h}_{168}^{} g^{\lambda \nu } Q^{\mu \rho }{}_{\alpha } T^{\alpha }- \bar{h}_{169}^{} g^{\lambda \nu } Q_{\alpha }{}^{\mu \rho } T^{\alpha }+ \bar{h}_{170}^{} g^{\lambda \mu } Q^{\rho \alpha }{}_{\alpha } T^{\nu }\nonumber\\
   &- \bar{h}_{171}^{} g^{\lambda \nu } Q^{\mu \alpha }{}_{\alpha } T^{\rho }+ \bar{h}_{172}^{} g^{\lambda \mu } Q_{\alpha }{}^{\kappa }{}_{\kappa } T^{\rho \nu \alpha }+ \bar{h}_{173}^{} g^{\lambda \mu } Q_{\alpha }{}^{\kappa }{}_{\kappa } T^{\nu \rho \alpha }- \bar{h}_{174}^{} g^{\lambda \mu } Q_{\alpha }{}^{\kappa }{}_{\kappa } T^{\alpha \nu \rho }- \bar{h}_{175}^{} g^{\mu \rho } Q_{\alpha }{}^{\alpha }{}_{\kappa } T^{\lambda \nu \kappa }\nonumber\\
   &+\bar{h}_{176}^{} g^{\nu \rho } Q_{\alpha }{}^{\mu }{}_{\kappa } T^{\lambda \alpha \kappa }- \bar{h}_{177}^{} g^{\mu \rho } Q_{\alpha }{}^{\alpha }{}_{\kappa } T^{\nu \lambda \kappa } - \bar{h}_{178}^{} g^{\mu \rho } Q_{\alpha }{}^{\lambda }{}_{\kappa } T^{\nu \alpha \kappa }+ \bar{h}_{179}^{} g^{\nu \rho } Q^{\mu }{}_{\alpha \kappa } T^{\alpha \lambda \kappa } + \bar{h}_{180}^{} g^{\nu \rho } Q_{\alpha }{}^{\mu }{}_{\kappa } T^{\alpha \lambda \kappa }\nonumber\\
   &-\bar{h}_{181}^{} g^{\mu \rho } Q^{\lambda }{}_{\alpha \kappa } T^{\alpha \nu \kappa } - \bar{h}_{182}^{} g^{\mu \rho } Q_{\alpha }{}^{\lambda }{}_{\kappa } T^{\alpha \nu \kappa }-\bar{h}_{183}^{} g^{\mu \rho } Q_{\alpha }{}^{\alpha }{}_{\kappa } T^{\kappa \lambda \nu }+ \bar{h}_{184}^{} g^{\nu \rho } Q_{\alpha }{}^{\mu }{}_{\kappa } T^{\kappa \lambda \alpha }- \bar{h}_{185}^{} g^{\mu \rho } Q_{\alpha }{}^{\lambda }{}_{\kappa } T^{\kappa \nu \alpha }\nonumber\\
   &+\bar{h}_{186}^{} g^{\nu \rho } Q_{\alpha }{}^{\mu \alpha } T^{\lambda } - \bar{h}_{187}^{} g^{\mu \rho } Q_{\alpha }{}^{\lambda \alpha } T^{\nu } + \bar{h}_{188}^{} g^{\nu \rho } Q^{\lambda \mu }{}_{\alpha } T^{\alpha }+ \bar{h}_{189}^{} g^{\nu \rho } Q^{\mu \lambda }{}_{\alpha } T^{\alpha } + \bar{h}_{190}^{} g^{\nu \rho } Q_{\alpha }{}^{\lambda \mu } T^{\alpha }-\bar{h}_{191}^{} g^{\mu \rho } Q^{\lambda \alpha }{}_{\alpha } T^{\nu }\nonumber\\
   &+\bar{h}_{192}^{} g^{\nu \rho } Q^{\mu \alpha }{}_{\alpha } T^{\lambda } - \bar{h}_{193}^{} g^{\mu \rho } Q_{\alpha }{}^{\kappa }{}_{\kappa } T^{\lambda \nu \alpha } - \bar{h}_{194}^{} g^{\mu \rho } Q_{\alpha }{}^{\kappa }{}_{\kappa } T^{\nu \lambda \alpha }- \bar{h}_{195}^{} g^{\mu \rho } Q_{\alpha }{}^{\kappa }{}_{\kappa } T^{\alpha \lambda \nu }- \bar{h}_{196}^{} g^{\lambda \rho } Q_{\alpha }{}^{\mu }{}_{\kappa } T^{\alpha \nu \kappa }\nonumber\\
   &-\bar{h}_{197}^{} g^{\lambda \rho } Q^{\mu }{}_{\alpha \kappa } T^{\alpha \nu \kappa } + \bar{h}_{198}^{} g^{\lambda \rho } Q_{\alpha }{}^{\alpha }{}_{\kappa } T^{\kappa \mu \nu }- \bar{h}_{199}^{} g^{\lambda \rho } Q_{\alpha }{}^{\mu }{}_{\kappa } T^{\kappa \nu \alpha } - \bar{h}_{200}^{} g^{\lambda \rho } Q_{\alpha }{}^{\alpha }{}_{\kappa } T^{\mu \nu \kappa }+\bar{h}_{201}^{} g^{\lambda \rho } Q_{\alpha }{}^{\mu }{}_{\kappa } T^{\nu \alpha \kappa }\nonumber\\
   & + \bar{h}_{202}^{} g^{\lambda \rho } Q^{\mu \nu }{}_{\alpha } T^{\alpha } + \bar{h}_{203}^{} g^{\lambda \rho } Q_{\alpha }{}^{\mu \alpha } T^{\nu }+ \bar{h}_{204}^{} g^{\lambda \rho } Q_{\alpha }{}^{\kappa }{}_{\kappa } T^{\alpha \mu \nu } - \bar{h}_{205}^{} g^{\lambda \rho } Q_{\alpha }{}^{\kappa }{}_{\kappa } T^{\mu \nu \alpha } + \bar{h}_{206}^{} g^{\lambda \rho } Q^{\mu \alpha }{}_{\alpha } T^{\nu }\nonumber\\
   & + \bar{h}_{207}^{} g^{\lambda \mu } g^{\nu \rho } Q_{\alpha \kappa \theta } T^{\kappa \alpha \theta }+ \bar{h}_{208}^{} g^{\lambda \mu } g^{\nu \rho } Q_{\kappa \alpha }{}^{\kappa } T^{\alpha }+\bar{h}_{209}^{} g^{\lambda \mu } g^{\nu \rho } Q_{\alpha }{}^{\kappa }{}_{\kappa } T^{\alpha }\,,
\end{align}
\begin{align}
    X^{\lambda\mu\nu}=&\;\bar{h}_{1}^{} \bigl(\tilde{R}^{\nu \mu \alpha \kappa } T^{\lambda }{}_{\alpha \kappa } -  \tilde{R}_{\alpha }{}^{\kappa \mu \nu } T^{\lambda \alpha }{}_{\kappa }\bigr) + \bar{h}_{2}^{} \bigl(\tilde{R}^{\nu \alpha \mu \kappa } T^{\lambda }{}_{\alpha \kappa } -  \tilde{R}_{\alpha }{}^{\nu \mu \kappa } T^{\lambda \alpha }{}_{\kappa }\bigr) - \bar{h}_{3}^{} \bigl( \tilde{R}_{\alpha }{}^{\kappa \mu \nu } T^{\alpha \lambda }{}_{\kappa } +  \tilde{R}^{\lambda \nu \alpha \kappa } T^{\mu }{}_{\alpha \kappa }\bigr) \nonumber\\
    &-\bar{h}_{4}^{} \bigl(\tilde{R}_{\alpha }{}^{\nu \mu \kappa } T^{\alpha \lambda }{}_{\kappa } +  \tilde{R}^{\lambda \alpha \nu \kappa } T^{\mu }{}_{\alpha \kappa }\bigr) - \bar{h}_{5}^{} \bigl( \tilde{R}_{\alpha }{}^{\nu \mu \kappa } T_{\kappa }{}^{\lambda \alpha } +  \tilde{R}^{\nu \alpha \lambda \kappa } T^{\mu }{}_{\alpha \kappa }\bigr) + \bar{h}_{6}^{} \bigl(\tilde{R}^{\nu \mu \alpha \kappa } T_{\alpha }{}^{\lambda }{}_{\kappa } -  \tilde{R}_{\alpha }{}^{\kappa \lambda \nu } T^{\mu \alpha }{}_{\kappa }\bigr)  \nonumber\\
    &+ \bar{h}_{7}^{} \bigl(g^{\lambda \mu } \tilde{R}^{\nu \alpha \kappa \theta } T_{\alpha \kappa \theta } -  \tilde{R}_{\alpha }{}^{\lambda \mu \nu } T^{\alpha }\bigr) + \bar{h}_{8}^{} \bigl(g^{\lambda \mu } \tilde{R}^{\nu \alpha \kappa \theta } T_{\kappa \alpha \theta } + \tilde{R}_{\alpha }{}^{\nu \lambda \mu } T^{\alpha }\bigr) + \bar{h}_{9}^{} \bigl(g^{\lambda \mu } \tilde{R}_{\alpha }{}^{\kappa \nu \theta } T^{\alpha }{}_{\kappa \theta } -  \tilde{R}^{\lambda \nu \mu \alpha } T_{\alpha }\bigr) \nonumber\\
    & + \bar{h}_{10}^{} \bigl(g^{\lambda \mu } \tilde{R}_{\alpha }{}^{\kappa \nu \theta } T_{\theta }{}^{\alpha }{}_{\kappa } -  \tilde{R}^{\nu \mu \lambda \alpha } T_{\alpha }\bigr) - \bar{h}_{11}^{} \bigl( \tilde{R}^{\lambda \nu \alpha \kappa } T_{\alpha }{}^{\mu }{}_{\kappa } +  \tilde{R}_{\alpha }{}^{\kappa \lambda \nu } T^{\alpha \mu }{}_{\kappa }\bigr) + \bar{h}_{12}^{} \bigl( \tilde{R}_{\alpha }{}^{\lambda \nu \kappa } T^{\alpha \mu }{}_{\kappa }- \tilde{R}^{\lambda \alpha \nu \kappa } T_{\alpha }{}^{\mu }{}_{\kappa }\bigr) \nonumber\\
    & + \bar{h}_{13}^{} \bigl(\tilde{R}_{\alpha }{}^{\nu \lambda \kappa } T^{\alpha \mu }{}_{\kappa } -  \tilde{R}^{\lambda \alpha \nu \kappa } T_{\kappa }{}^{\mu }{}_{\alpha }\bigr) + \bar{h}_{14}^{} \bigl(\tilde{R}_{\alpha }{}^{\nu \lambda \kappa } T_{\kappa }{}^{\mu \alpha }- \tilde{R}^{\nu \alpha \lambda \kappa } T_{\kappa }{}^{\mu }{}_{\alpha }\bigr) - \bar{h}_{15}^{} \bigl(\tilde{R}^{\alpha \lambda } + \tilde{R}^{\lambda \alpha }\bigr) T_{\alpha }{}^{\mu \nu }  \nonumber\\
    &+ \bar{h}_{16}^{} \bigl(\tilde{R}^{\nu \alpha } T_{\alpha }{}^{\lambda \mu } -  \tilde{R}^{\alpha \lambda } T^{\nu \mu }{}_{\alpha }\bigr) + \bar{h}_{17}^{} \bigl(\tilde{R}^{\alpha \nu } T_{\alpha }{}^{\lambda \mu } -  \tilde{R}^{\lambda \alpha } T^{\nu \mu }{}_{\alpha }\bigr) - \bar{h}_{18}^{} \bigl(\tilde{R}^{\alpha \nu } + \tilde{R}^{\nu \alpha }\bigr) T^{\lambda \mu }{}_{\alpha } - \bar{h}_{19}^{} \bigl(\tilde{R}^{\alpha \nu } + \tilde{R}^{\nu \alpha }\bigr) T^{\mu \lambda }{}_{\alpha }  \nonumber\\
    &+ \bar{h}_{20}^{} g^{\lambda \mu } \bigl(\tilde{R}^{\alpha \nu } + \tilde{R}^{\nu \alpha }\bigr) T_{\alpha } + \bar{h}_{21}^{} \bigl( \tilde{R}^{\lambda \nu } T^{\mu }- g^{\lambda \mu } \tilde{R}^{\alpha \kappa } T_{\alpha }{}^{\nu }{}_{\kappa }\bigr) + \bar{h}_{22}^{} \bigl( \tilde{R}^{\nu \lambda } T^{\mu }- g^{\lambda \mu } \tilde{R}^{\alpha \kappa } T_{\kappa }{}^{\nu }{}_{\alpha }\bigr)  \nonumber\\
    &+ \bar{h}_{23}^{} \bigl(g^{\lambda \mu } \tilde{R}^{\alpha \kappa } T^{\nu }{}_{\alpha \kappa } + \tilde{R}^{\nu \mu } T^{\lambda }\bigr) -2 \bar{h}_{24}^{} \tilde{R} T^{\lambda \mu \nu } + \bar{h}_{25}^{} \tilde{R} \bigl( T^{\nu \lambda \mu }- T^{\mu \lambda \nu }\bigr) + \bar{h}_{26}^{} \bigl( g^{\lambda \mu } \tilde{R} T^{\nu }- g^{\lambda \nu } \tilde{R} T^{\mu }\bigr) \nonumber\\
    & - \bar{h}_{27}^{} \bigl( \tilde{R}^{\nu \lambda \alpha \kappa } T_{\alpha }{}^{\mu }{}_{\kappa } +  \tilde{R}_{\alpha }{}^{\kappa \lambda \nu } T_{\kappa }{}^{\mu \alpha }\bigr) + \bar{h}_{28}^{} \bigl(\tilde{R}^{\nu \alpha \lambda \kappa } T_{\alpha }{}^{\mu }{}_{\kappa } -  \tilde{R}_{\alpha }{}^{\lambda \nu \kappa } T_{\kappa }{}^{\mu \alpha }\bigr) + \bar{h}_{29}^{} \bigl(\tilde{R}^{\nu \alpha \mu \kappa } T_{\alpha }{}^{\lambda }{}_{\kappa } + \tilde{R}_{\alpha }{}^{\lambda \nu \kappa } T^{\mu \alpha }{}_{\kappa }\bigr)  \nonumber\\
    &+ \bar{h}_{30}^{} \bigl(\tilde{R}^{\nu \alpha \mu \kappa } T_{\kappa }{}^{\lambda }{}_{\alpha } + \tilde{R}_{\alpha }{}^{\nu \lambda \kappa } T^{\mu \alpha }{}_{\kappa }\bigr) + \bar{h}_{31}^{} \bigl(\tilde{R}_{\alpha }{}^{\kappa \mu \nu } T_{\kappa }{}^{\lambda \alpha } + \tilde{R}^{\nu \lambda \alpha \kappa } T^{\mu }{}_{\alpha \kappa }\bigr) + \bar{h}_{32}^{} \bigl(g^{\lambda \mu } \tilde{R}_{\alpha }{}^{\nu \kappa \theta } T^{\alpha }{}_{\kappa \theta } -  \tilde{R}^{\lambda \alpha \mu \nu } T_{\alpha }\bigr)  \nonumber\\
    &+ \bar{h}_{33}^{} \bigl(g^{\lambda \mu } \tilde{R}_{\alpha }{}^{\nu \kappa \theta } T_{\kappa }{}^{\alpha }{}_{\theta } + \tilde{R}^{\nu \alpha \lambda \mu } T_{\alpha }\bigr) + \bar{h}_{34}^{} \bigl(g^{\lambda \mu } \tilde{R}_{\alpha }{}^{\kappa \nu \theta } T_{\kappa }{}^{\alpha }{}_{\theta } -  \tilde{R}^{\nu \lambda \mu \alpha } T_{\alpha }\bigr) - \bar{h}_{35}^{} \bigl( \hat{R}^{\lambda \alpha } T_{\alpha }{}^{\mu \nu } +  \hat{R}_{\alpha }{}^{\lambda } T^{\alpha \mu \nu }\bigr)  \nonumber\\
    &- \bar{h}_{36}^{} \bigl( \hat{R}^{\nu \alpha } T^{\lambda \mu }{}_{\alpha } +  \hat{R}_{\alpha }{}^{\nu } T^{\lambda \mu \alpha }\bigr) + \bar{h}_{37}^{} \bigl(\hat{R}^{\nu \alpha } T_{\alpha }{}^{\lambda \mu } -  \hat{R}_{\alpha }{}^{\lambda } T^{\nu \mu \alpha }\bigr) + \bar{h}_{38}^{} \bigl(\hat{R}_{\alpha }{}^{\nu } T^{\alpha \lambda \mu } -  \hat{R}^{\lambda \alpha } T^{\nu \mu }{}_{\alpha }\bigr)  \nonumber\\
    &- \bar{h}_{39}^{} \bigl( \hat{R}^{\nu \alpha } T^{\mu \lambda }{}_{\alpha } +  \hat{R}_{\alpha }{}^{\nu } T^{\mu \lambda \alpha }\bigr) + \bar{h}_{40}^{} g^{\lambda \mu } \bigl(\hat{R}^{\nu \alpha } T_{\alpha } + \hat{R}_{\alpha }{}^{\nu } T^{\alpha }\bigr) + \bar{h}_{41}^{} \bigl(\hat{R}^{\lambda \nu } T^{\mu }- g^{\lambda \mu } \hat{R}_{\alpha }{}^{\kappa } T^{\alpha \nu }{}_{\kappa }\bigr)  \nonumber\\
    &+ \bar{h}_{42}^{} \bigl(\hat{R}^{\nu \lambda } T^{\mu }- g^{\lambda \mu } \hat{R}_{\alpha }{}^{\kappa } T_{\kappa }{}^{\nu \alpha }\bigr) + \bar{h}_{43}^{} \bigl(g^{\lambda \mu } \hat{R}_{\alpha }{}^{\kappa } T^{\nu \alpha }{}_{\kappa } + \hat{R}^{\nu \mu } T^{\lambda }\bigr) - \bar{h}_{44}^{} \bigl( \bar{R}^{\nu \alpha } T_{\alpha }{}^{\lambda \mu } +  \bar{R}^{\lambda \alpha } T^{\nu \mu }{}_{\alpha }\bigr)  \nonumber\\
    &+ \bar{h}_{45}^{} \bigl(g^{\lambda \mu } \bar{R}^{\alpha \kappa } T^{\nu }{}_{\alpha \kappa } -  \bar{R}^{\mu \nu } T^{\lambda }\bigr) + \bar{h}_{46}^{} \bigl(g^{\lambda \mu } \bar{R}^{\alpha \kappa } T_{\alpha }{}^{\nu }{}_{\kappa } -  \bar{R}^{\lambda \nu } T^{\mu }\bigr)- \bar{h}_{108}^{} Q^{\lambda \alpha }{}_{\kappa } \tilde{R}^{\kappa }{}_{\alpha }{}^{\mu \nu } - \bar{h}_{109}^{} Q^{\lambda \alpha }{}_{\kappa } \tilde{R}^{\kappa \nu \mu }{}_{\alpha }\nonumber\\
    & - \bar{h}_{110}^{} Q^{\lambda \alpha \kappa } \tilde{R}^{\nu }{}_{\alpha }{}^{\mu }{}_{\kappa } - \bar{h}_{111}^{} Q^{\alpha \lambda \kappa } \tilde{R}_{\alpha \kappa }{}^{\mu \nu } - \bar{h}_{112}^{} Q^{\alpha \lambda \kappa } \tilde{R}_{\alpha }{}^{\nu \mu }{}_{\kappa }- \bar{h}_{113}^{} Q^{\alpha \lambda }{}_{\kappa } \tilde{R}^{\kappa }{}_{\alpha }{}^{\mu \nu } - \bar{h}_{114}^{} Q_{\alpha }{}^{\alpha }{}_{\kappa } \tilde{R}^{\kappa \lambda \mu \nu }\nonumber\\
   &  - \bar{h}_{115}^{} Q^{\alpha \lambda }{}_{\kappa } \tilde{R}^{\kappa \nu \mu }{}_{\alpha } - \bar{h}_{116}^{} Q_{\alpha }{}^{\alpha \kappa } \tilde{R}^{\lambda }{}_{\kappa }{}^{\mu \nu } - \bar{h}_{117}^{} Q_{\alpha }{}^{\alpha \kappa } \tilde{R}^{\lambda \nu \mu }{}_{\kappa } - \bar{h}_{118}^{} Q^{\alpha \lambda \kappa } \tilde{R}^{\nu }{}_{\alpha }{}^{\mu }{}_{\kappa } - \bar{h}_{119}^{} Q^{\alpha \lambda \kappa } \tilde{R}^{\nu }{}_{\kappa }{}^{\mu }{}_{\alpha }\nonumber\\
   & + \bar{h}_{120}^{} Q^{\alpha \lambda \kappa } \tilde{R}^{\nu \mu }{}_{\alpha \kappa } + \bar{h}_{121}^{} Q^{\nu \alpha }{}_{\kappa } \tilde{R}^{\kappa }{}_{\alpha }{}^{\lambda \mu } - \bar{h}_{122}^{} Q^{\nu \alpha }{}_{\kappa } \tilde{R}^{\kappa \lambda \mu }{}_{\alpha } - \bar{h}_{123}^{} Q^{\nu \alpha \kappa } \tilde{R}^{\lambda }{}_{\alpha }{}^{\mu }{}_{\kappa } + \bar{h}_{124}^{} Q^{\alpha \nu \kappa } \tilde{R}_{\alpha \kappa }{}^{\lambda \mu }\nonumber\\
   & - \bar{h}_{125}^{} Q^{\alpha \nu \kappa } \tilde{R}_{\alpha }{}^{\lambda \mu }{}_{\kappa } + \bar{h}_{126}^{} Q_{\alpha }{}^{\nu \kappa } \tilde{R}_{\kappa }{}^{\alpha \lambda \mu } + \bar{h}_{127}^{} Q_{\alpha }{}^{\alpha }{}_{\kappa } \tilde{R}^{\kappa \nu \lambda \mu }- \bar{h}_{128}^{} Q_{\alpha }{}^{\nu \kappa } \tilde{R}_{\kappa }{}^{\lambda \mu \alpha } + \bar{h}_{129}^{} Q_{\alpha }{}^{\alpha \kappa } \tilde{R}^{\nu }{}_{\kappa }{}^{\lambda \mu }\nonumber\\
   &  - \bar{h}_{130}^{} Q_{\alpha }{}^{\alpha \kappa } \tilde{R}^{\nu \lambda \mu }{}_{\kappa } - \bar{h}_{131}^{} Q^{\alpha \nu \kappa } \tilde{R}^{\lambda }{}_{\alpha }{}^{\mu }{}_{\kappa } - \bar{h}_{132}^{} Q^{\alpha \nu \kappa } \tilde{R}^{\lambda }{}_{\kappa }{}^{\mu }{}_{\alpha } + \bar{h}_{133}^{} Q^{\alpha \nu \kappa } \tilde{R}^{\lambda \mu }{}_{\alpha \kappa } - \bar{h}_{134}^{} Q^{\nu \alpha }{}_{\kappa } \tilde{R}^{\kappa \mu \lambda }{}_{\alpha }\nonumber\\
   & - \bar{h}_{135}^{} Q^{\nu \alpha \kappa } \tilde{R}^{\mu }{}_{\alpha }{}^{\lambda }{}_{\kappa } - \bar{h}_{136}^{} Q^{\alpha \nu \kappa } \tilde{R}_{\alpha }{}^{\mu \lambda }{}_{\kappa } - \bar{h}_{137}^{} Q_{\alpha }{}^{\nu \kappa } \tilde{R}_{\kappa }{}^{\mu \lambda \alpha }- \bar{h}_{138}^{} Q_{\alpha }{}^{\alpha \kappa } \tilde{R}^{\nu \mu \lambda }{}_{\kappa } - \bar{h}_{139}^{} Q^{\alpha \nu \kappa } \tilde{R}^{\mu }{}_{\alpha }{}^{\lambda }{}_{\kappa }\nonumber\\
   &  - \bar{h}_{140}^{} Q^{\alpha \nu \kappa } \tilde{R}^{\mu }{}_{\kappa }{}^{\lambda }{}_{\alpha } + \bar{h}_{141}^{} Q^{\alpha \nu \kappa } \tilde{R}^{\mu \lambda }{}_{\alpha \kappa } - \bar{h}_{142}^{} g^{\lambda \mu } Q^{\alpha \kappa \theta } \tilde{R}_{\alpha \kappa }{}^{\nu }{}_{\theta }- \bar{h}_{143}^{} g^{\lambda \mu } Q^{\alpha \kappa }{}_{\theta } \tilde{R}_{\kappa \alpha }{}^{\nu \theta } - \bar{h}_{144}^{} g^{\lambda \mu } Q^{\alpha \kappa }{}_{\theta } \tilde{R}_{\kappa }{}^{\theta \nu }{}_{\alpha }\nonumber\\
   &  + \bar{h}_{145}^{} g^{\lambda \mu } Q^{\alpha \kappa }{}_{\theta } \tilde{R}_{\kappa }{}^{\nu }{}_{\alpha }{}^{\theta } + \bar{h}_{146}^{} g^{\lambda \mu } Q^{\alpha \kappa \theta } \tilde{R}^{\nu }{}_{\kappa \alpha \theta } - \bar{h}_{147}^{} Q^{\alpha \kappa }{}_{\kappa } \tilde{R}_{\alpha }{}^{\lambda \mu \nu } - \bar{h}_{148}^{} Q^{\alpha \kappa }{}_{\kappa } \tilde{R}^{\lambda }{}_{\alpha }{}^{\mu \nu } - \bar{h}_{149}^{} Q^{\alpha \kappa }{}_{\kappa } \tilde{R}^{\lambda \nu \mu }{}_{\alpha }\nonumber\\
   & + \bar{h}_{150}^{} Q^{\alpha \kappa }{}_{\kappa } \tilde{R}_{\alpha }{}^{\nu \lambda \mu } + \bar{h}_{151}^{} Q^{\alpha \kappa }{}_{\kappa } \tilde{R}^{\nu }{}_{\alpha }{}^{\lambda \mu } - \bar{h}_{152}^{} Q^{\alpha \kappa }{}_{\kappa } \tilde{R}^{\nu \lambda \mu }{}_{\alpha } - \bar{h}_{153}^{} Q^{\alpha \kappa }{}_{\kappa } \tilde{R}^{\nu \mu \lambda }{}_{\alpha } - \bar{h}_{154}^{} Q_{\alpha }{}^{\nu \alpha } \tilde{R}^{\lambda \mu }+ \bar{h}_{155}^{} Q^{\nu \mu \alpha } \tilde{R}^{\lambda }{}_{\alpha }\nonumber\\
   &  - \bar{h}_{156}^{} Q_{\alpha }{}^{\nu \alpha } \tilde{R}^{\mu \lambda } + \bar{h}_{157}^{} Q^{\nu \mu \alpha } \tilde{R}_{\alpha }{}^{\lambda } - \bar{h}_{158}^{} Q_{\alpha }{}^{\lambda \nu } \tilde{R}^{\mu \alpha }- \bar{h}_{159}^{} Q^{\lambda \nu \alpha } \tilde{R}^{\mu }{}_{\alpha } - \bar{h}_{160}^{} Q_{\alpha }{}^{\lambda \nu } \tilde{R}^{\alpha \mu }- \bar{h}_{161}^{} Q^{\lambda \nu \alpha } \tilde{R}_{\alpha }{}^{\mu }\nonumber\\
   & + \bar{h}_{162}^{} Q_{\alpha }{}^{\lambda \alpha } \tilde{R}^{\nu \mu } - \bar{h}_{163}^{} Q^{\nu \lambda \alpha } \tilde{R}^{\mu }{}_{\alpha } - \bar{h}_{164}^{} Q^{\nu \lambda \alpha } \tilde{R}_{\alpha }{}^{\mu }+ \bar{h}_{165}^{} g^{\lambda \mu } Q_{\kappa \alpha }{}^{\kappa } \tilde{R}^{\nu \alpha } + \bar{h}_{166}^{} g^{\lambda \mu } Q_{\kappa \alpha }{}^{\kappa } \tilde{R}^{\alpha \nu } + \bar{h}_{167}^{} g^{\lambda \mu } Q_{\alpha }{}^{\nu }{}_{\kappa } \tilde{R}^{\alpha \kappa }\nonumber\\
   & + \bar{h}_{168}^{} g^{\lambda \mu } Q_{\kappa }{}^{\nu }{}_{\alpha } \tilde{R}^{\alpha \kappa } + \bar{h}_{169}^{} g^{\lambda \mu } Q^{\nu }{}_{\alpha \kappa } \tilde{R}^{\alpha \kappa }+ \bar{h}_{170}^{} g^{\lambda \mu } Q_{\alpha }{}^{\kappa }{}_{\kappa } \tilde{R}^{\alpha \nu } + \bar{h}_{171}^{} g^{\lambda \mu } Q_{\alpha }{}^{\kappa }{}_{\kappa } \tilde{R}^{\nu \alpha } - \bar{h}_{172}^{} Q^{\nu \alpha }{}_{\alpha } \tilde{R}^{\lambda \mu }  \nonumber\\
   &- \bar{h}_{173}^{} Q^{\nu \alpha }{}_{\alpha } \tilde{R}^{\mu \lambda } + \bar{h}_{174}^{} Q^{\lambda \alpha }{}_{\alpha } \tilde{R}^{\nu \mu } - \bar{h}_{175}^{} Q_{\alpha }{}^{\nu \alpha } \hat{R}^{\lambda \mu }+ \bar{h}_{176}^{} Q^{\nu \mu \alpha } \hat{R}^{\lambda }{}_{\alpha } - \bar{h}_{177}^{} Q_{\alpha }{}^{\nu \alpha } \hat{R}^{\mu \lambda } + \bar{h}_{178}^{} Q^{\nu \mu \alpha } \hat{R}_{\alpha }{}^{\lambda }\nonumber\\
   &- \bar{h}_{179}^{} Q^{\alpha \lambda \nu } \hat{R}^{\mu }{}_{\alpha } - \bar{h}_{180}^{} Q^{\lambda \nu \alpha } \hat{R}^{\mu }{}_{\alpha } - \bar{h}_{181}^{} Q^{\alpha \lambda \nu } \hat{R}_{\alpha }{}^{\mu } - \bar{h}_{182}^{} Q^{\lambda \nu \alpha } \hat{R}_{\alpha }{}^{\mu } + \bar{h}_{183}^{} Q_{\alpha }{}^{\lambda \alpha } \hat{R}^{\nu \mu } - \bar{h}_{184}^{} Q^{\nu \lambda \alpha } \hat{R}^{\mu }{}_{\alpha }\nonumber\\
   & - \bar{h}_{185}^{} Q^{\nu \lambda \alpha } \hat{R}_{\alpha }{}^{\mu } + \bar{h}_{186}^{} g^{\lambda \mu } Q_{\kappa }{}^{\alpha \kappa } \hat{R}^{\nu }{}_{\alpha } + \bar{h}_{187}^{} g^{\lambda \mu } Q_{\kappa \alpha }{}^{\kappa } \hat{R}^{\alpha \nu } + \bar{h}_{188}^{} g^{\lambda \mu } Q^{\alpha \nu \kappa } \hat{R}_{\alpha \kappa } + \bar{h}_{189}^{} g^{\lambda \mu } Q^{\kappa \nu }{}_{\alpha } \hat{R}^{\alpha }{}_{\kappa }\nonumber\\
   &+\bar{h}_{190}^{} g^{\lambda \mu } Q^{\nu \alpha \kappa } \hat{R}_{\alpha \kappa } + \bar{h}_{191}^{} g^{\lambda \mu } Q^{\alpha \kappa }{}_{\kappa } \hat{R}_{\alpha }{}^{\nu } + \bar{h}_{192}^{} g^{\lambda \mu } Q^{\alpha \kappa }{}_{\kappa } \hat{R}^{\nu }{}_{\alpha }- \bar{h}_{193}^{} Q^{\nu \alpha }{}_{\alpha } \hat{R}^{\lambda \mu } - \bar{h}_{194}^{} Q^{\nu \alpha }{}_{\alpha } \hat{R}^{\mu \lambda }  \nonumber\\
   & + \bar{h}_{195}^{} Q^{\lambda \alpha }{}_{\alpha } \hat{R}^{\nu \mu }- \bar{h}_{196}^{} Q^{\lambda \nu \alpha } \bar{R}^{\mu }{}_{\alpha } - \bar{h}_{197}^{} Q^{\alpha \lambda \nu } \bar{R}^{\mu }{}_{\alpha } - \bar{h}_{198}^{} Q_{\alpha }{}^{\lambda \alpha } \bar{R}^{\mu \nu } - \bar{h}_{199}^{} Q^{\nu \lambda \alpha } \bar{R}^{\mu }{}_{\alpha } + \bar{h}_{200}^{} Q_{\alpha }{}^{\nu \alpha } \bar{R}^{\lambda \mu }\nonumber\\
   &- \bar{h}_{201}^{} Q^{\nu \mu \alpha } \bar{R}^{\lambda }{}_{\alpha } + \bar{h}_{202}^{} g^{\lambda \mu } Q^{\alpha \nu \kappa } \bar{R}_{\alpha \kappa } - \bar{h}_{203}^{} g^{\lambda \mu } Q_{\alpha }{}^{\alpha \kappa } \bar{R}^{\nu }{}_{\kappa } - \bar{h}_{204}^{} Q^{\lambda \alpha }{}_{\alpha } \bar{R}^{\mu \nu } + \bar{h}_{205}^{} Q^{\nu \alpha }{}_{\alpha } \bar{R}^{\lambda \mu }\nonumber\\
   & - \bar{h}_{206}^{} g^{\lambda \mu } Q^{\alpha \kappa }{}_{\kappa } \bar{R}^{\nu }{}_{\alpha } + \bar{h}_{207}^{} Q^{\nu \lambda \mu } \tilde{R} + \bar{h}_{208}^{} g^{\lambda \mu } Q_{\alpha }{}^{\nu \alpha } \tilde{R} + \bar{h}_{209}^{} g^{\lambda \mu } Q^{\nu \alpha }{}_{\alpha } \tilde{R}\,,
\end{align}
\begin{align}
    Z^{\lambda \mu\nu}=&\;\bar{h}_{47}^{} Q_{\alpha }{}^{\lambda }{}_{\kappa } \bigl(\tilde{R}^{\nu \mu \alpha \kappa }- \tilde{R}^{\alpha \kappa \mu \nu }\bigr) + \bar{h}_{48}^{} \bigl(Q^{\lambda }{}_{\alpha \kappa } \tilde{R}^{\nu \alpha \mu \kappa }- Q_{\alpha }{}^{\nu }{}_{\kappa } \tilde{R}^{\alpha \lambda \mu \kappa }\bigr) + \bar{h}_{49}^{} Q_{\alpha }{}^{\lambda }{}_{\kappa } \bigl(\tilde{R}^{\nu \alpha \mu \kappa }- \tilde{R}^{\alpha \nu \mu \kappa }\bigr)+ 2 \bar{h}_{106}^{} g^{\lambda \mu } Q^{\nu \alpha }{}_{\alpha } \tilde{R}\nonumber\\
   &+ \bar{h}_{50}^{} Q_{\alpha }{}^{\lambda }{}_{\kappa } \bigl(\tilde{R}^{\nu \kappa \mu \alpha }- \tilde{R}^{\alpha \mu \nu \kappa }\bigr) + \bar{h}_{51}^{} Q^{\nu }{}_{\alpha \kappa } \bigl(\tilde{R}^{\lambda \alpha \mu \kappa }- \tilde{R}^{\alpha \lambda \mu \kappa }\bigr) + \bar{h}_{52}^{} \bigl(g^{\lambda \nu } Q_{\alpha \kappa \theta } \tilde{R}^{\mu \kappa \alpha \theta }- Q_{\alpha }{}^{\alpha }{}_{\kappa } \tilde{R}^{\kappa \lambda \mu \nu }\bigr) + 2 \bar{h}_{103}^{} Q^{\nu \lambda \mu } \tilde{R}\nonumber\\
   &+ \bar{h}_{53}^{} Q_{\alpha }{}^{\lambda }{}_{\kappa } \bigl(\tilde{R}^{\mu \nu \alpha \kappa }- \tilde{R}^{\kappa \alpha \mu \nu }\bigr) + \bar{h}_{54}^{} \bigl(Q^{\lambda }{}_{\alpha \kappa } \tilde{R}^{\mu \alpha \nu \kappa }- Q_{\alpha }{}^{\nu }{}_{\kappa } \tilde{R}^{\kappa \lambda \mu \alpha }\bigr) + \bar{h}_{55}^{} Q_{\alpha }{}^{\lambda }{}_{\kappa } \bigl(\tilde{R}^{\mu \alpha \nu \kappa }- \tilde{R}^{\kappa \nu \mu \alpha }\bigr)+2\bar{h}_{104}^{} Q^{\lambda \mu \nu } \tilde{R}\nonumber\\
   & + \bar{h}_{56}^{} \bigl(g^{\lambda \nu } Q_{\alpha \kappa \theta } \tilde{R}^{\alpha \kappa \mu \theta } -  Q_{\alpha }{}^{\alpha }{}_{\kappa } \tilde{R}^{\nu \lambda \mu \kappa }\bigr) + \bar{h}_{57}^{} Q_{\alpha }{}^{\lambda }{}_{\kappa } \bigl(\tilde{R}^{\mu \kappa \nu \alpha }- \tilde{R}^{\kappa \mu \nu \alpha }\bigr) + \bar{h}_{58}^{} \bigl(Q^{\lambda }{}_{\alpha \kappa } \tilde{R}^{\alpha \nu \mu \kappa } -  Q_{\alpha }{}^{\nu }{}_{\kappa } \tilde{R}^{\lambda \alpha \mu \kappa }\bigr) \nonumber\\
   &+ \bar{h}_{59}^{} \bigl(g^{\lambda \nu } Q_{\alpha \kappa \theta } \tilde{R}^{\kappa \mu \alpha \theta } -  Q_{\alpha }{}^{\alpha }{}_{\kappa } \tilde{R}^{\lambda \kappa \mu \nu }\bigr) + \bar{h}_{60}^{} \bigl(Q^{\lambda }{}_{\alpha \kappa } \tilde{R}^{\alpha \mu \nu \kappa } -  Q_{\alpha }{}^{\nu }{}_{\kappa } \tilde{R}^{\lambda \kappa \mu \alpha }\bigr) + \bar{h}_{61}^{} \bigl(g^{\lambda \nu } Q_{\alpha \kappa \theta } \tilde{R}^{\kappa \alpha \mu \theta } -  Q_{\alpha }{}^{\alpha }{}_{\kappa } \tilde{R}^{\lambda \nu \mu \kappa }\bigr) \nonumber\\
   &+ \bar{h}_{62}^{} \bigl(Q_{\alpha }{}^{\nu }{}_{\kappa } \tilde{R}^{\lambda \mu \alpha \kappa }- Q^{\lambda }{}_{\alpha \kappa } \tilde{R}^{\alpha \kappa \mu \nu }\bigr) + \bar{h}_{63}^{} \bigl(g^{\lambda \nu } Q_{\alpha \kappa \theta } \tilde{R}^{\kappa \theta \mu \alpha } -  Q_{\alpha }{}^{\alpha }{}_{\kappa } \tilde{R}^{\lambda \mu \nu \kappa }\bigr) + \bar{h}_{64}^{} \bigl(g^{\lambda \mu } Q_{\alpha \kappa \theta } \tilde{R}^{\nu \kappa \alpha \theta }- Q_{\alpha }{}^{\kappa }{}_{\kappa } \tilde{R}^{\alpha \lambda \mu \nu }\bigr) \nonumber\\
   &+ \bar{h}_{65}^{} \bigl(g^{\lambda \mu } Q_{\alpha \kappa \theta } \tilde{R}^{\alpha \kappa \nu \theta } -  Q_{\alpha }{}^{\kappa }{}_{\kappa } \tilde{R}^{\nu \lambda \mu \alpha }\bigr) + \bar{h}_{66}^{} \bigl(g^{\lambda \mu } Q_{\alpha \kappa \theta } \tilde{R}^{\kappa \nu \alpha \theta } -  Q_{\alpha }{}^{\kappa }{}_{\kappa } \tilde{R}^{\lambda \alpha \mu \nu }\bigr) + \bar{h}_{67}^{} \bigl(g^{\lambda \mu } Q_{\alpha \kappa \theta } \tilde{R}^{\kappa \alpha \nu \theta } -  Q_{\alpha }{}^{\kappa }{}_{\kappa } \tilde{R}^{\lambda \nu \mu \alpha }\bigr) \nonumber\\
   &+ \bar{h}_{68}^{} \bigl(g^{\lambda \mu } Q_{\alpha \kappa \theta } \tilde{R}^{\kappa \theta \nu \alpha } -  Q_{\alpha }{}^{\kappa }{}_{\kappa } \tilde{R}^{\lambda \mu \nu \alpha }\bigr) + \bar{h}_{69}^{} Q^{\alpha \lambda \mu } \bigl(\tilde{R}_{\alpha }{}^{\nu } + \tilde{R}^{\nu }{}_{\alpha }\bigr) + \bar{h}_{70}^{} \bigl(Q^{\lambda \mu \alpha } \tilde{R}_{\alpha }{}^{\nu } + Q^{\alpha \lambda \nu } \tilde{R}^{\mu }{}_{\alpha }\bigr) - \bar{h}_{155}^{} \tilde{R}_{\alpha }{}^{\mu } T^{\alpha \lambda \nu }\nonumber\\
   &+ \bar{h}_{71}^{} \bigl(Q^{\alpha \lambda \nu } \tilde{R}_{\alpha }{}^{\mu } + Q^{\lambda \mu \alpha } \tilde{R}^{\nu }{}_{\alpha }\bigr) + \bar{h}_{72}^{} Q^{\nu \lambda \alpha } \bigl(\tilde{R}_{\alpha }{}^{\mu } + \tilde{R}^{\mu }{}_{\alpha }\bigr) + \bar{h}_{73}^{} Q^{\lambda \nu \alpha } \bigl(\tilde{R}_{\alpha }{}^{\mu } + \tilde{R}^{\mu }{}_{\alpha }\bigr) + \bar{h}_{74}^{} g^{\lambda \nu } Q_{\kappa }{}^{\alpha \kappa } \bigl(\tilde{R}_{\alpha }{}^{\mu } + \tilde{R}^{\mu }{}_{\alpha }\bigr)\nonumber\\
   & + \bar{h}_{75}^{} \bigl(g^{\lambda \nu } Q^{\alpha \mu \kappa } \tilde{R}_{\alpha \kappa } + Q_{\alpha }{}^{\lambda \alpha } \tilde{R}^{\nu \mu }\bigr) + \bar{h}_{76}^{} \bigl(g^{\lambda \nu } Q^{\kappa \mu \alpha } \tilde{R}_{\alpha \kappa } + Q_{\alpha }{}^{\lambda \alpha } \tilde{R}^{\mu \nu }\bigr) + \bar{h}_{77}^{} \bigl(g^{\lambda \nu } Q^{\mu \alpha \kappa } \tilde{R}_{\alpha \kappa } + Q_{\alpha }{}^{\nu \alpha } \tilde{R}^{\lambda \mu }\bigr) \nonumber\\
   &+ \bar{h}_{78}^{} \bigl(g^{\lambda \nu } Q^{\alpha \kappa }{}_{\kappa } \tilde{R}_{\alpha }{}^{\mu } + g^{\lambda \mu } Q_{\kappa }{}^{\alpha \kappa } \tilde{R}^{\nu }{}_{\alpha }\bigr) + \bar{h}_{79}^{} \bigl(g^{\lambda \mu } Q_{\kappa }{}^{\alpha \kappa } \tilde{R}_{\alpha }{}^{\nu } + g^{\lambda \nu } Q^{\alpha \kappa }{}_{\kappa } \tilde{R}^{\mu }{}_{\alpha }\bigr) + \bar{h}_{80}^{} g^{\lambda \mu } Q^{\alpha \kappa }{}_{\kappa } \bigl(\tilde{R}_{\alpha }{}^{\nu } + \tilde{R}^{\nu }{}_{\alpha }\bigr) \nonumber\\
   &+ \bar{h}_{81}^{} \bigl(g^{\lambda \mu } Q^{\alpha \nu \kappa } \tilde{R}_{\alpha \kappa } + Q^{\lambda \alpha }{}_{\alpha } \tilde{R}^{\nu \mu }\bigr) + \bar{h}_{82}^{} \bigl(g^{\lambda \mu } Q^{\kappa \nu \alpha } \tilde{R}_{\alpha \kappa } + Q^{\lambda \alpha }{}_{\alpha } \tilde{R}^{\mu \nu }\bigr) + \bar{h}_{83}^{} \bigl(g^{\lambda \mu } Q^{\nu \alpha \kappa } \tilde{R}_{\alpha \kappa } + Q^{\nu \alpha }{}_{\alpha } \tilde{R}^{\lambda \mu }\bigr) \nonumber\\
   &+ \bar{h}_{84}^{} Q^{\alpha \lambda \mu } \bigl(\hat{R}_{\alpha }{}^{\nu } + \hat{R}^{\nu }{}_{\alpha }\bigr) + \bar{h}_{85}^{} \bigl(Q^{\lambda \mu \alpha } \hat{R}_{\alpha }{}^{\nu } + Q^{\alpha \lambda \nu } \hat{R}^{\mu }{}_{\alpha }\bigr) + \bar{h}_{86}^{} \bigl(Q^{\alpha \lambda \nu } \hat{R}_{\alpha }{}^{\mu } + Q^{\lambda \mu \alpha } \hat{R}^{\nu }{}_{\alpha }\bigr) + \bar{h}_{87}^{} Q^{\nu \lambda \alpha } \bigl(\hat{R}_{\alpha }{}^{\mu } + \hat{R}^{\mu }{}_{\alpha }\bigr) \nonumber\\
   &+ \bar{h}_{88}^{} Q^{\lambda \nu \alpha } \bigl(\hat{R}_{\alpha }{}^{\mu } + \hat{R}^{\mu }{}_{\alpha }\bigr) + \bar{h}_{89}^{} g^{\lambda \nu } Q_{\kappa }{}^{\alpha \kappa } \bigl(\hat{R}_{\alpha }{}^{\mu } + \hat{R}^{\mu }{}_{\alpha }\bigr) + \bar{h}_{90}^{} \bigl(g^{\lambda \nu } Q^{\alpha \mu \kappa } \hat{R}_{\alpha \kappa } + Q_{\alpha }{}^{\lambda \alpha } \hat{R}^{\nu \mu }\bigr)+\bar{h}_{108}^{} \tilde{R}^{\lambda \mu }{}_{\alpha \kappa } T^{\nu \alpha \kappa } \nonumber\\
   &+ \bar{h}_{91}^{} \bigl(g^{\lambda \nu } Q^{\kappa \mu \alpha } \hat{R}_{\alpha \kappa } + Q_{\alpha }{}^{\lambda \alpha } \hat{R}^{\mu \nu }\bigr) + \bar{h}_{92}^{} \bigl(g^{\lambda \nu } Q^{\mu \alpha \kappa } \hat{R}_{\alpha \kappa } + Q_{\alpha }{}^{\nu \alpha } \hat{R}^{\lambda \mu }\bigr) + \bar{h}_{93}^{} \bigl(g^{\lambda \nu } Q^{\alpha \kappa }{}_{\kappa } \hat{R}_{\alpha }{}^{\mu } + g^{\lambda \mu } Q_{\kappa }{}^{\alpha \kappa } \hat{R}^{\nu }{}_{\alpha }\bigr)\nonumber\\
   & + \bar{h}_{94}^{} \bigl(g^{\lambda \mu } Q_{\kappa }{}^{\alpha \kappa } \hat{R}_{\alpha }{}^{\nu } + g^{\lambda \nu } Q^{\alpha \kappa }{}_{\kappa } \hat{R}^{\mu }{}_{\alpha }\bigr) + \bar{h}_{95}^{} g^{\lambda \mu } Q^{\alpha \kappa }{}_{\kappa } \bigl(\hat{R}_{\alpha }{}^{\nu } + \hat{R}^{\nu }{}_{\alpha }\bigr) + \bar{h}_{96}^{} \bigl(g^{\lambda \mu } Q^{\alpha \nu \kappa } \hat{R}_{\alpha \kappa } + Q^{\lambda \alpha }{}_{\alpha } \hat{R}^{\nu \mu }\bigr)\nonumber\\
   & + \bar{h}_{97}^{} \bigl(g^{\lambda \mu } Q^{\kappa \nu \alpha } \hat{R}_{\alpha \kappa } + Q^{\lambda \alpha }{}_{\alpha } \hat{R}^{\mu \nu }\bigr) + \bar{h}_{98}^{} \bigl(g^{\lambda \mu } Q^{\nu \alpha \kappa } \hat{R}_{\alpha \kappa } + Q^{\nu \alpha }{}_{\alpha } \hat{R}^{\lambda \mu }\bigr) + \bar{h}_{99}^{} \bigl(Q^{\lambda \mu }{}_{\alpha } \bar{R}^{\nu \alpha }- Q_{\alpha }{}^{\lambda \nu } \bar{R}^{\mu \alpha }\bigr) \nonumber\\
   &+ \bar{h}_{100}^{} \bigl(g^{\lambda \nu } Q_{\alpha }{}^{\mu }{}_{\kappa } \bar{R}^{\alpha \kappa } -  Q_{\alpha }{}^{\lambda \alpha } \bar{R}^{\mu \nu }\bigr) + \bar{h}_{101}^{} \bigl(g^{\lambda \mu } Q_{\alpha }{}^{\nu }{}_{\kappa } \bar{R}^{\alpha \kappa } -  Q^{\lambda \alpha }{}_{\alpha } \bar{R}^{\mu \nu }\bigr) + \bar{h}_{102}^{} \bigl(g^{\lambda \nu } Q_{\alpha }{}^{\kappa }{}_{\kappa } \bar{R}^{\mu \alpha } -  g^{\lambda \mu } Q_{\alpha }{}^{\alpha }{}_{\kappa } \bar{R}^{\nu \kappa }\bigr) \nonumber\\
   &+\bar{h}_{105}^{} \bigl(g^{\mu \nu } Q_{\alpha }{}^{\lambda \alpha } + g^{\lambda \nu } Q_{\alpha }{}^{\mu \alpha }\bigr) \tilde{R}+\bar{h}_{107}^{} \bigl(g^{\lambda \mu } Q_{\alpha }{}^{\nu \alpha } + g^{\lambda \nu } Q^{\mu \alpha }{}_{\alpha }\bigr) \tilde{R}+\bar{h}_{109}^{} \tilde{R}^{\lambda }{}_{\alpha }{}^{\mu }{}_{\kappa } T^{\nu \alpha \kappa } + \bar{h}_{110}^{} \tilde{R}^{\alpha \lambda \mu }{}_{\kappa } T^{\nu }{}_{\alpha }{}^{\kappa } \nonumber\\
   &+\bar{h}_{111}^{} \tilde{R}^{\nu \lambda }{}_{\alpha \kappa } T^{\mu \alpha \kappa } + \bar{h}_{112}^{} \tilde{R}^{\nu }{}_{\alpha }{}^{\lambda }{}_{\kappa } T^{\mu \alpha \kappa } + \bar{h}_{113}^{} \tilde{R}^{\lambda \nu }{}_{\alpha \kappa } T^{\mu \alpha \kappa }+ \bar{h}_{114}^{} g^{\lambda \nu } \tilde{R}^{\mu }{}_{\alpha \kappa \theta } T^{\alpha \kappa \theta } + \bar{h}_{115}^{} \tilde{R}^{\lambda }{}_{\alpha }{}^{\nu }{}_{\kappa } T^{\mu \alpha \kappa } \nonumber\\
   &+\bar{h}_{116}^{} g^{\lambda \nu } \tilde{R}^{\alpha \mu }{}_{\kappa \theta } T_{\alpha }{}^{\kappa \theta } + \bar{h}_{117}^{} g^{\lambda \nu } \tilde{R}^{\alpha }{}_{\kappa }{}^{\mu }{}_{\theta } T_{\alpha }{}^{\kappa \theta } + \bar{h}_{118}^{} \tilde{R}^{\alpha \nu \lambda }{}_{\kappa } T^{\mu }{}_{\alpha }{}^{\kappa } + \bar{h}_{119}^{} \tilde{R}^{\alpha \lambda \nu }{}_{\kappa } T^{\mu }{}_{\alpha }{}^{\kappa } - \bar{h}_{120}^{} \tilde{R}^{\alpha }{}_{\kappa }{}^{\lambda \nu } T^{\mu }{}_{\alpha }{}^{\kappa }\nonumber\\
   & + \bar{h}_{121}^{} \tilde{R}^{\lambda \mu }{}_{\alpha \kappa } T^{\alpha \nu \kappa } + \bar{h}_{122}^{} \tilde{R}^{\lambda }{}_{\alpha }{}^{\mu }{}_{\kappa } T^{\alpha \nu \kappa } + \bar{h}_{123}^{} \tilde{R}^{\alpha \lambda \mu }{}_{\kappa } T_{\alpha }{}^{\nu \kappa }+ \bar{h}_{124}^{} \tilde{R}^{\nu \lambda }{}_{\alpha \kappa } T^{\alpha \mu \kappa }+ \bar{h}_{125}^{} \tilde{R}^{\nu }{}_{\alpha }{}^{\lambda }{}_{\kappa } T^{\alpha \mu \kappa } + \bar{h}_{126}^{} \tilde{R}^{\lambda \nu }{}_{\alpha \kappa } T^{\alpha \mu \kappa }\nonumber\\
   & +\bar{h}_{127}^{} g^{\lambda \nu } \tilde{R}^{\mu }{}_{\alpha \kappa \theta } T^{\kappa \alpha \theta } + \bar{h}_{128}^{} \tilde{R}^{\lambda }{}_{\alpha }{}^{\nu }{}_{\kappa } T^{\alpha \mu \kappa } + \bar{h}_{129}^{} g^{\lambda \nu } \tilde{R}^{\alpha \mu }{}_{\kappa \theta } T^{\kappa }{}_{\alpha }{}^{\theta }+ \bar{h}_{130}^{} g^{\lambda \nu } \tilde{R}^{\alpha }{}_{\kappa }{}^{\mu }{}_{\theta } T^{\kappa }{}_{\alpha }{}^{\theta }+ \bar{h}_{147}^{} g^{\lambda \mu } \tilde{R}^{\nu }{}_{\alpha \kappa \theta } T^{\alpha \kappa \theta } \nonumber\\
   &+\bar{h}_{132}^{} \tilde{R}^{\alpha \lambda \nu }{}_{\kappa } T_{\alpha }{}^{\mu \kappa } - \bar{h}_{133}^{} \tilde{R}^{\alpha }{}_{\kappa }{}^{\lambda \nu } T_{\alpha }{}^{\mu \kappa } + \bar{h}_{134}^{} \tilde{R}^{\lambda }{}_{\alpha }{}^{\mu }{}_{\kappa } T^{\kappa \nu \alpha } + \bar{h}_{142}^{} \tilde{R}^{\nu \lambda \mu }{}_{\alpha } T^{\alpha } + \bar{h}_{135}^{} \tilde{R}^{\alpha \lambda \mu }{}_{\kappa } T^{\kappa \nu }{}_{\alpha }+ \bar{h}_{136}^{} \tilde{R}^{\nu }{}_{\alpha }{}^{\lambda }{}_{\kappa } T^{\kappa \mu \alpha }\nonumber\\
   &+\bar{h}_{137}^{} \tilde{R}^{\lambda }{}_{\alpha }{}^{\nu }{}_{\kappa } T^{\kappa \mu \alpha } + \bar{h}_{138}^{} g^{\lambda \nu } \tilde{R}^{\alpha }{}_{\kappa }{}^{\mu }{}_{\theta } T^{\theta }{}_{\alpha }{}^{\kappa } + \bar{h}_{139}^{} \tilde{R}^{\alpha \nu \lambda }{}_{\kappa } T^{\kappa \mu }{}_{\alpha }+ \bar{h}_{140}^{} \tilde{R}^{\alpha \lambda \nu }{}_{\kappa } T^{\kappa \mu }{}_{\alpha } - \bar{h}_{141}^{} \tilde{R}^{\alpha }{}_{\kappa }{}^{\lambda \nu } T^{\kappa \mu }{}_{\alpha } + \bar{h}_{143}^{} \tilde{R}^{\lambda \nu \mu }{}_{\alpha } T^{\alpha }\nonumber\\
   &+\bar{h}_{144}^{} \tilde{R}^{\lambda \mu \nu }{}_{\alpha } T^{\alpha } + \bar{h}_{131}^{} \tilde{R}^{\alpha \nu \lambda }{}_{\kappa } T_{\alpha }{}^{\mu \kappa }- \bar{h}_{145}^{} \tilde{R}^{\lambda }{}_{\alpha }{}^{\mu \nu } T^{\alpha } - \bar{h}_{146}^{} \tilde{R}^{\alpha \lambda \mu \nu } T_{\alpha }+\bar{h}_{148}^{} g^{\lambda \mu } \tilde{R}^{\alpha \nu }{}_{\kappa \theta } T_{\alpha }{}^{\kappa \theta } + \bar{h}_{149}^{} g^{\lambda \mu } \tilde{R}^{\alpha }{}_{\kappa }{}^{\nu }{}_{\theta } T_{\alpha }{}^{\kappa \theta }\nonumber\\
   &+\bar{h}_{150}^{} g^{\lambda \mu } \tilde{R}^{\nu }{}_{\alpha \kappa \theta } T^{\kappa \alpha \theta } + \bar{h}_{151}^{} g^{\lambda \mu } \tilde{R}^{\alpha \nu }{}_{\kappa \theta } T^{\kappa }{}_{\alpha }{}^{\theta } + \bar{h}_{152}^{} g^{\lambda \mu } \tilde{R}^{\alpha }{}_{\kappa }{}^{\nu }{}_{\theta } T^{\kappa }{}_{\alpha }{}^{\theta }+ \bar{h}_{153}^{} g^{\lambda \mu } \tilde{R}^{\alpha }{}_{\kappa }{}^{\nu }{}_{\theta } T^{\theta }{}_{\alpha }{}^{\kappa } - \bar{h}_{154}^{} g^{\lambda \nu } \tilde{R}_{\alpha \kappa } T^{\alpha \mu \kappa} \nonumber\\
   & -\bar{h}_{156}^{} g^{\lambda \nu } \tilde{R}_{\alpha \kappa } T^{\kappa \mu \alpha } - \bar{h}_{157}^{} \tilde{R}^{\mu }{}_{\alpha } T^{\alpha \lambda \nu } - \bar{h}_{158}^{} \tilde{R}_{\alpha }{}^{\nu } T^{\lambda \mu \alpha } - \bar{h}_{159}^{} \tilde{R}_{\alpha }{}^{\mu } T^{\nu \lambda \alpha } - \bar{h}_{160}^{} \tilde{R}^{\nu }{}_{\alpha } T^{\lambda \mu \alpha }-\bar{h}_{161}^{} \tilde{R}^{\mu }{}_{\alpha } T^{\nu \lambda \alpha }\nonumber\\
   & + \bar{h}_{162}^{} g^{\lambda \nu } \tilde{R}_{\alpha \kappa } T^{\mu \alpha \kappa } - \bar{h}_{163}^{} \tilde{R}_{\alpha }{}^{\mu } T^{\lambda \nu \alpha }- \bar{h}_{177}^{} g^{\lambda \nu } \hat{R}_{\alpha \kappa } T^{\kappa \mu \alpha }+ \bar{h}_{165}^{} g^{\lambda \nu } \tilde{R}_{\alpha }{}^{\mu } T^{\alpha }+ \bar{h}_{166}^{} g^{\lambda \nu } \tilde{R}^{\mu }{}_{\alpha } T^{\alpha } + \bar{h}_{167}^{} \tilde{R}^{\nu \mu } T^{\lambda }   \nonumber\\
   &+\bar{h}_{168}^{} \tilde{R}^{\mu \nu } T^{\lambda } + \bar{h}_{169}^{} \tilde{R}^{\lambda \mu } T^{\nu } + \bar{h}_{170}^{} g^{\lambda \mu } \tilde{R}^{\nu }{}_{\alpha } T^{\alpha } + \bar{h}_{171}^{} g^{\lambda \mu } \tilde{R}_{\alpha }{}^{\nu } T^{\alpha } - \bar{h}_{172}^{} g^{\lambda \mu } \tilde{R}_{\alpha \kappa } T^{\alpha \nu \kappa }- \bar{h}_{173}^{} g^{\lambda \mu } \tilde{R}_{\alpha \kappa } T^{\kappa \nu \alpha }  \nonumber\\
   &+ \bar{h}_{174}^{} g^{\lambda \mu } \tilde{R}_{\alpha \kappa } T^{\nu \alpha \kappa }-\bar{h}_{175}^{} g^{\lambda \nu } \hat{R}_{\alpha \kappa } T^{\alpha \mu \kappa } - \bar{h}_{176}^{} \hat{R}_{\alpha }{}^{\mu } T^{\alpha \lambda \nu } - \bar{h}_{164}^{} \tilde{R}^{\mu }{}_{\alpha } T^{\lambda \nu \alpha }- \bar{h}_{178}^{} \hat{R}^{\mu }{}_{\alpha } T^{\alpha \lambda \nu }- \bar{h}_{179}^{} \hat{R}_{\alpha }{}^{\nu } T^{\lambda \mu \alpha }  \nonumber\\
   &-\bar{h}_{180}^{} \hat{R}_{\alpha }{}^{\mu } T^{\nu \lambda \alpha } - \bar{h}_{181}^{} \hat{R}^{\nu }{}_{\alpha } T^{\lambda \mu \alpha } - \bar{h}_{182}^{} \hat{R}^{\mu }{}_{\alpha } T^{\nu \lambda \alpha } + \bar{h}_{183}^{} g^{\lambda \nu } \hat{R}_{\alpha \kappa } T^{\mu \alpha \kappa } - \bar{h}_{184}^{} \hat{R}_{\alpha }{}^{\mu } T^{\lambda \nu \alpha } - \bar{h}_{185}^{} \hat{R}^{\mu }{}_{\alpha } T^{\lambda \nu \alpha } \nonumber\\
   & + \bar{h}_{186}^{} g^{\lambda \nu } \hat{R}_{\alpha }{}^{\mu } T^{\alpha } + \bar{h}_{187}^{} g^{\lambda \nu } \hat{R}^{\mu }{}_{\alpha } T^{\alpha } + \bar{h}_{188}^{} \hat{R}^{\nu \mu } T^{\lambda } + \bar{h}_{189}^{} \hat{R}^{\mu \nu } T^{\lambda } + \bar{h}_{190}^{} \hat{R}^{\lambda \mu } T^{\nu }+ \bar{h}_{191}^{} g^{\lambda \mu } \hat{R}^{\nu }{}_{\alpha } T^{\alpha } + \bar{h}_{192}^{} g^{\lambda \mu } \hat{R}_{\alpha }{}^{\nu } T^{\alpha } \nonumber\\
   &-\bar{h}_{193}^{} g^{\lambda \mu } \hat{R}_{\alpha \kappa } T^{\alpha \nu \kappa } - \bar{h}_{194}^{} g^{\lambda \mu } \hat{R}_{\alpha \kappa } T^{\kappa \nu \alpha } + \bar{h}_{195}^{} g^{\lambda \mu } \hat{R}_{\alpha \kappa } T^{\nu \alpha \kappa } + \bar{h}_{196}^{} \bar{R}^{\lambda }{}_{\alpha } T^{\nu \mu \alpha } + \bar{h}_{197}^{} \bar{R}^{\nu }{}_{\alpha } T^{\lambda \mu \alpha } + \bar{h}_{198}^{} g^{\lambda \nu } \bar{R}_{\alpha \kappa } T^{\mu \alpha \kappa }\nonumber\\
   &+\bar{h}_{199}^{} \bar{R}^{\lambda }{}_{\alpha } T^{\mu \nu \alpha } + \bar{h}_{200}^{} g^{\lambda \nu } \bar{R}_{\alpha \kappa } T^{\alpha \mu \kappa } - \bar{h}_{201}^{} \bar{R}^{\lambda }{}_{\alpha } T^{\alpha \mu \nu } - \bar{h}_{202}^{} \bar{R}^{\lambda \nu } T^{\mu } + \bar{h}_{203}^{} g^{\lambda \nu } \bar{R}^{\mu }{}_{\alpha } T^{\alpha } + \bar{h}_{204}^{} g^{\lambda \mu } \bar{R}_{\alpha \kappa } T^{\nu \alpha \kappa } \nonumber\\
   &+\bar{h}_{205}^{} g^{\lambda \mu } \bar{R}_{\alpha \kappa } T^{\alpha \nu \kappa } + \bar{h}_{206}^{} g^{\lambda \mu } \bar{R}^{\nu }{}_{\alpha } T^{\alpha } - \bar{h}_{207}^{} \tilde{R} T^{\lambda \mu \nu } + \bar{h}_{208}^{} g^{\lambda \nu } \tilde{R} T^{\mu } + \bar{h}_{209}^{} g^{\lambda \mu } \tilde{R} T^{\nu }\,.
\end{align}
Note that in the particular cases of Riemann-Cartan geometry and Weyl-Cartan geometry, the symmetric and symmetric traceless components of the anholonomic connection respectively vanish, which trivialises the same types of components of the connection field equations.

On the other hand, the tensor $E^{\mu\nu}$ provides the tetrad field equations and can be written as
\begin{equation}
    E^{\mu\nu}=2 G^{\mu\nu} +\tilde{\mathcal{L}} \,g^{\mu \nu } + V^{\mu \nu } + 2N_{\alpha}{}^{\mu}{}_{\beta}X^{\alpha[\beta\nu]}-2\nabla_\alpha X^{\mu[\alpha\nu]}\,,
\end{equation}
where $\tilde{\mathcal{L}}$ represents the Lagrangian density given by the quadratic and cubic order invariants in Expression~\eqref{cubicmodel} and the tensor $V^{\mu\nu}$ reads


\section{Lagrangian coefficients of the solutions}\label{appe3}

From the action~\eqref{cubicmodel} of cubic MAG, the gravitational-wave solutions obtained in Sec.~\ref{sec:sol} are derived by imposing different relations between the Lagrangian coefficients, as outlined below.

First, the avoidance of Ostrogradsky instabilities in the vector and axial sectors requires imposing relations~\eqref{sti}-\eqref{stf}.

A further restriction has shown to provide Reissner-Nordström-like solutions with spin, dilation and shear charges~\cite{Bahamonde:2024efl}, which additionally demands:


Focusing then on the search of gravitational wave solutions under the profile described in Sec.~\ref{sec:wave_settings}, the resolution of the field equations carried out in Sec.~\ref{subsecRC} and Sec.~\ref{subsecWC} for the Riemann-Cartan and Weyl-Cartan geometries provides the conditions~\eqref{RC} and~\eqref{WC}. Likewise, a similar procedure followed in Sec.~\ref{subsecGeneralMAG} for the general metric-affine geometry initially sets the following cumbersome expressions in the Lagrangian coefficients:
\begin{align}
    h_{42} =&\, \frac{1}{18}d_{1}+\frac{1}{2}h_{23}+\frac{29}{36}h_{25}+\frac{1}{6}h_{160}-\frac{2}{3}h_{5}-\frac{1}{6}h_{9}-\frac{3}{2}h_{33}-h_{41}\,,\label{genMAG1}\\
    h_{43} =&\, \frac{1}{36}\left(2d_{1}+29h_{25}+6h_{160}-24h_{5}-6h_{9}-54h_{33}\right),\\
    h_{45} =&\, 3h_{28}-\frac{1}{16}d_{1}-\frac{1}{2}h_{25}-h_{44}\,,\\
    h_{92} =& -\frac{144655}{4783872}d_{1}+\frac{2295}{6229}h_{149}+\frac{117035}{49832}h_{101}-\frac{192255}{199328}h_{152}+\frac{405}{6229}h_{189}+\frac{15}{12458}h_{191}+\frac{1635}{12458}h_{192}+\frac{5455}{99664}h_{169}\nonumber\\
    &-\frac{125715}{199328}h_{151}-\frac{220589}{24916}h_{104}+\frac{2865}{199328}h_{154}+\frac{390}{6229}h_{186}+\frac{1605}{6229}h_{179}+\frac{1995}{12458}h_{178}-\frac{18095}{49832}h_{158}-\frac{111435}{24916}h_{105}\nonumber\\
    &+\frac{4525}{49832}h_{160}-\frac{825}{12458}h_{181}-\frac{71377}{12458}h_{103}+\frac{2295}{6229}h_{147}-\frac{8895}{49832}h_{145}-\frac{1185}{12458}h_{182}-\frac{82705}{99664}h_{156}+\frac{14475}{24916}h_{146}\nonumber\\
    &+\frac{72235}{49832}h_{102}+\frac{88125}{199328}h_{153}-\frac{8895}{24916}h_{144}-\frac{795}{6229}h_{180}+\frac{4145}{99664}h_{171}-\frac{210}{6229}h_{184}+\frac{14475}{49832}h_{148}-\frac{2400}{6229}h_{60}\nonumber\\
    &-\frac{2541197}{49832}h_{65}-\frac{3}{2}h_{53}-\frac{6629}{6229}h_{94}-\frac{12058}{18687}h_{95}-\frac{8325}{49832}h_{33}+\frac{1200}{6229}h_{58}+\frac{83435}{49832}h_{99}-\frac{1200}{6229}h_{90}+\frac{57861}{12458}h_{61}\nonumber\\
    &+\frac{1224315}{49832}h_{66}-\frac{110605}{1195968}h_{25}-\frac{45}{24916}h_{28}+\frac{1500}{6229}h_{54}-\frac{76691}{24916}h_{97}-\frac{8229}{6229}h_{91}+\frac{20487}{12458}h_{55}-\frac{59061}{12458}h_{59}\nonumber\\
    &-\frac{16887}{12458}h_{57}-\frac{3}{2}h_{56}+\frac{1071703}{149496}a_{2}-\frac{170343}{24916}N_{3}a_{2}+\frac{1172503}{298992}a_{6}-\frac{170343}{49832}N_{3}a_{6}\,,\\
    h_{96}=&-h_{97}\,,\\
    h_{123} =&\,\frac{2}{3}h_{75}-h_{121}-\frac{2}{3}h_{74}-\frac{2}{3}a_{2}-2N_{3}a_{2}-\frac{1}{3}a_{6}-N_{3}a_{6}\,,\\
    h_{132} =&-\frac{3}{4}h_{33}\,,\\
    h_{135} =& \frac{1}{4}\left(3h_{33}+6h_{34}+4h_{132}\right),\\
    h_{174} =& \,\frac{533313}{797312}d_{1}+\frac{34641}{12458}h_{149}+\frac{67437}{24916}h_{101}-\frac{273969}{99664}h_{152}-\frac{1890}{6229}h_{189}-\frac{6264}{6229}h_{191}-\frac{10044}{6229}h_{192}+\frac{20493}{49832}h_{169}\nonumber\\
    &-\frac{479061}{99664}h_{151}+\frac{99441}{12458}h_{104}+\frac{317223}{99664}h_{154}+\frac{10638}{6229}h_{186}+\frac{4968}{6229}h_{179}+\frac{7803}{6229}h_{178}+\frac{100359}{24916}h_{158}-\frac{51435}{12458}h_{105}\nonumber\\
    &+\frac{8217}{12458}h_{160}+\frac{8154}{6229}h_{181}+\frac{27432}{6229}h_{103}+\frac{34641}{12458}h_{147}+\frac{7263}{12458}h_{145}-\frac{9693}{6229}h_{182}-\frac{388395}{49832}h_{156}+\frac{4914}{6229}h_{146}\nonumber\\
    &-\frac{60579}{24916}h_{102}+\frac{367443}{99664}h_{153}+\frac{7263}{6229}h_{144}-\frac{8748}{6229}h_{180}+\frac{6939}{49832}h_{171}+\frac{7209}{6229}h_{184}+\frac{2457}{6229}h_{148}-\frac{13716}{6229}h_{60}\nonumber\\
    &-\frac{266319}{24916}h_{65}-\frac{2286}{6229}h_{94}+\frac{762}{6229}h_{95}-\frac{55323}{24916}h_{33}+\frac{6858}{6229}h_{58}-\frac{28575}{24916}h_{99}-\frac{6858}{6229}h_{90}+\frac{10287}{12458}h_{61}-\frac{91059}{24916}h_{66}\nonumber\\
    &-\frac{57933}{199328}h_{25}-\frac{485757}{12458}h_{28}+\frac{17145}{12458}h_{54}+\frac{381}{12458}h_{97}-\frac{11430}{6229}h_{91}+\frac{10287}{12458}h_{55}+\frac{10287}{12458}h_{57}-\frac{17145}{12458}h_{59}\nonumber\\
    &+\frac{361037}{24916}a_{2}-\frac{53541}{6229}N_{3}a_{2}+\frac{457049}{49832}a_{6}-\frac{53541}{12458}N_{3}a_{6}\,,\\
    h_{185} =& \,\frac{15164593}{3587904}d_{1}-\frac{177}{6229}h_{149}+\frac{1947}{12458}h_{101}+\frac{302363}{49832}h_{152}+\frac{1068}{6229}h_{189}-\frac{11054}{6229}h_{191}-\frac{8918}{6229}h_{192}+\frac{24631}{74748}h_{169}\nonumber\\
    &+\frac{252103}{49832}h_{151}+\frac{2871}{6229}h_{104}+\frac{314723}{49832}h_{154}-\frac{1740}{6229}h_{186}-\frac{4765}{6229}h_{179}-\frac{6367}{6229}h_{178}+\frac{816893}{37374}h_{158}-\frac{1485}{6229}h_{105}\nonumber\\
    &-\frac{8}{6229}h_{160}+\frac{3757}{6229}h_{181}+\frac{1584}{6229}h_{103}-\frac{177}{6229}h_{147}+\frac{9872}{6229}h_{145}+\frac{7435}{6229}h_{182}-\frac{841841}{74748}h_{156}-\frac{20218}{6229}h_{146}\nonumber\\
    &-\frac{1749}{12458}h_{102}+\frac{243583}{49832}h_{153}+\frac{19744}{6229}h_{144}+\frac{6901}{6229}h_{180}-\frac{22255}{74748}h_{171}+\frac{10520}{6229}h_{184}-\frac{10109}{6229}h_{148}-\frac{792}{6229}h_{60}\nonumber\\
    &-\frac{7689}{12458}h_{65}-\frac{132}{6229}h_{94}+\frac{44}{6229}h_{95}+\frac{2817}{12458}h_{33}+\frac{396}{6229}h_{58}-\frac{825}{12458}h_{99}-\frac{396}{6229}h_{90}+\frac{297}{6229}h_{61}-\frac{2629}{12458}h_{66}\nonumber\\
    &+\frac{697603}{896976}h_{25}-\frac{880395}{6229}h_{28}+\frac{495}{6229}h_{54}+\frac{11}{6229}h_{97}-\frac{660}{6229}h_{91}+\frac{297}{6229}h_{55}+\frac{297}{6229}h_{57}-\frac{495}{6229}h_{59}+\frac{2841047}{112122}a_{2}\nonumber\\
    &-\frac{101088}{6229}N_{3}a_{2}+\frac{2865995}{224244}a_{6}-\frac{50544}{6229}N_{3}a_{6}\,,\\
    h_{187} =&-\frac{197}{192}d_{1}+\frac{3}{2}h_{145}-3h_{146}+3h_{144}+\frac{45}{2}h_{28}+\frac{15}{2}h_{33}-\frac{3}{2}h_{148}-\frac{9}{4}h_{151}+h_{181}+h_{182}-\frac{5}{6}h_{160}+h_{184}+\frac{43}{48}h_{25}\nonumber\\
    &+\frac{9}{4}h_{154}-\frac{9}{4}h_{153}-\frac{7}{2}h_{158}+\frac{9}{4}h_{152}+\frac{7}{4}h_{156}+\frac{73}{6}a_{2}-12N_{3}a_{2}+\frac{73}{12}a_{6}-6N_{3}a_{6}\,,\\
    h_{188} =&\, \frac{2977}{576}d_{1}+\frac{5}{6}h_{160}+\frac{1}{4}h_{169}+h_{186}-\frac{1}{4}h_{171}-171h_{28}-\frac{161}{12}h_{156}-\frac{149}{144}h_{25}+h_{144}-h_{146}+\frac{1}{2}h_{145}-\frac{1}{2}h_{148}\nonumber\\
    &+\frac{19}{8}h_{151}-\frac{1}{8}h_{152}-\frac{1}{8}h_{154}+\frac{19}{8}h_{153}-\frac{15}{2}h_{33}+\frac{161}{6}h_{158}+\frac{161}{18}a_{2}+\frac{161}{36}a_{6}\,,\\
    h_{190} =& -\frac{205229}{1195968}d_{1}-\frac{5685}{12458}h_{149}-\frac{12213}{24916}h_{101}-\frac{136665}{99664}h_{152}-\frac{6181}{6229}h_{189}+\frac{13151}{6229}h_{191}+\frac{13247}{6229}h_{192}-\frac{3875}{149496}h_{169}\nonumber\\
    &-\frac{127185}{99664}h_{151}-\frac{18009}{12458}h_{104}-\frac{241377}{99664}h_{154}-\frac{1338}{6229}h_{186}-\frac{1194}{6229}h_{179}-\frac{1266}{6229}h_{178}-\frac{1883}{6229}h_{158}+\frac{9315}{12458}h_{105}\nonumber\\
    &-\frac{2731}{24916}h_{160}-\frac{741}{6229}h_{181}-\frac{4968}{6229}h_{103}-\frac{5685}{12458}h_{147}+\frac{165}{24916}h_{145}+\frac{1314}{6229}h_{182}+\frac{7462}{6229}h_{156}-\frac{3987}{12458}h_{146}\nonumber\\
    &+\frac{10971}{24916}h_{102}-\frac{264009}{99664}h_{153}+\frac{165}{12458}h_{144}+\frac{1290}{6229}h_{180}-\frac{11029}{149496}h_{171}-\frac{717}{6229}h_{184}-\frac{3987}{24916}h_{148}+\frac{2484}{6229}h_{60}\nonumber\\
    &+\frac{48231}{24916}h_{65}+\frac{414}{6229}h_{94}-\frac{138}{6229}h_{95}+\frac{1953}{6229}h_{33}-\frac{1242}{6229}h_{58}+\frac{5175}{24916}h_{99}+\frac{1242}{6229}h_{90}-\frac{1863}{12458}h_{61}+\frac{16491}{24916}h_{66}\nonumber\\
    &-\frac{32195}{74748}h_{25}+\frac{89277}{24916}h_{28}-\frac{3105}{12458}h_{54}-\frac{69}{12458}h_{97}+\frac{2070}{6229}h_{91}-\frac{1863}{12458}h_{55}-\frac{1863}{12458}h_{57}+\frac{3105}{12458}h_{59}-\frac{438097}{37374}a_{2}\nonumber\\
    &+\frac{62226}{6229}N_{3}a_{2}-\frac{464179}{74748}a_{6}+\frac{31113}{6229}N_{3}a_{6}\,,\\
    h_{193} =&-\frac{664451}{5381856}d_{1}+\frac{13140}{6229}h_{149}+\frac{2478}{6229}h_{101}+\frac{86027}{298992}h_{152}-\frac{1471}{18687}h_{189}+\frac{11854}{18687}h_{191}-\frac{28462}{18687}h_{192}+\frac{14971}{149496}h_{169}\nonumber\\
    &-\frac{1148189}{298992}h_{151}+\frac{7308}{6229}h_{104}+\frac{600971}{298992}h_{154}+\frac{18424}{18687}h_{186}-\frac{17134}{18687}h_{179}+\frac{13103}{18687}h_{178}+\frac{2863}{112122}h_{158}-\frac{3780}{6229}h_{105}\nonumber\\
    &+\frac{81743}{112122}h_{160}+\frac{2768}{6229}h_{181}+\frac{4032}{6229}h_{103}+\frac{13140}{6229}h_{147}+\frac{35819}{37374}h_{145}-\frac{27032}{18687}h_{182}-\frac{193375}{224244}h_{156}+\frac{1333}{18687}h_{146}\nonumber\\
    &-\frac{2226}{6229}h_{102}+\frac{1004899}{298992}h_{153}+\frac{35819}{18687}h_{144}-\frac{5651}{6229}h_{180}-\frac{2875}{149496}h_{171}+\frac{4454}{18687}h_{184}+\frac{1333}{37374}h_{148}-\frac{2016}{6229}h_{60}\nonumber\\
    &-\frac{9786}{6229}h_{65}-\frac{336}{6229}h_{94}+\frac{112}{6229}h_{95}-\frac{74939}{12458}h_{33}+\frac{1008}{6229}h_{58}-\frac{1050}{6229}h_{99}-\frac{1008}{6229}h_{90}+\frac{756}{6229}h_{61}-\frac{3346}{6229}h_{66}\nonumber\\
    &+\frac{295817}{2690928}h_{25}-\frac{67311}{24916}h_{28}+\frac{1260}{6229}h_{54}+\frac{28}{6229}h_{97}-\frac{1680}{6229}h_{91}+\frac{756}{6229}h_{55}-\frac{1260}{6229}h_{59}+\frac{756}{6229}h_{57}-\frac{237041}{336366}a_{2}\nonumber\\
    &+\frac{14496}{6229}N_{3}a_{2}-\frac{46529}{672732}a_{6}+\frac{7248}{6229}N_{3}a_{6}\,,\\
    h_{194} =& \,\frac{5062795}{10763712}d_{1}-\frac{19887}{12458}h_{149}-\frac{5487}{24916}h_{101}-\frac{175921}{149496}h_{152}-\frac{5081}{18687}h_{189}+\frac{15017}{18687}h_{191}+\frac{42229}{18687}h_{192}+\frac{417}{24916}h_{169}\nonumber\\
    &+\frac{130543}{149496}h_{151}-\frac{8091}{12458}h_{104}-\frac{350485}{149496}h_{154}-\frac{3970}{18687}h_{186}+\frac{11932}{18687}h_{179}-\frac{8477}{18687}h_{178}+\frac{362033}{112122}h_{158}+\frac{4185}{12458}h_{105}\nonumber\\
    &-\frac{138331}{224244}h_{160}+\frac{2917}{6229}h_{181}-\frac{2232}{6229}h_{103}-\frac{19887}{12458}h_{147}-\frac{10069}{74748}h_{145}+\frac{15854}{18687}h_{182}-\frac{256571}{224244}h_{156}-\frac{47081}{37374}h_{146}\nonumber\\
    &+\frac{4929}{24916}h_{102}-\frac{689645}{149496}h_{153}-\frac{10069}{37374}h_{144}+\frac{3017}{6229}h_{180}-\frac{1533}{24916}h_{171}+\frac{9325}{18687}h_{184}-\frac{47081}{74748}h_{148}+\frac{1116}{6229}h_{60}\nonumber\\
    &+\frac{21669}{24916}h_{65}+\frac{186}{6229}h_{94}-\frac{62}{6229}h_{95}+\frac{65399}{12458}h_{33}-\frac{558}{6229}h_{58}+\frac{2325}{24916}h_{99}+\frac{558}{6229}h_{90}-\frac{837}{12458}h_{61}+\frac{7409}{24916}h_{66}\nonumber\\
    &-\frac{1378967}{2690928}h_{25}-\frac{122745}{6229}h_{28}-\frac{1395}{12458}h_{54}-\frac{31}{12458}h_{97}+\frac{930}{6229}h_{91}-\frac{837}{12458}h_{55}-\frac{837}{12458}h_{57}+\frac{1395}{12458}h_{59}\nonumber\\
    &+\frac{667876}{168183}a_{2}-\frac{24042}{6229}N_{3}a_{2}+\frac{615145}{336366}a_{6}-\frac{12021}{6229}N_{3}a_{6}\,,\\
    h_{197} =& -\frac{1992767}{5381856}d_{1}+\frac{3285}{6229}h_{149}+\frac{1239}{12458}h_{101}-\frac{281231}{149496}h_{152}+\frac{4304}{18687}h_{189}-\frac{6380}{18687}h_{191}+\frac{2228}{18687}h_{192}-\frac{1193}{24916}h_{169}\nonumber\\
    &-\frac{229951}{149496}h_{151}+\frac{1827}{6229}h_{104}-\frac{216863}{149496}h_{154}+\frac{4606}{18687}h_{186}+\frac{5060}{18687}h_{179}+\frac{15895}{37374}h_{178}-\frac{114100}{56061}h_{158}-\frac{945}{6229}h_{105}\nonumber\\
    &-\frac{40297}{112122}h_{160}+\frac{692}{6229}h_{181}+\frac{1008}{6229}h_{103}+\frac{3285}{6229}h_{147}-\frac{7202}{18687}h_{145}-\frac{6758}{18687}h_{182}+\frac{45143}{56061}h_{156}+\frac{23692}{18687}h_{146}\nonumber\\
    &-\frac{1113}{12458}h_{102}+\frac{39185}{149496}h_{153}-\frac{14404}{18687}h_{144}+\frac{289}{12458}h_{180}+\frac{1697}{24916}h_{171}-\frac{8230}{18687}h_{184}+\frac{11846}{18687}h_{148}-\frac{504}{6229}h_{60}\nonumber\\
    &-\frac{4893}{12458}h_{65}-\frac{84}{6229}h_{94}+\frac{28}{6229}h_{95}+\frac{20999}{6229}h_{33}+\frac{252}{6229}h_{58}-\frac{525}{12458}h_{99}-\frac{252}{6229}h_{90}+\frac{189}{6229}h_{61}-\frac{1673}{12458}h_{66}\nonumber\\
    &-\frac{288679}{672732}h_{25}+\frac{162105}{12458}h_{28}+\frac{315}{6229}h_{54}+\frac{7}{6229}h_{97}-\frac{420}{6229}h_{91}+\frac{189}{6229}h_{55}-\frac{315}{6229}h_{59}+\frac{189}{6229}h_{57}+\frac{1}{2}h_{74}\nonumber\\
    &-\frac{1}{2}h_{75}-2h_{78}+\frac{600929}{672732}a_{2}+\frac{145305}{24916}N_{3}a_{2}+\frac{696185}{1345464}a_{6}+\frac{145305}{49832}N_{3}a_{6}\,.\label{genMAG16}
\end{align}
The remaining equations to be solved for determining the complete form of the torsion and nonmetricity tensors of the solutions are then reduced to Eqs.~\eqref{deq1}-\eqref{deq3}, which depend on the parameters
\begin{align}
    b_{1}=&\,\frac{240741}{24916}h_{33}-\frac{13745501}{2391936}d_{1}-\frac{80919}{12458}h_{147}+\frac{46995}{6229}h_{182}-\frac{129117}{12458}h_{145}+\frac{1263157}{49832}h_{156}+\frac{92982}{6229}h_{146}-\frac{80919}{12458}h_{149}\nonumber\\
    &+\frac{1140843}{99664}h_{151}+\frac{46491}{6229}h_{148}-\frac{57855}{6229}h_{184}+\frac{33252}{6229}h_{180}-\frac{71598}{6229}h_{181}+\frac{572107}{37374}h_{97}+\frac{14843}{37374}h_{95}-\frac{200219}{24916}h_{99}\nonumber\\
    &-\frac{5766}{6229}h_{58}+\frac{15}{2}h_{53}-\frac{345015}{12458}h_{61}-\frac{3777115}{24916}h_{66}+\frac{1922}{6229}h_{94}+\frac{15}{2}h_{56}+\frac{3399507}{12458}h_{28}+\frac{30682}{6229}h_{90}+\frac{21720}{6229}h_{179}\nonumber\\
    &-\frac{1020985}{24916}h_{158}-\frac{19509}{6229}h_{178}+\frac{267489}{12458}h_{105}-\frac{1465905}{99664}h_{152}-\frac{1021101}{99664}h_{153}-\frac{72896}{6229}h_{103}-\frac{60738}{6229}h_{186}-\frac{14415}{12458}h_{54}\nonumber\\
    &-\frac{129117}{6229}h_{144}-\frac{280943}{24916}h_{101}+\frac{40973}{12458}h_{104}+\frac{24379}{49832}h_{171}-\frac{173311}{24916}h_{102}-\frac{47443}{49832}h_{169}+\frac{42393}{6229}h_{57}-\frac{730055}{597984}h_{25}\nonumber\\
    &-\frac{51042}{6229}h_{55}+\frac{350781}{12458}h_{59}+\frac{4658961}{24916}h_{65}+\frac{34526}{6229}h_{91}+\frac{11532}{6229}h_{60}-\frac{2170137}{99664}h_{154}+\frac{44112}{6229}h_{191}+\frac{99084}{6229}h_{192}\nonumber\\
    &+\frac{27486}{6229}h_{189}-\frac{17699}{12458}h_{160}-\frac{1792837}{18687}a_{2}+\frac{1927665}{24916}a_{2}N_{3}-\frac{926690}{18687}a_{6}+\frac{1927665}{49832}a_{6}N_{3}\,,\label{par1}\\
    b_{2}=&\,4130784h_{33}- 1244397d_{1}-227584h_{95}  - 1536192h_{57}- 2048256h_{58} + 2133600h_{99}- 2560320h_{54}+ 6799072h_{66}\nonumber\\
    &+ 14500080h_{156} + 2894976h_{182} - 1084608h_{145} - 5173056h_{147} + 2048256h_{90} - 1536192h_{61}+ 4523232h_{102}\nonumber\\
    &- 8193024h_{103} - 2435328h_{181} + 540708h_{25} + 72539712h_{28} - 1227072h_{160} + 7680960h_{105} - 1467648h_{146}\nonumber\\
    &- 7493472h_{158} - 2330496h_{178} - 1483776h_{179} - 3177216h_{186} - 5921496h_{154} - 765072h_{169} - 259056h_{171}\nonumber\\
    &+ 19885152h_{65} + 4096512h_{60} + 2612736h_{180} - 2153088h_{184} - 733824h_{148} - 5173056h_{149} - 6858936h_{153} \nonumber\\
    &- 5035296h_{101}+ 5114088h_{152} + 564480h_{189} + 1870848h_{191} + 2999808h_{192} + 2560320h_{59} - 1536192h_{55}\nonumber\\
    & + 8942472h_{151}- 14849856h_{104} - 2169216h_{144} + 682752h_{94} + 3413760h_{91} - 56896h_{97}- 26027232a_{2}\nonumber\\
    & + 15990912N_{3}a_{2}- 16598064a_{6} + 7995456N_{3}a_{6}\,,\\
    b_{3}=&\,21527424h_{33}-3787232d_{1}-161455680h_{61}-796514688h_{66}+281451136h_{97}+39466944h_{57}-98069376h_{101}\nonumber\\
    &+50230656h_{56}-25115328h_{146}+83717760h_{156}+4783872h_{94}+50230656h_{53}+179395200h_{105}-17939520h_{54}\nonumber\\
    &+5182528h_{25}+279856512h_{28}+239193600h_{104}-14351616h_{58}-47838720h_{99}-31095168h_{102}-23321376h_{154}\nonumber\\
    &-7175808h_{181}-32291136h_{149}-60994368h_{55}-110029056h_{91}+160259712h_{103}+28703232h_{60}+3587904h_{144}\nonumber\\
    &-43453504h_{95}-7773792h_{160}+52024608h_{152}+7175808h_{191}+7175808h_{192}+168631488h_{59}+1784384256h_{65}\nonumber\\
    &-2391936h_{171}-119596800h_{90}-4783872h_{169}-16743552h_{158}-14351616h_{178}-14351616h_{179}-14351616h_{186}\nonumber\\
    &+55612512h_{151}-41260896h_{153}+14351616h_{180}-7175808h_{184}-12557664h_{148}+14351616h_{182}+1793952h_{145}\nonumber\\
    &-32291136h_{147}-340452224a_{2}+290620224N_{3}a_{2}-195341440a_{6}+145310112N_{3}a_{6}\,.\label{par3}
\end{align}
Thus, the existence of solutions with nontrivial contributions in the transverse space of the general metric-affine geometry requires the previous parameters to vanish, which means the last set of relations:
\begin{align}
    h_{53}=&-\frac{361}{237}d_{1}-\frac{3098}{237}h_{101}-\frac{2700}{79}h_{104}+\frac{114}{79}h_{144}-\frac{247}{79}h_{160}-\frac{1311}{79}h_{153}+\frac{1653}{79}h_{152}-\frac{76}{79}h_{171}+\frac{2660}{79}h_{156}\nonumber\\
    &-\frac{798}{79}h_{146}-\frac{1026}{79}h_{149}-\frac{228}{79}h_{184}-\frac{228}{79}h_{181}-\frac{3124}{237}h_{103}-\frac{456}{79}h_{186}-\frac{456}{79}h_{179}-\frac{456}{79}h_{178}-\frac{532}{79}h_{158}\nonumber\\
    &+\frac{4600}{237}h_{105}-\frac{741}{79}h_{154}+\frac{228}{79}h_{192}+\frac{456}{79}h_{180}+\frac{3286}{237}h_{102}+\frac{2584}{711}h_{97}-\frac{517}{237}h_{95}+\frac{1690}{237}h_{99}-\frac{456}{79}h_{58}\nonumber\\
    &-h_{56}-\frac{1026}{79}h_{147}+\frac{33464}{711}h_{65}-\frac{263}{79}h_{55}+\frac{208}{79}h_{59}-\frac{570}{79}h_{54}+\frac{228}{79}h_{191}+\frac{684}{79}h_{33}+\frac{304}{79}h_{91}+\frac{912}{79}h_{60}\nonumber\\
    &-\frac{152}{79}h_{169}+\frac{456}{79}h_{182}+\frac{1767}{79}h_{151}-\frac{399}{79}h_{148}+\frac{57}{79}h_{145}+\frac{32728}{711}h_{66}+\frac{20}{79}h_{61}+\frac{152}{79}h_{94}-\frac{421}{79}h_{57}\nonumber\\
    &+\frac{494}{237}h_{25}+\frac{8892}{79}h_{28}-\frac{23179}{474}a_{2}+\frac{4143}{158}N_{3}a_{2}-\frac{32755}{948}a_{6}+\frac{4143}{316}N_{3}a_{6}\,,\label{genMAG17}\\
    h_{90}=&-\frac{1273}{1896}d_{1}-\frac{997}{158}h_{101}-\frac{976}{79}h_{104}+\frac{201}{316}h_{144}-\frac{871}{632}h_{160}-\frac{4623}{632}h_{153}+\frac{5829}{632}h_{152}-\frac{67}{158}h_{171}+\frac{2345}{158}h_{156}\nonumber\\
    &-\frac{1407}{316}h_{146}-\frac{1809}{316}h_{149}-\frac{201}{158}h_{184}-\frac{201}{158}h_{181}-\frac{663}{158}h_{103}-\frac{201}{79}h_{186}-\frac{201}{79}h_{179}-\frac{201}{79}h_{178}-\frac{469}{158}h_{158}\nonumber\\
    &+\frac{1525}{158}h_{105}-\frac{2613}{632}h_{154}+\frac{201}{158}h_{192}+\frac{201}{79}h_{180}+\frac{879}{158}h_{102}+\frac{613}{158}h_{97}-\frac{1213}{948}h_{95}+\frac{205}{79}h_{99}-\frac{201}{79}h_{58}\nonumber\\
    &-\frac{1809}{316}h_{147}+\frac{8221}{237}h_{65}-\frac{603}{316}h_{55}+\frac{795}{316}h_{59}-\frac{1005}{316}h_{54}+\frac{201}{158}h_{191}+\frac{603}{158}h_{33}+\frac{55}{79}h_{91}+\frac{402}{79}h_{60}-\frac{67}{79}h_{169}\nonumber\\
    &+\frac{201}{79}h_{182}+\frac{6231}{632}h_{151}-\frac{1407}{632}h_{148}+\frac{201}{632}h_{145}+\frac{6007}{474}h_{66}-\frac{393}{316}h_{61}+\frac{67}{79}h_{94}-\frac{603}{316}h_{57}+\frac{871}{948}h_{25}+\frac{7839}{158}h_{28}\nonumber\\
    &-\frac{22169}{948}a_{2}+\frac{1062}{79}N_{3}a_{2}-\frac{30611}{1896}a_{6}+\frac{531}{79}N_{3}a_{6}\,,\\
    h_{189}=&\,\frac{93855}{20224}d_{1}+\frac{1016}{79}h_{58}-\frac{25019}{1896}h_{99}-\frac{305435}{1896}h_{65}+\frac{762}{79}h_{55}-\frac{2159}{158}h_{59}+\frac{1270}{79}h_{54}-\frac{13377}{632}h_{33}-\frac{330073}{5688}h_{66}\nonumber\\
    &-\frac{100541}{1264}h_{156}+\frac{5927}{316}h_{146}+\frac{2365}{79}h_{149}+\frac{666}{79}h_{184}+\frac{1411}{158}h_{181}+\frac{4699}{158}h_{103}+\frac{1174}{79}h_{186}+\frac{937}{79}h_{179}+\frac{2111}{158}h_{178}\nonumber\\
    &-\frac{2032}{237}h_{91}-\frac{2032}{79}h_{60}+\frac{5603}{1264}h_{169}-\frac{2269}{158}h_{182}-\frac{130487}{2528}h_{151}+\frac{5927}{632}h_{148}+\frac{485}{632}h_{145}+\frac{1143}{158}h_{61}+\frac{2525}{1264}h_{171}\nonumber\\
    &+\frac{15197}{632}h_{158}-\frac{15367}{316}h_{105}+\frac{64445}{2528}h_{154}-\frac{1569}{158}h_{192}-\frac{1095}{79}h_{180}-\frac{53467}{1896}h_{102}-\frac{39751}{2844}h_{97}+\frac{14351}{2844}h_{95}\nonumber\\
    &+\frac{2365}{79}h_{147}-\frac{1253}{158}h_{191}+\frac{60325}{1896}h_{101}+\frac{22479}{316}h_{104}+\frac{485}{316}h_{144}+\frac{13603}{1896}h_{160}+\frac{97817}{2528}h_{153}-\frac{107507}{2528}h_{152}\nonumber\\
    &-\frac{1016}{237}h_{94}+\frac{762}{79}h_{57}-\frac{21699}{5056}h_{25}-\frac{97497}{316}h_{28}+\frac{248305}{1896}a_{2}-\frac{12183}{158}N_{3}a_{2}+\frac{333649}{3792}a_{6}-\frac{12183}{316}N_{3}a_{6}\,.\label{genMAG19}
\end{align}

Hence, a total of $158$ Lagrangian coefficients are fixed by this procedure, which leads to a cubic MAG action with $64$ independent coefficients, in addition to the gravitational constant of GR.

\newpage

\bibliographystyle{utphys}
\bibliography{references}

@book{Blagojevic:2013xpa,
    editor = "Blagojevi\'c, Milutin and Hehl, Friedrich W.",
    title = "{Gauge Theories of Gravitation}: {A Reader with Commentaries}",
    isbn = "978-1-84816-726-1",
    publisher = "World Scientific",
    address = "Singapore",
    year = "2013"
}

@book{ponomarev2017gauge,
	year = 2017,
	publisher = {Nauka, Moscow},
	author = "{V. N. Ponomarev, A. O. Barvinsky, and Yu. N. Obukhov}",
	title = {Gauge Approach and Quantization Methods in Gravity Theory}}

@incollection{obukhov2023poincare,
  title={Poincar{\'e} gauge gravity primer},
  author={Obukhov, Yuri N},
  booktitle={Modified and Quantum Gravity: From Theory to Experimental Searches on All Scales},
  pages={105--143},
  year={2023},
  publisher={Springer}
}

@article{Cabral:2020fax,
    author = "Cabral, Francisco and Lobo, Francisco S.N. and Rubiera-Garcia, Diego",
    title = "{Fundamental Symmetries and Spacetime Geometries in Gauge Theories of Gravity: Prospects for Unified Field Theories}",
    eprint = "2012.06356",
    archivePrefix = "arXiv",
    primaryClass = "gr-qc",
    doi = "10.3390/universe6120238",
    journal = "Universe",
    volume = "6",
    number = "12",
    pages = "238",
    year = "2020"
}

@article{LIGOScientific:2017ycc,
    author = "{B. P. Abbott et al. (LIGO Scientific and Virgo Collaborations)}",
    title = "{GW170814: A Three-Detector Observation of Gravitational Waves from a Binary Black Hole Coalescence}",
    eprint = "1709.09660",
    archivePrefix = "arXiv",
    primaryClass = "gr-qc",
    doi = "10.1103/PhysRevLett.119.141101",
    journal = "Phys. Rev. Lett.",
    volume = "119",
    number = "14",
    pages = "141101",
    year = "2017"
}

@article{TheLIGOScientific:2017qsa,
    author = "{B. P. Abbott et al. (LIGO Scientific and Virgo Collaborations)}",
    title = "{GW170817: Observation of Gravitational Waves from a Binary Neutron Star Inspiral}",
    eprint = "1710.05832",
    archivePrefix = "arXiv",
    primaryClass = "gr-qc",
    reportNumber = "LIGO-P170817",
    doi = "10.1103/PhysRevLett.119.161101",
    journal = "Phys. Rev. Lett.",
    volume = "119",
    number = "16",
    pages = "161101",
    year = "2017"
}

@article{LIGOScientific:2018dkp,
    author = "{B. P. Abbott et al. (LIGO Scientific and Virgo Collaborations)}",
    title = "{Tests of General Relativity with GW170817}",
    eprint = "1811.00364",
    archivePrefix = "arXiv",
    primaryClass = "gr-qc",
    reportNumber = "LIGO-P1800059",
    doi = "10.1103/PhysRevLett.123.011102",
    journal = "Phys. Rev. Lett.",
    volume = "123",
    number = "1",
    pages = "011102",
    year = "2019"
}

@article{Brinkmann:1925fr,
    author = "Brinkmann, H. W.",
    title = "{Einstein spaces which are mapped conformally on each other}",
    doi = "10.1007/BF01208647",
    journal = "Math. Ann.",
    volume = "94",
    pages = "119--145",
    year = "1925"
}

@article{Einstein:1937qu,
    author = "Einstein, Albert and Rosen, N.",
    title = "{On Gravitational waves}",
    doi = "10.1016/S0016-0032(37)90583-0",
    journal = "J. Franklin Inst.",
    volume = "223",
    pages = "43--54",
    year = "1937"
}

@book{misner1973gravitation,
  title={Gravitation},
  author={Misner, C.W. and Thorne, K.S. and Wheeler, J.A.},
  number={pt. 3},
  isbn={9780716703440},
  lccn={78156043},
  series={Gravitation},
  year={1973},
  publisher={W. H. Freeman}
}

@article{blanco2013structure,
  title="{Structure of second-order symmetric Lorentzian manifolds}",
  author={Blanco, Oihane F and S{\'a}nchez, Miguel and Senovilla, Jos{\'e} M. M.},
  journal={Journal of the European Mathematical Society (EMS Publishing)},
  volume={15},
  number={2},
  year={2013}
}

@article{Lasenby:2019gmi,
  author = "Lasenby, Anthony N.",
  title = "{Geometric Algebra, Gravity and Gravitational Waves}",
  eprint = "1912.05960",
  archivePrefix = "arXiv",
  primaryClass = "gr-qc",
  doi = "10.1007/s00006-019-0991-y",
  journal = "Adv. Appl. Clifford Algebras",
  volume = "29",
  number = "4",
  pages = "79",
  year = "2019"
}

@book{Stephani:2003tm,
    author = "Stephani, Hans and Kramer, D. and MacCallum, Malcolm A.H. and Hoenselaers, Cornelius and Herlt, Eduard",
    title = "{Exact solutions of Einstein's field equations}",
    doi = "10.1017/CBO9780511535185",
    isbn = "978-0-521-46702-5, 978-0-511-05917-9",
    publisher = "Cambridge Univ. Press",
    address = "Cambridge",
    series = "Cambridge Monographs on Mathematical Physics",
    year = "2003"
}

@book{Griffiths:2009dfa,
    author = "Griffiths, Jerry B. and Podolský, Jiri",
    title = "{Exact Space-Times in Einstein's General Relativity}",
    doi = "10.1017/CBO9780511635397",
    isbn = "978-1-139-48116-8",
    publisher = "Cambridge University Press",
    address = "Cambridge",
    series = "Cambridge Monographs on Mathematical Physics",
    year = "2009"
}

@article{Bahamonde:2023piz,
    author = "Bahamonde, Sebastian and Gigante Valcarcel, Jorge",
    title = "{Algebraic classification of the gravitational field in Weyl-Cartan spacetimes}",
    eprint = "2305.05501",
    archivePrefix = "arXiv",
    primaryClass = "gr-qc",
    doi = "10.1103/PhysRevD.108.044037",
    journal = "Phys. Rev. D",
    volume = "108",
    number = "4",
    pages = "044037",
    year = "2023"
}

@article{Bahamonde:2024svi,
    author = "Bahamonde, Sebastian and Gigante Valcarcel, Jorge and Senovilla, Jos{\'e} M. M.",
    title = "{Algebraic classification of the gravitational field in general metric-affine geometries}",
    eprint = "2409.07153",
    archivePrefix = "arXiv",
    primaryClass = "gr-qc",
    doi = "10.1103/PhysRevD.110.124053",
    journal = "Phys. Rev. D",
    volume = "110",
    number = "12",
    pages = "124053",
    year = "2024"
}

@article{Hehl:1994ue,
      author         = "Hehl, Friedrich W. and McCrea, J. Dermott and Mielke,
                        Eckehard W. and Ne'eman, Yuval",
      title          = "{Metric-Affine gauge theory of gravity: Field equations,
                        Noether identities, world spinors, and breaking of
                        dilation invariance}",
      journal        = "Phys. Rept.",
      volume         = "258",
      year           = "1995",
      pages          = "1-171",
      doi            = "10.1016/0370-1573(94)00111-F",
      eprint         = "gr-qc/9402012",
      archivePrefix  = "arXiv",
      primaryClass   = "gr-qc",
      reportNumber   = "TAUP-N192-94, TAUP-192-94",
      SLACcitation   = "%%CITATION = GR-QC/9402012;%%"
}

@inproceedings{Gronwald:1995em,
    author = "Gronwald, Frank and Hehl, Friedrich W.",
    title = "{On the gauge aspects of gravity}",
    booktitle = "{International School of Cosmology and Gravitation: 14th Course: Quantum Gravity}",
    eprint = "gr-qc/9602013",
    archivePrefix = "arXiv",
    pages = "148--198",
    month = "5",
    year = "1995"
}

@article{McCrea:1992wa,
    author = "McCrea, J. D.",
    doi = "10.1088/0264-9381/9/2/018",
    journal = "Class. Quant. Grav.",
    pages = "553--568",
    title = "{Irreducible decompositions of non-metricity, torsion, curvature and Bianchi identities in metric-affine spacetimes}",
    volume = "9",
    year = "1992"
}

@article{Hehl:1976kj,
    author = "Hehl, F.W. and von der Heyde, P. and Kerlick, G.D. and Nester, J.M.",
    doi = "10.1103/RevModPhys.48.393",
    journal = "Rev. Mod. Phys.",
    pages = "393--416",
    title = "{General Relativity with Spin and Torsion: Foundations and Prospects}",
    volume = "48",
    year = "1976"
}

@article{Obukhov:1987tz,
    author = "Obukhov, Yu.N. and Ponomarev, V.N. and Zhytnikov, V.V.",
    doi = "10.1007/BF00763457",
    journal = "Gen. Rel. Grav.",
    pages = "1107--1142",
    title = "{Quadratic Poincar\'e Gauge Theory of Gravity: A Comparison With the General Relativity Theory}",
    volume = "21",
    year = "1989"
}

@article{Neville:1978bk,
    author = "Neville, Donald E.",
    title = "{Gravity Lagrangian with ghost-free curvature-squared terms}",
    reportNumber = "Print-78-1155 (TEMPLE)",
    doi = "10.1103/PhysRevD.18.3535",
    journal = "Phys. Rev. D",
    volume = "18",
    pages = "3535",
    year = "1978"
}

@article{Sezgin:1979zf,
    author = "Sezgin, E. and van Nieuwenhuizen, P.",
    title = "{New ghost-free gravity Lagrangians with propagating torsion}",
    reportNumber = "ITP-SB-79-97",
    doi = "10.1103/PhysRevD.21.3269",
    journal = "Phys. Rev. D",
    volume = "21",
    pages = "3269",
    year = "1980"
}

@article{Sezgin:1981xs,
    author = "Sezgin, E.",
    doi = "10.1103/PhysRevD.24.1677",
    journal = "Phys. Rev. D",
    pages = "1677--1680",
    reportNumber = "ITP-SB-80-59",
    title = "{Class of ghost-free gravity Lagrangians with massive or massless propagating torsion}",
    volume = "24",
    year = "1981"
}

@article{Miyamoto:1983bf,
    author = "Miyamoto, S. and Nakano, T. and Ohtani, T. and Tamura, Y.",
    title = "{Linear approximation for the massless Lorentz gauge field}",
    doi = "10.1143/PTP.69.1236",
    journal = "Prog. Theor. Phys.",
    volume = "69",
    pages = "1236--1240",
    year = "1983"
}

@article{Fukui:1984gn,
    author = "Fukui, Masayasu and Masukawa, Junnichi",
    title = "{Massless torsion fields. II. The case $\alpha + 2a/3 = 0$}",
    reportNumber = "OUAM 84-6-1",
    doi = "10.1143/PTP.73.75",
    journal = "Prog. Theor. Phys.",
    volume = "73",
    pages = "75",
    year = "1985"
}

@article{Fukuma:1984cz,
    author = "Fukuma, Kazumi and Miyamoto, Shikao and Nakano, Tadao and Ohtani, Teruya and Tamura, Yoshinobu",
    doi = "10.1143/PTP.73.874",
    journal = "Prog. Theor. Phys.",
    pages = "874",
    reportNumber = "Print-84-0837 (OSAKA CITY)",
    title = "{Massless Lorentz gauge field consistent with Einstein’s gravitation theory. The case $\alpha + 3a/2 = \beta - 2a/3 = \gamma + 3a/2 = 0$}",
    volume = "73",
    year = "1985"
}

@article{Battiti:1985mu,
    author = "Battiti, R. and Toller, M.",
    doi = "10.1007/BF02746948",
    journal = "Lett. Nuovo Cim.",
    pages = "35",
    reportNumber = "Print-85-0691 (TRENTO)",
    title = "{Zero-mass normal modes in linearized Poincaré gauge theories}",
    volume = "44",
    year = "1985"
}

@article{Kuhfuss:1986rb,
    author = "Kuhfuss, R. and Nitsch, J.",
    doi = "10.1007/BF00763447",
    journal = "Gen. Rel. Grav.",
    pages = "1207",
    reportNumber = "MPA-223",
    title = "{Propagating Modes in Gauge Field Theories of Gravity}",
    volume = "18",
    year = "1986"
}

@article{Blagojevic:1986dm,
    author = "Blagojevi\'c, M. and Vasili\'c, M.",
    doi = "10.1103/PhysRevD.35.3748",
    journal = "Phys. Rev. D",
    pages = "3748",
    reportNumber = "Print-86-0954 (BELGRADE)",
    title = "{Extra gauge symmetries in a weak-field approximation of an $R + T^2 + R^2$ theory of gravity}",
    volume = "35",
    year = "1987"
}

@article{Baikov:1992uh,
    author = "Baikov, P. and Hayashi, M. and Nelipa, N. and Ostapchenko, S.",
    title = "{Ghost and tachyon free gauge invariant, Poincaré, affine and projective Lagrangians}",
    reportNumber = "SMC-12-91",
    doi = "10.1007/BF00759092",
    journal = "Gen. Rel. Grav.",
    volume = "24",
    pages = "867--880",
    year = "1992"
}

@article{Yo:1999ex,
    author = "Yo, Hwei-Jang and Nester, James M.",
    title = "{Hamiltonian analysis of Poincaré gauge theory scalar modes}",
    eprint = "gr-qc/9902032",
    archivePrefix = "arXiv",
    doi = "10.1142/S021827189900033X",
    journal = "Int. J. Mod. Phys. D",
    volume = "08",
    pages = "459--479",
    year = "1999"
}

@article{Yo:2001sy,
    author = "Yo, Hwei-Jang and Nester, James M.",
    archivePrefix = "arXiv",
    doi = "10.1142/S0218271802001998",
    eprint = "gr-qc/0112030",
    journal = "Int. J. Mod. Phys. D",
    pages = "747--780",
    title = "{Hamiltonian analysis of Poincar\'e gauge theory: higher spin modes}",
    volume = "11",
    year = "2002"
}

@article{Lin:2018awc,
    author = "Lin, Yun-Cherng and Hobson, Michael P. and Lasenby, Anthony N.",
    title = "{Ghost and tachyon free Poincar\'e gauge theories: A systematic approach}",
    eprint = "1812.02675",
    archivePrefix = "arXiv",
    primaryClass = "gr-qc",
    doi = "10.1103/PhysRevD.99.064001",
    journal = "Phys. Rev. D",
    volume = "99",
    number = "6",
    pages = "064001",
    year = "2019"
}

@article{BeltranJimenez:2019acz,
    author = "Beltr\'an Jim\'enez, Jose and Delhom, Adria",
    title = "{Ghosts in metric-affine higher order curvature gravity}",
    eprint = "1901.08988",
    archivePrefix = "arXiv",
    primaryClass = "gr-qc",
    doi = "10.1140/epjc/s10052-019-7149-x",
    journal = "Eur. Phys. J. C",
    volume = "79",
    number = "8",
    pages = "656",
    year = "2019"
}

@article{Jimenez:2019qjc,
    author = "Beltrán Jiménez, Jose and Maldonado Torralba, Francisco José",
    title = "{Revisiting the stability of quadratic Poincaré gauge gravity}",
    eprint = "1910.07506",
    archivePrefix = "arXiv",
    primaryClass = "gr-qc",
    doi = "10.1140/epjc/s10052-020-8163-8",
    journal = "Eur. Phys. J. C",
    volume = "80",
    number = "7",
    pages = "611",
    year = "2020"
}

@article{Percacci:2020ddy,
    author = "Percacci, R. and Sezgin, E.",
    title = "{New class of ghost- and tachyon-free metric affine gravities}",
    eprint = "1912.01023",
    archivePrefix = "arXiv",
    primaryClass = "hep-th",
    reportNumber = "MI-TH-1941",
    doi = "10.1103/PhysRevD.101.084040",
    journal = "Phys. Rev. D",
    volume = "101",
    number = "8",
    pages = "084040",
    year = "2020"
}

@article{BeltranJimenez:2020sqf,
    author = "Beltr\'an Jim\'enez, Jose and Delhom, Adri\`a",
    title = "{Instabilities in metric-affine theories of gravity with higher order curvature terms}",
    eprint = "2004.11357",
    archivePrefix = "arXiv",
    primaryClass = "gr-qc",
    doi = "10.1140/epjc/s10052-020-8143-z",
    journal = "Eur. Phys. J. C",
    volume = "80",
    number = "6",
    pages = "585",
    year = "2020"
}

@article{Lin:2020phk,
    author = "Lin, Yun-Cherng and Hobson, Michael P. and Lasenby, Anthony N.",
    title = "{Ghost and tachyon free Weyl gauge theories: A systematic approach}",
    eprint = "2005.02228",
    archivePrefix = "arXiv",
    primaryClass = "gr-qc",
    doi = "10.1103/PhysRevD.104.024034",
    journal = "Phys. Rev. D",
    volume = "104",
    number = "2",
    pages = "024034",
    year = "2021"
}

@article{Marzo:2021iok,
    author = "Marzo, Carlo",
    title = "{Radiatively stable ghost and tachyon freedom in metric affine gravity}",
    eprint = "2110.14788",
    archivePrefix = "arXiv",
    primaryClass = "hep-th",
    doi = "10.1103/PhysRevD.106.024045",
    journal = "Phys. Rev. D",
    volume = "106",
    number = "2",
    pages = "024045",
    year = "2022"
}

@article{Baldazzi:2021kaf,
    author = "Baldazzi, A. and Melichev, O. and Percacci, R.",
    title = "{Metric-Affine Gravity as an effective field theory}",
    eprint = "2112.10193",
    archivePrefix = "arXiv",
    primaryClass = "gr-qc",
    doi = "10.1016/j.aop.2022.168757",
    journal = "Annals Phys.",
    volume = "438",
    pages = "168757",
    year = "2022"
}

@article{Jimenez-Cano:2022sds,
    author = "Jim\'enez-Cano, Alejandro and Maldonado Torralba, Francisco Jos\'e",
    title = "{Vector stability in quadratic metric-affine theories}",
    eprint = "2205.05674",
    archivePrefix = "arXiv",
    primaryClass = "gr-qc",
    doi = "10.1088/1475-7516/2022/09/044",
    journal = "JCAP",
    volume = "09",
    pages = "044",
    year = "2022"
}

@article{Barker:2024dhb,
    author = "Barker, Will and Zell, Sebastian",
    title = "{Consistent particle physics in metric-affine gravity from extended projective symmetry}",
    eprint = "2402.14917",
    archivePrefix = "arXiv",
    primaryClass = "hep-th",
    month = "2",
    year = "2024"
}

@article{Marzo:2024pyn,
    author = "Marzo, Carlo",
    title = "{Can MAG be a predictive EFT? Radiative stability and ghost resurgence in massive vector models}",
    eprint = "2403.15003",
    archivePrefix = "arXiv",
    primaryClass = "hep-th",
    doi = "10.1088/1361-6382/adc9f1",
    journal = "Class. Quant. Grav.",
    volume = "42",
    number = "9",
    pages = "095007",
    year = "2025"
}

@article{Bahamonde:2024sqo,
    author = "Bahamonde, Sebastian and Gigante Valcarcel, Jorge",
    title = "{Stability of Poincar\'e gauge theory with cubic order invariants}",
    eprint = "2402.08937",
    archivePrefix = "arXiv",
    primaryClass = "gr-qc",
    doi = "10.1103/PhysRevD.109.104075",
    journal = "Phys. Rev. D",
    volume = "109",
    number = "10",
    pages = "104075",
    year = "2024"
}

@article{Bahamonde:2024efl,
    author = "Bahamonde, Sebastian and Gigante Valcarcel, Jorge",
    title = "{Stability in cubic metric-affine gravity}",
    eprint = "2411.12954",
    archivePrefix = "arXiv",
    primaryClass = "gr-qc",
    doi = "10.1103/PhysRevD.111.084058",
    journal = "Phys. Rev. D",
    volume = "111",
    number = "8",
    pages = "084058",
    year = "2025"
}

@article{adamowicz1980plane,
  title={Plane waves in gauge theories of gravitation},
  author={Adamowicz, W},
  journal={General Relativity and Gravitation},
  volume={12},
  number={9},
  pages={677},
  year={1980},
  publisher={Springer}
}

@article{Babourova:1998ct,
    author = "Babourova, O. V. and Frolov, B. N. and Klimova, E. A.",
    title = "{Plane torsion waves in quadratic gravitational theories}",
    eprint = "gr-qc/9805005",
    archivePrefix = "arXiv",
    doi = "10.1088/0264-9381/16/4/005",
    journal = "Class. Quant. Grav.",
    volume = "16",
    pages = "1149--1162",
    year = "1999"
}

@article{Garcia:2000yi,
    author = "García, Alberto and Macías, Alfredo and Puetzfeld, Dirk and Socorro, Jose",
    title = "{Plane fronted waves in metric-affine gravity}",
    eprint = "gr-qc/0005038",
    archivePrefix = "arXiv",
    doi = "10.1103/PhysRevD.62.044021",
    journal = "Phys. Rev. D",
    volume = "62",
    pages = "044021",
    year = "2000"
}

@article{Obukhov:2006gy,
    author = "Obukhov, Yuri N.",
    title = "{Plane waves in metric-affine gravity}",
    eprint = "gr-qc/0601074",
    archivePrefix = "arXiv",
    doi = "10.1103/PhysRevD.73.024025",
    journal = "Phys. Rev. D",
    volume = "73",
    pages = "024025",
    year = "2006"
}

@article{Blagojevic:2017wzf,
    author = "Blagojevi\'c, M. and Cvetkovi\'c, B.",
    title = "{Generalized pp waves in Poincar\'e gauge theory}",
    eprint = "1702.04367",
    archivePrefix = "arXiv",
    primaryClass = "gr-qc",
    doi = "10.1103/PhysRevD.95.104018",
    journal = "Phys. Rev. D",
    volume = "95",
    number = "10",
    pages = "104018",
    year = "2017"
}

@article{Obukhov:2017pxa,
    author = "Obukhov, Yuri N.",
    title = "{Gravitational waves in Poincar{\'e} gauge gravity theory}",
    eprint = "1702.05185",
    archivePrefix = "arXiv",
    primaryClass = "gr-qc",
    doi = "10.1103/PhysRevD.95.084028",
    journal = "Phys. Rev. D",
    volume = "95",
    number = "8",
    pages = "084028",
    year = "2017"
}

@article{Blagojevic:2017ssv,
    author = "Blagojevi\'c, Milutin and Cvetkovi\'c, Branislav and Obukhov, Yuri N.",
    title = "{Generalized plane waves in Poincar\'e gauge theory of gravity}",
    eprint = "1708.08766",
    archivePrefix = "arXiv",
    primaryClass = "gr-qc",
    doi = "10.1103/PhysRevD.96.064031",
    journal = "Phys. Rev. D",
    volume = "96",
    number = "6",
    pages = "064031",
    year = "2017"
}

@article{Jimenez-Cano:2020lea,
    author = "Jim\'enez-Cano, Alejandro and Obukhov, Yuri N.",
    title = "{Gravitational waves in metric-affine gravity theory}",
    eprint = "2010.14528",
    archivePrefix = "arXiv",
    primaryClass = "gr-qc",
    doi = "10.1103/PhysRevD.103.024018",
    journal = "Phys. Rev. D",
    volume = "103",
    number = "2",
    pages = "024018",
    year = "2021"
}

@article{Jimenez-Cano:2022arz,
    author = "Jim{\'e}nez-Cano, Alejandro",
    title = "{Review of gravitational wave solutions in quadratic metric-affine gauge gravity}",
    eprint = "2203.03936",
    archivePrefix = "arXiv",
    primaryClass = "gr-qc",
    doi = "10.1142/S0219887822400047",
    journal = "Int. J. Geom. Meth. Mod. Phys.",
    volume = "19",
    number = "Supp01",
    pages = "2240004",
    year = "2022"
}

@article{Khodadi:2026zoi,
    author = "Khodadi, Mohsen and Saridakis, Emmanuel N.",
    title = "{Atomic clocks and gravitational waves as probes of non-metricity}",
    eprint = "2601.19407",
    archivePrefix = "arXiv",
    primaryClass = "gr-qc",
    month = "1",
    year = "2026"
}

@article{Pravda:2002us,
    author = "Pravda, Vojtech and Pravdová, A. and Coley, A. and Milson, R.",
    title = "{All space-times with vanishing curvature invariants}",
    eprint = "gr-qc/0209024",
    archivePrefix = "arXiv",
    doi = "10.1088/0264-9381/19/23/318",
    journal = "Class. Quant. Grav.",
    volume = "19",
    pages = "6213--6236",
    year = "2002"
}

@article{Pirani:1956tn,
    author = "Pirani, F. A. E.",
    title = "{On the physical significance of the Riemann tensor}",
    doi = "10.1007/s10714-009-0787-9",
    journal = "Acta Phys. Polon.",
    volume = "15",
    pages = "389--405",
    year = "1956"
}

@article{Pirani:1956wr,
    author = "Pirani, F. A. E.",
    title = "{Invariant formulation of gravitational radiation theory}",
    doi = "10.1103/PhysRev.105.1089",
    journal = "Phys. Rev.",
    volume = "105",
    pages = "1089--1099",
    year = "1957"
}

@article{Szekeres:1965ux,
    author = "Szekeres, P.",
    title = "{The gravitational compass}",
    doi = "10.1063/1.1704788",
    journal = "J. Math. Phys.",
    volume = "6",
    pages = "1387--1391",
    year = "1965"
}

@article{Bicak:1999hb,
    author = "Bi\v{c}ák, Jiri and Podolský, Jiri",
    title = "{Gravitational waves in vacuum space-times with cosmological constant. 2. Deviation of geodesics and interpretation of nontwisting type N solutions}",
    eprint = "gr-qc/9907049",
    archivePrefix = "arXiv",
    doi = "10.1063/1.532982",
    journal = "J. Math. Phys.",
    volume = "40",
    pages = "4506--4517",
    year = "1999"
}

@article{Podolsky:2012he,
    author = "Podolský, Jiri and \v{S}varc, Robert",
    title = "{Interpreting spacetimes of any dimension using geodesic deviation}",
    eprint = "1201.4790",
    archivePrefix = "arXiv",
    primaryClass = "gr-qc",
    doi = "10.1103/PhysRevD.85.044057",
    journal = "Phys. Rev. D",
    volume = "85",
    pages = "044057",
    year = "2012"
}

@article{deReyNeto:2003mp,
    author = "de Rey Neto, Edgard C.",
    title = "{Geodesic deviation in pp-wave spacetimes of quadratic curvature gravity}",
    eprint = "gr-qc/0309128",
    archivePrefix = "arXiv",
    doi = "10.1103/PhysRevD.68.124013",
    journal = "Phys. Rev. D",
    volume = "68",
    pages = "124013",
    year = "2003"
}

@article{Takeda:2021hgo,
    author = "Takeda, Hiroki and Morisaki, Soichiro and Nishizawa, Atsushi",
    title = "{Search for scalar-tensor mixed polarization modes of gravitational waves}",
    eprint = "2105.00253",
    archivePrefix = "arXiv",
    primaryClass = "gr-qc",
    reportNumber = "LIGO-P2100137",
    doi = "10.1103/PhysRevD.105.084019",
    journal = "Phys. Rev. D",
    volume = "105",
    number = "8",
    pages = "084019",
    year = "2022"
}

@article{Wu:2024yno,
    author = "Wu, Jie and Li, Jin",
    title = "{Prospects of constraining on the polarizations of gravitational waves from binary black holes using space- and ground-based detectors}",
    eprint = "2407.13590",
    archivePrefix = "arXiv",
    primaryClass = "gr-qc",
    doi = "10.1103/PhysRevD.110.084057",
    journal = "Phys. Rev. D",
    volume = "110",
    number = "8",
    pages = "084057",
    year = "2024"
}

@article{Liang:2024sfn,
    author = "Liang, Dicong and Chen, Siyuan and Zhang, Chao and Shao, Lijing",
    title = "{Unveiling the existence of nontensorial gravitational-wave polarizations from individual supermassive black hole binaries with pulsar timing arrays}",
    eprint = "2404.16680",
    archivePrefix = "arXiv",
    primaryClass = "gr-qc",
    doi = "10.1103/PhysRevD.110.084040",
    journal = "Phys. Rev. D",
    volume = "110",
    number = "8",
    pages = "084040",
    year = "2024"
}

\end{document}